\documentclass[12pt, twocolumn, trackchanges, twocolappendix, ]{aastex631}

\usepackage{color}
\usepackage{amsmath}
\usepackage{float}
\usepackage{gensymb}
\usepackage{upgreek}
\usepackage{longtable}
\usepackage{threeparttablex}
\usepackage{booktabs}
\usepackage{placeins}
\usepackage{url}
\usepackage{enumitem} 

\submitjournal{\apj}

\begin{document}
\title{Morphology of 35 Repeating Fast Radio Burst Sources at Microsecond Time Scales with CHIME/FRB}
\shorttitle{}
\shortauthors{}

\author[0000-0002-8376-1563]{Alice P.~Curtin}
  \affiliation{Department of Physics, McGill University, 3600 rue University, Montr\'eal, QC H3A 2T8, Canada}
  \affiliation{Trottier Space Institute, McGill University, 3550 rue University, Montr\'eal, QC H3A 2A7, Canada}
\author[0000-0003-3154-3676]{Ketan R Sand}
  \affiliation{Department of Physics, McGill University, 3600 rue University, Montr\'eal, QC H3A 2T8, Canada}
  \affiliation{Trottier Space Institute, McGill University, 3550 rue University, Montr\'eal, QC H3A 2A7, Canada}
\author[0000-0002-4795-697X]{Ziggy Pleunis}
  \affiliation{Anton Pannekoek Institute for Astronomy, University of Amsterdam, Science Park 904, 1098 XH Amsterdam, The Netherlands}
  \affiliation{ASTRON, Netherlands Institute for Radio Astronomy, Oude Hoogeveensedijk 4, 7991 PD Dwingeloo, The Netherlands}
\author[0009-0009-0938-1595]{Naman Jain}
  \affiliation{McGill}
\author[0000-0001-9345-0307]{Victoria Kaspi}
  \affiliation{Department of Physics, McGill University, 3600 rue University, Montr\'eal, QC H3A 2T8, Canada}
  \affiliation{Trottier Space Institute, McGill University, 3550 rue University, Montr\'eal, QC H3A 2A7, Canada}
\author[0000-0002-2551-7554]{Daniele Michilli}
  \affiliation{MIT Kavli Institute for Astrophysics and Space Research, Massachusetts Institute of Technology, 77 Massachusetts Ave, Cambridge, MA 02139, USA}
  \affiliation{Department of Physics, Massachusetts Institute of Technology, 77 Massachusetts Ave, Cambridge, MA 02139, USA}
\author[0000-0001-8384-5049]{Emmanuel Fonseca}
  \affiliation{Department of Physics and Astronomy, West Virginia University, PO Box 6315, Morgantown, WV 26506, USA }
  \affiliation{Center for Gravitational Waves and Cosmology, West Virginia University, Chestnut Ridge Research Building, Morgantown, WV 26505, USA}
\author[0000-0002-6823-2073]{Kaitlyn Shin}
  \affiliation{MIT Kavli Institute for Astrophysics and Space Research, Massachusetts Institute of Technology, 77 Massachusetts Ave, Cambridge, MA 02139, USA}
  \affiliation{Department of Physics, Massachusetts Institute of Technology, 77 Massachusetts Ave, Cambridge, MA 02139, USA}
\author[0000-0003-0510-0740]{Kenzie Nimmo}
  \affiliation{MIT Kavli Institute for Astrophysics and Space Research, Massachusetts Institute of Technology, 77 Massachusetts Ave, Cambridge, MA 02139, USA}
\author[0000-0002-1800-8233]{Charanjot Brar}
  \affiliation{National Research Council of Canada, Herzberg Astronomy and Astrophysics, 5071 West Saanich Road, Victoria, BC V9E2E7, Canada}
\author[0000-0003-4098-5222]{Fengqiu Adam Dong}
  \affiliation{Department of Physics and Astronomy, University of British Columbia, 6224 Agricultural Road, Vancouver, BC V6T 1Z1 Canada}
  \affiliation{National Radio Astronomy Observatory, 520 Edgemont Rd, Charlottesville, VA 22903, USA}
\author[0000-0003-3734-8177]{Gwendolyn M.~Eadie}
  \affiliation{David A. Dunlap Department of Astronomy and Astrophysics, 50 St. George Street, University of Toronto, ON M5S 3H4, Canada}
  \affiliation{Department of Statistical Sciences}
\author[0000-0002-3382-9558]{B.~M.~Gaensler}
  \affiliation{Department of Astronomy and Astrophysics, University of California Santa Cruz, 1156 High Street, Santa Cruz, CA 95060, USA}
  \affiliation{Dunlap Institute for Astronomy and Astrophysics, 50 St. George Street, University of Toronto, ON M5S 3H4, Canada}
  \affiliation{David A. Dunlap Department of Astronomy and Astrophysics, 50 St. George Street, University of Toronto, ON M5S 3H4, Canada}
\author[0000-0002-3654-4662]{Antonio Herrera-Martin}
  \affiliation{David A. Dunlap Department of Astronomy and Astrophysics, 50 St. George Street, University of Toronto, ON M5S 3H4, Canada}
  \affiliation{Department of Statistical Science, University of Toronto, Ontario Power Building, 700 University Avenue, 9th Floor, Toronto, ON M5G 1Z5, Toronto, Ontario, Canada}
\author[0000-0003-2405-2967]{Adaeze L.~Ibik}
  \affiliation{Dunlap Institute for Astronomy and Astrophysics, 50 St. George Street, University of Toronto, ON M5S 3H4, Canada}
  \affiliation{David A. Dunlap Department of Astronomy and Astrophysics, 50 St. George Street, University of Toronto, ON M5S 3H4, Canada}
\author[0000-0003-3457-4670]{Ronniy C.~Joseph}
  \affiliation{Department of Physics, McGill University, 3600 rue University, Montr\'eal, QC H3A 2T8, Canada}
  \affiliation{Trottier Space Institute, McGill University, 3550 rue University, Montr\'eal, QC H3A 2A7, Canada}
\author[0000-0003-4810-7803]{Jane Kaczmarek}
  \affiliation{CSIRO Space \& Astronomy, Parkes Observatory, P.O. Box 276, Parkes NSW 2870, Australia}
\author[0000-0002-4209-7408]{Calvin Leung}
  \affiliation{Department of Astronomy, University of California, Berkeley, CA 94720, United States}
  \affiliation{NASA Hubble Fellowship Program~(NHFP) Einstein Fellow}
\author[0000-0002-7164-9507]{Robert Main}
  \affiliation{Department of Physics, McGill University, 3600 rue University, Montr\'eal, QC H3A 2T8, Canada}
\author[0000-0002-4279-6946]{Kiyoshi W.~Masui}
  \affiliation{MIT Kavli Institute for Astrophysics and Space Research, Massachusetts Institute of Technology, 77 Massachusetts Ave, Cambridge, MA 02139, USA}
  \affiliation{Department of Physics, Massachusetts Institute of Technology, 77 Massachusetts Ave, Cambridge, MA 02139, USA}
\author[0000-0001-7348-6900]{Ryan Mckinven}
  \affiliation{Department of Physics, McGill University, 3600 rue University, Montr\'eal, QC H3A 2T8, Canada}
  \affiliation{Trottier Space Institute, McGill University, 3550 rue University, Montr\'eal, QC H3A 2A7, Canada}
\author[0000-0002-0772-9326]{Juan Mena-Parra}
  \affiliation{Dunlap Institute for Astronomy and Astrophysics, 50 St. George Street, University of Toronto, ON M5S 3H4, Canada}
  \affiliation{David A. Dunlap Department of Astronomy and Astrophysics, 50 St. George Street, University of Toronto, ON M5S 3H4, Canada}
\author[0000-0002-3616-5160]{Cherry Ng}
  \affiliation{Laboratoire de Physique et Chimie de l’Environnement et de l’Espace (LPC2E) UMR7328, Université d’Orléans, CNRS, F-45071 Orléans, France}
\author[0000-0002-8897-1973]{Ayush Pandhi}
  \affiliation{David A. Dunlap Department of Astronomy and Astrophysics, 50 St. George Street, University of Toronto, ON M5S 3H4, Canada}
  \affiliation{Dunlap Institute for Astronomy and Astrophysics, 50 St. George Street, University of Toronto, ON M5S 3H4, Canada}
\author[0000-0002-8912-0732]{Aaron B.~Pearlman}
  \affiliation{Department of Physics, McGill University, 3600 rue University, Montr\'eal, QC H3A 2T8, Canada}
  \affiliation{Trottier Space Institute, McGill University, 3550 rue University, Montr\'eal, QC H3A 2A7, Canada}
\author[0000-0001-7694-6650]{Masoud Rafiei-Ravandi}
  \affiliation{Department of Physics, McGill University, 3600 rue University, Montr\'eal, QC H3A 2T8, Canada}
\author[0000-0002-4623-5329]{Mawson W.~Sammons}
  \affiliation{Department of Physics, McGill University, 3600 rue University, Montr\'eal, QC H3A 2T8, Canada}
  \affiliation{Trottier Space Institute, McGill University, 3550 rue University, Montr\'eal, QC H3A 2A7, Canada}
\author[0000-0002-7374-7119]{Paul Scholz}
  \affiliation{Department of Physics and Astronomy, York University, 4700 Keele Street, Toronto, ON MJ3 1P3, Canada}
  \affiliation{Dunlap Institute for Astronomy and Astrophysics, 50 St. George Street, University of Toronto, ON M5S 3H4, Canada}
\author[0000-0002-2088-3125]{Kendrick Smith}
  \affiliation{Perimeter Institute of Theoretical Physics, 31 Caroline Street North, Waterloo, ON N2L 2Y5, Canada}
\author[0000-0001-9784-8670]{Ingrid Stairs}
  \affiliation{Department of Physics and Astronomy, University of British Columbia, 6224 Agricultural Road, Vancouver, BC V6T 1Z1 Canada}
\newcommand{\allacks}{
A.P.C is a Vanier Canada Graduate Scholar
K.R.S. acknowledges support from Fonds de Recherche du Quebec – Nature et Technologies (FRQNT) Doctoral Research Award
Z.P. is supported by an NWO Veni fellowship (VI.Veni.222.295).
V.M.K. holds the Lorne Trottier Chair in Astrophysics \& Cosmology, a Distinguished James McGill Professorship, and receives support from an NSERC Discovery grant (RGPIN 228738-13).
E.F. is supported by the National Science Foundation grant AST-2407399.
K.S. is supported by the NSF Graduate Research Fellowship Program.
K.N. is an MIT Kavli Fellow.
F.A.D is supported by a Jansky Fellowship
G.M.E. holds a Collaborative Research Team grant from the Canadian Statistical Sciences Institute (CANSSI), which is supported by Natural Sciences and Engineering Research Council of Canada (NSERC), and an NSERC Discovery Grant RGPIN2020-04554.
A.H.M. is supported by a Collaborative Research Team grant from the Canadian Statistical Sciences Institute(CANSSI)
C. L. is supported by NASA through the NASA Hubble Fellowship grant HST-HF2-51536.001-A awarded by the Space Telescope Science Institute, which is operated by the Association of Universities for Research in Astronomy, Inc., under NASA contract NAS5-26555.
K.W.M. holds the Adam J. Burgasser Chair in Astrophysics.
A.P. is funded by the NSERC Canada Graduate Scholarships -- Doctoral program.
A.B.P. is a Banting Fellow, a McGill Space Institute~(MSI) Fellow, and a Fonds de Recherche du Quebec -- Nature et Technologies~(FRQNT) postdoctoral fellow.
M.W.S. acknowledges support from the Trottier Space Institute Fellowship program.
P.S. acknowledges the support of an NSERC Discovery Grant (RGPIN-2024-06266).
FRB research at UBC is funded by an NSERC Discovery Grant and by the Canadian Institute for Advanced Research. The baseband recording system on CHIME/FRB is funded in part by a CFI John R. Evans Leaders Fund award to IHs.
}

\correspondingauthor{Alice P. Curtin}
\email{alice.curtin@mail.mcgill.ca}

\begin{abstract}
The Canadian Hydrogen Intensity Mapping Experiment Fast Radio Burst (CHIME/FRB) project has discovered the most repeating fast radio burst (FRB) sources of any telescope. However,  most of the physical conclusions derived from this sample are based on data with a time resolution of $\sim$1 ms. In this work, we present for the first time a morphological analysis of the raw voltage data \citep{Thompson2017} for 124 bursts from 35 of CHIME/FRB’s repeating sources. We do not find any significant correlations amongst fluence, dispersion measure (DM), burst rate, and burst duration. 
Performing the first large-scale morphological comparison at timescales down to microseconds between our repeating sources and 125 non-repeating FRBs, we find that repeaters are narrower in frequency and broader in duration than non-repeaters, supporting previous findings. However, we find that the duration-normalized sub-burst widths of the two populations are consistent, possibly suggesting a shared physical emission mechanism. Additionally, we find that the spectral fluences of the two are consistent. When combined with the larger bandwidths and previously found larger DMs of non-repeaters, this suggests that non-repeaters may have higher intrinsic specific energies than repeating FRBs. We do not find any consistent increase or decrease in the DM ($\lessapprox 1$ pc cm$^{-3}$ yr$^{-1}$) and scattering timescales ($\lessapprox 2$ ms yr$^{-1}$) of our sources over $\sim2-4$ year periods. 

\end{abstract}


\section{Introduction} 
\label{sec:intro}

Fast radio bursts (FRBs) are brief ($\upmu\textrm{s} - \textrm{ms}$) bursts of radio emission originating from primarily extragalactic distances with radio luminosities $\sim10^{36}-10^{44}$  erg s$^{-1}$ \citep{lbm+07}. FRBs are typically divided into non-repeating (or ``one-off") FRBs \citep{chimefrbcatalog} and repeating FRBs \citep{ssh+16a,abb+19b,abb+19c,RN3,fab+20}. While most FRBs' repeatability is seemingly uncorrelated, one source shows definitive periodic windows of burst activity \citep{aab+20}.

Despite the large number of published FRBs, the progenitors and emission mechanisms of these sources remain uncertain. 
Theoretical models generally involve neutron stars (NS), and fall into two main categories: those that suggest emission arises from a synchrotron maser process driven by shocks far from the central compact object \citep{mms19}, and those that propose the emission originates closer to the central compact object, such as within the magnetosphere \citep{lu2018radiation, Ioka2020}. Recent evidence has increasingly supported the idea of a magnetospheric origin for at least some sources \citep{aab+20, nimmo2021_micro, nimmo2022burst,  mckinven2024pulsarPPASwing, 2024NimmoScintillation}.

It also remains uncertain whether repeating and non-repeating FRBs represent two distinct classes of objects. Modeling of the FRB population has shown that the Canadian Hydrogen Intensity Mapping Experiment Fast Radio Burst (CHIME/FRB) sample is consistent with all FRBs being repeaters \citep{ClancyModelling}. However, there are distinct differences between the morphological properties of these two classes. A large-scale study of the morphological features of repeating versus non-repeating FRBs using CHIME/FRB was first presented by \citet{pleunismorph}. They confirmed that repeating FRBs often exhibit a pattern referred to as the ``sad-trombone effect'' wherein the emission marches downward in frequency with time \citep[and herein referred to as downward drifting; ][]{gsp+18, Hessels_2019}. Additionally, the temporal duration of repeating FRBs tends to be larger while the bandwidth is narrower than that of non-repeaters. 

The results of \citet{pleunismorph} were based on intensity data from CHIME/FRB with a time resolution of 0.983~s. However, there is evidence that FRBs have complex morphology that can only be seen at high-time resolution \citep[e.g.,][]{fmm+23}.
High-time resolution observations ($\sim\upmu$s) have demonstrated the potential to:

\begin{itemize}[noitemsep, nosep]
    \item \textbf{Improve Dispersion Measure Estimations:} Given the extragalactic distances of FRBs, the emitted radio waves will be dispersed by ionized electrons along the line-of-sight (LOS), resulting in a frequency-dependent dispersion sweep of the radio emission at the location of the telescope. This dispersion sweep can be used to calculate a dispersion measure (DM) per source and is directly proportional to the LOS integrated electron density between the source and the observer. This DM consists of components from the Milky Way, intergalactic medium (IGM), host galaxy, and local environment. The contributions from the Milky Way, IGM, and host galaxy are not expected to vary significantly with time, such that temporal variations in the DM of a source can typically be attributed to the local host environment. Microsecond features within a burst enable precise alignment of burst characteristics, facilitating accurate per-burst DM determinations. This allows us to disentangle the true DM from frequency-dependent features such as downward drifting of the burst in frequency with time. Such measurements are crucial for investigating DM variations over time in repeating FRBs, providing valuable insights into the FRB's local environment \citep[e.g.,][]{yz16, Hessels_2019, pg18, Sand2023R3}. 
    \item \textbf{Investigate potential intra-burst DM variations}: Observations at high time resolution may reveal DM variations between different components of a single burst, as possibly identified by \citet{fmm+23}, \citet{Hewitt_2023}, and \citet{2024Sand}. Intra-burst DM variations might suggest a highly dense, local environment such as a supernovae remnant, bow shock and/or different emitting heights for the various sub-components.
    \item \textbf{Identify and distinguish downward drifting:} Downward drifting can often be indistinguishable from small dispersion delays and/or scattering at $\sim$ms time resolution \citep{2024Sand}. This feature could arise either from emission in a NS magnetosphere \citep{2019ApJ...876L..15W} or through the deceleration of shock waves farther from the source \citep{mms20}.
    \item \textbf{Measure Scattering Timescales: } Measuring accurate scattering timescales for sources gives insight into their local and host environments along with that of intervening plasma in galaxy halos and the intergalactic medium \citep{ckr+22}.

    \item \textbf{Detect $\upmu$s or ns-scale features: } Some FRBs exhibit extremely short timescale features known as micro features and nano-shots \citep{ffb18, cms+20, nimmo2021_micro, majid_nano_2021, nimmo2022burst, Hewitt_2023}. Such features again strongly support magnetospheric emission models for at least the sources that exhibit them \citep{nimmo2021_micro}. 
    \item \textbf{Search for Periodicity:} By distinguishing various components within a burst, we can search for quasi-periodic or periodic structures, as shown by \citet{abb+21b}. \citet{2024NatAs...8..230K} found that all radio-loud magnetars show signs of quasi-periodic structure. They suggested that this structure in the emission was directly related to the magentar's rotational spin period. Thus, $\sim$ millisecond-duration periodic features in FRB emission would strongly lend support for a magnetospheric emission mechanism for FRBs.

\end{itemize}

\citet{aaa+23_basecat} recently published the complex raw voltages for 140 of the FRBs from the first CHIME/FRB catalog \citep{chimefrbcatalog}. This dataset enabled accurate flux and fluence measurements for these bursts, a critical improvement as CHIME/FRB intensity fluxes/fluences are only lower limits \citep{chimefrbcatalog}. \citet{2024Sand} analyzed the morphology of these bursts at timescales down to 2.56 $\upmu \textrm{s}$. They found that non-repeating FRBs also show complex morphology similar to repeaters. Their sample consisted of 125 bursts from thus-far non-repeaters and 12 bursts from repeaters, the largest sample of individual FRB sources studied at such high time resolutions. However, no work has studied a similarly large number of repeating sources at such high-time resolution. 

In this paper, we present the first large-scale study of the morphology of repeating FRBs at time resolutions down to 2.56 $\upmu \textrm{s}$ using the CHIME/FRB experiment. In Section \ref{sec: chime and analysis}, we discuss the CHIME/FRB instrument, its complex voltage recording system, and the analysis pipeline used to determine burst properties. In Section  \ref{Sec:Sources}, we present the sources analyzed in this work. In Section \ref{sec: Results}, we present a variety analyses related to the burst properties. We start by searching for correlations amongst relevant morphological parameters for our repeating FRB dataset. We then compare the morphological parameters of the repeating FRBs with those of the non-repeating FRBs studied by \citet{2024Sand}. Finally, 
we explore whether any of our repeaters show changing environments with time. In Section \ref{sec: discussion}, we discuss our results in the context of the FRB population. We end in Section \ref{sec: summary} by summarizing our work.

\section{CHIME/FRB Instrument \& Analysis Pipelines}
\label{sec: chime and analysis}
\subsection{Instrument}
\label{sec:chime}
The CHIME telescope is located near Penticton, British Columbia and consists of four fixed cyclindrical parabolic reflectors oriented in the N-S direction. Each cylinder has 256 dual polarization feeds operating from 400 to 800 MHz. CHIME's total field-of-view (FOV) is $\sim250$ deg$^2$, with a N-S FOV of $\sim120\degree$ and a frequency-dependent E-W FOV of $\sim1.3-2.5\degree$ \citep[see ][for more details]{nvp+17}. The output of CHIME's 1024 antenna signals are first sent to an FX-correlator which digitizes the output and then forms 1024 intensity beams with a time resolution of 0.983 ms and frequency resolution of 24.4 kHz on the sky \citep{Chikada1984, nvp+17, 2022chimeoverview, Bridget2023FluxCalibration}. These formed beams are searched for FRB signals by first performing a radio frequency interference (RFI) removal algorithm \citep{2022MasoudRFI} and then iteratively searching over DM-space for possible astrophysical signals. FRB candidates are passed to a machine learning algorithm that further removes RFI. The final set of FRBs \citep[e.g., those published by][]{chimefrbcatalog} are verified by CHIME members. Further details on this pipeline can be found in \citet{abb+18}. 

CHIME/FRB is equipped with a complex voltage recording system (referred to as the `baseband system') that constantly buffers $\sim$20 seconds of channelized  (1024 channels over 400 MHz) data with a time resolution of 2.56 $\upmu \textrm{s}$  \citep{2021DanielleBaseband}. When an FRB with a S/N $\gtrsim$ 12 is identified by the real-time system, the system triggers $\sim$100 ms of data to be saved around the de-dispersed pulse for offline analysis. 

Capturing the complex voltages offers significant advantages in addition to that of providing a high-time resolution dataset. One key benefit is the ability to directly beamform\footnote{Beamforming refers to the process in which the raw voltages are phase-referenced to a given location.} the data which greatly improves the initial localization of the source and avoids the spectral artifacts of the FFT. For CHIME/FRB, initial real-time localizations can span several degrees in RA and approximately 0.5 degrees in Dec. However, with this system, localizations on the order of arcminutes (and sometimes even $\sim$10 arcsec) are achievable depending on the S/N of the burst \citep{Michilli_2023, 2023ChimeBasebandCatalog}. Additionally, by correcting for the beam response of CHIME/FRB, these data enable significantly more accurate determinations of fluxes and fluences for bursts \citep{2023ChimeBasebandCatalog, 2024Sand}. 

\subsection{Analysis Pipeline}
The pipeline deployed in this work to analyze the complex raw voltages and determine the burst parameters has been described in detail by \citet{Sand2023R3, 2024Sand}. The output of this pipeline is: a structure-maximized DM, a flux/fluence for each burst, the number of distinct sub-bursts for an FRB at the given time resolution, the width of each sub-burst, the full duration and bandwidth of the burst, a DM determined using a 2D spectro-temporal fit, and a scattering time measured using the same 2D spectro-temporal fit (if applicable). Below, we provide a brief overview of each step of this pipeline from initial collection of the complex raw voltages to the final burst fits.

\noindent \textbf{Step 1:} The complex raw voltages are first beamformed to the previously published best-position for each repeater \citep{Michilli_2023, RN3} and coherently de-dispersed to the best known DM of the source. RFI masking is also performed. 

\noindent \textbf{Step 2:}  The data are downsampled from a time resolution of 2.56 $\upmu \textrm{s}$ such that a minimum peak S/N of $\sim 10-15$ is reached for each burst. The S/N limit is chosen on a burst-to-burst basis given burst characteristics. If a burst has not reached the desired S/N threshold at a time resolution of 0.655 ms, we either a) remove the burst from our sample if it is particularly faint or b) include the burst but caution that the parameter fit (discussed below) may be less reliable. 

\noindent \textbf{Step 3:}  We determine a DM by maximizing S/N for the burst. We use this DM as an input to search for a structure-optimized DM using the \texttt{DM phase} routine \citep{dm_phase}. We do not fit for the DM of individual sub-bursts, but instead assume one DM value per event. 

\noindent \textbf{Step 4:} We determine a profile for the burst by summing over all non-RFI channels. We then smooth the profile using a locally weighted scatter plot smoothing method (LOWESS) implemented by \texttt{statsmodel} in python \citep{seabold2010statsmodels}. We then determine the number of peaks within the burst (e.g., number of sub-bursts) using the \texttt{find\_peaks} algorithm implemented by \texttt{SciPy} \citep{scipy_Virtanen_2020}. In certain cases, we manually supply the number of peaks when the \texttt{find\_peaks} algorithm does not pick out the same number of peaks as determined by eye. 

\noindent \textbf{Step 5a:} To determine the initial temporal profile parameters, we fit a sum of exponentially modified Gaussians \citep[EMGs; ][]{mckinnon_2014}, with one Gaussian per sub-burst using the nonlinear least-squares algorithm \texttt{curve\_fit} from \texttt{SciPy}. The same scattering timescale is assumed for each sub-burst. 

\noindent \textbf{Step 5b:} In certain cases, a visual inspection of the results of the least-squares algorithm for the initial profile parameters does not provide a sufficient fit to the data. In these cases, we use the \texttt{emcee} routine to employ a Markov Chain Monte Carlo (MCMC) algorithm to fit for the profile parameters \citep{emcee}.

\noindent \textbf{Step 6:} We provide these initial temporal profile parameters to \texttt{fitburst}, a 2D spectro-temporal fitting routine that uses a combination of EMGs to model the data \citep{fonseca_fitburst}. The data are first fit assuming no scattering contribution.  The output of this no-scattering fit is then re-fed into \texttt{fitburst} to fit for scattering using EMGs. The output of \texttt{fitburst} is a 1$\sigma$ width per sub-burst, a spectral index and spectral running per sub-burst, the arrival time of each sub-burst, the best-fit DM for all sub-bursts, and the best fit scattering timescale for all sub-bursts (if applicable). 

\noindent \textbf{Step 7:} There is significant degeneracy between scattering and downward drifting. Thus, for each burst, we visually decide between a no-scattering and scattering fit for the final parameters. In the majority of cases ($\sim80\%$), the no-scattering fit is chosen. Upper limits on the scattering are reported as the reported \texttt{fitburst} width of the narrowest sub-burst component for that given burst. Simulations by \citet{2024Sand} showed that this was a reasonable limit on the scattering timescale. The reference frequency for these scattering measurements is 600 MHz, the centre of the CHIME/FRB band. 

\noindent \textbf{Step 8}: We use the \texttt{fitburst} parameters to determine the duration of the burst. For a single component burst, we define the burst duration as the full-width half-maximum (FWHM) of that single component. For a multi-component burst, we define the burst duration as: $w_{\textrm{tot}} = (\textrm{TOA}_{f} + 1.2 w_{f}) - (\textrm{TOA}_{i} + 1.2 w_{i})$ where $f$ indicates the TOA for the final sub-burst and $i$ is the TOA for the first sub-burst. The factor of $\sim$1.2 translates the reported 1$\sigma$ Gaussian width to half of the FWHM of that sub-burst.

\noindent \textbf{Step 9}: We also use the \texttt{fitburst} parameters to determine the bandwidth of the burst. Assuming a running-power law model, we use the spectral indices and spectral-running to reconstruct the spectral model per sub-burst. Then, we determine the highest and lowest frequency per sub-burst using the full-width tenth-max of the running power-law. The bandwidth is then taken to be the difference between the highest frequency channel and the lowest frequency channel from all the sub-bursts. 

\noindent \textbf{Step 10}: We determine a flux and fluence for each burst using the method discussed by \citet{aaa+23_basecat}.

Certain FRBs in our sample show downward drifting or complex spectral structure that is not well fit using \texttt{fitburst}. For example, some downward drifting bursts show multiple distinct components that can easily by identified and fit (e.g., see FRB 20190611A and FRB 20201228A in Figure \ref{fig:burst_wfalls}) while bursts with a single downward drifting component cannot be well fit using Gaussian components (as \texttt{fitburst} tries to do, e.g., see FRB 20211104B in Figure \ref{fig:burst_wfalls}). Thus, the amplitude of the components fit using \texttt{fitburst} are not consistently reliable throughout our sample (and hence are not directly used in the analysis that follows). We visually confirm that all fits at least properly encompass the number of components, arrival times, DM, sub-component widths, and sub-component bandwidths. 

\subsection{Burst Rates}

For all sources in our sample with two or more bursts for which the complex raw voltages were captured, we update the burst rates. Circumpolar sources ($\delta > 70^{\circ}$) are detected both in an upper and lower transit overhead CHIME. Similar to \citet{RN3}, we only calculate rates for the upper transit of each source.

We follow the methodology outlined by \citet{RN3} to determine burst rates. In brief, we calculate the total exposure time for each source based on its improved position \citep{RN3, Michilli_2023}, and determine the number of bursts found within the FWHM of their detection beam at 600 MHz. We do not distinguish between periods of high activity or quiescence. We assume a Poissonian uncertainty on the burst rate, as this dominates over the uncertainty in exposure due to the improved source positions from the beam-formed data.

We calculate a spectral fluence threshold per source using the method outlined by \citet{jcf+19}. The fluence threshold accounts for three factors - day-to-day instrument gain variations, changes in synthesized beam response during source transit, and varying emission bandwidths and frequency centers. We simulate thresholds for different detection scenarios to estimate relative sensitivity between simulated and real detections, scaling the initial fluence threshold inferred from the real detection by the simulated thresholds. For repeaters, each realization of the simulation uses a randomly selected burst as the reference for determining relative sensitivity and initial threshold. Using the 95th percentile of the scaled threshold distribution per source, we scale each source's burst rate to a fluence threshold of 5 Jy ms assuming a power-law index of $-1.5$ for the cumulative energy distribution \citep{RN3}.

\subsection{Biases}

The CHIME/FRB system for recording the complex raw voltages has a number of biases in addition to those discussed for the real-time system by \citet{marcus_injection}. The full data can only be saved for sources with a DM $< 1000$ pc cm$^{-3}$ due to the size of the buffer for this system, with channels at the top of the band lost. Three sources in our sample, FRB 20200809E, FRB 20190417A, and FRB 20190216A, have a DM $>1000$ pc cm$^{-3}$. Second, only  $\sim 100$ ms of data are saved around a given event (after de-dispersion). Thus, events significantly longer than 100 ms, or events for which the initial DM was inaccurately determined by the real-time pipeline, can be clipped. Certain events also lack data at all frequency channels due to issues with the early recording system. Additionally, only events above a given S/N threshold will trigger this recording system. This threshold is different for known-repeating (S/N $\gtrsim 10$) and non-repeating FRBs (S/N $\gtrsim 12$). In both cases, faint bursts or bursts with particularly complex structure for which the peak S/N is low may be missed. While some of the selection biases of CHIME/FRB have been described by \citet{marcus_injection}, the selection biases against complex, narrow-band bursts (such as those studied here) has not been well studied. Lastly, \citet{2024Sand} found that bursts studied at time resolutions $<1$ ms tended to have higher peak flux densities. This suggested that CHIME/FRB may be missing ultra-narrow bursts that do not have such high peak flux densities due to the averaging of these bursts with noise at time resolutions of 0.983 ms \citep{RN3}.

\section{Repeating FRB Sources}
\label{Sec:Sources}
The CHIME/FRB collaboration previously published 45 repeating sources of FRBs \citep{abb+19b,abb+19c,RN3, fab+20,bgk+21,lac+22}. Here, we focus on the sources published in the above works that have at least one CHIME-detected burst with the complex raw voltages recorded. We do not include FRB 20180916B in our work as there has already been significant analysis of this FRB at high-time resolution with CHIME/FRB \citep{Sand2023R3}. We exclude two highly active, previously published repeaters detected by CHIME/FRB: FRB 20201124A and FRB 20220912A. Including highly active repeaters might bias certain comparisons as it will dominate the distribution shape, washing away the effect of the other sources. Additionally, highly active repeaters may represent a special sub-class of repeaters, which is not the main focus of this work. We also do not include bursts for which the S/N at a time resolution of 0.655 ms is $<5$. The remainder amounts to 124 bursts with recorded complex raw voltages from 35 repeating FRBs between 2018 August and 2023 September. The end date is chosen to be consistent with the cutoff for the second CHIME/FRB catalog \citep{SecondCHIMEFRBCatalog}. 

We present a sub-sample of the frequency versus time (hereafter referred to as `waterfall') plots in the Appendix in Figure \ref{fig:burst_wfalls}. We also list each burst and its fit parameters in Table \ref{Table: Burst Parameters}. We could not determine a flux/fluence measurement for four repeater bursts in our sample (FRB 20190626A, FRB 20191217A, FRB 20210302E and FRB 20220203A) as they were detected $>2\degree$ from the meridian of CHIME/FRB where our primary beam model is not sufficiently well understood. Thus, for all analysis requiring fluences, we do not include these bursts. We also could not determine a flux/fluence for FRB 20210130E due to its low S/N. In Table \ref{Table:Source rates}, we present the updated burst rates for our sources.

\section{Results}
\label{sec: Results}

\subsection{Burst Property Correlations} 
\label{ssec:Morphological Correlations}

\begin{figure*}[t]
    \centering
    \includegraphics[width=0.95\textwidth]{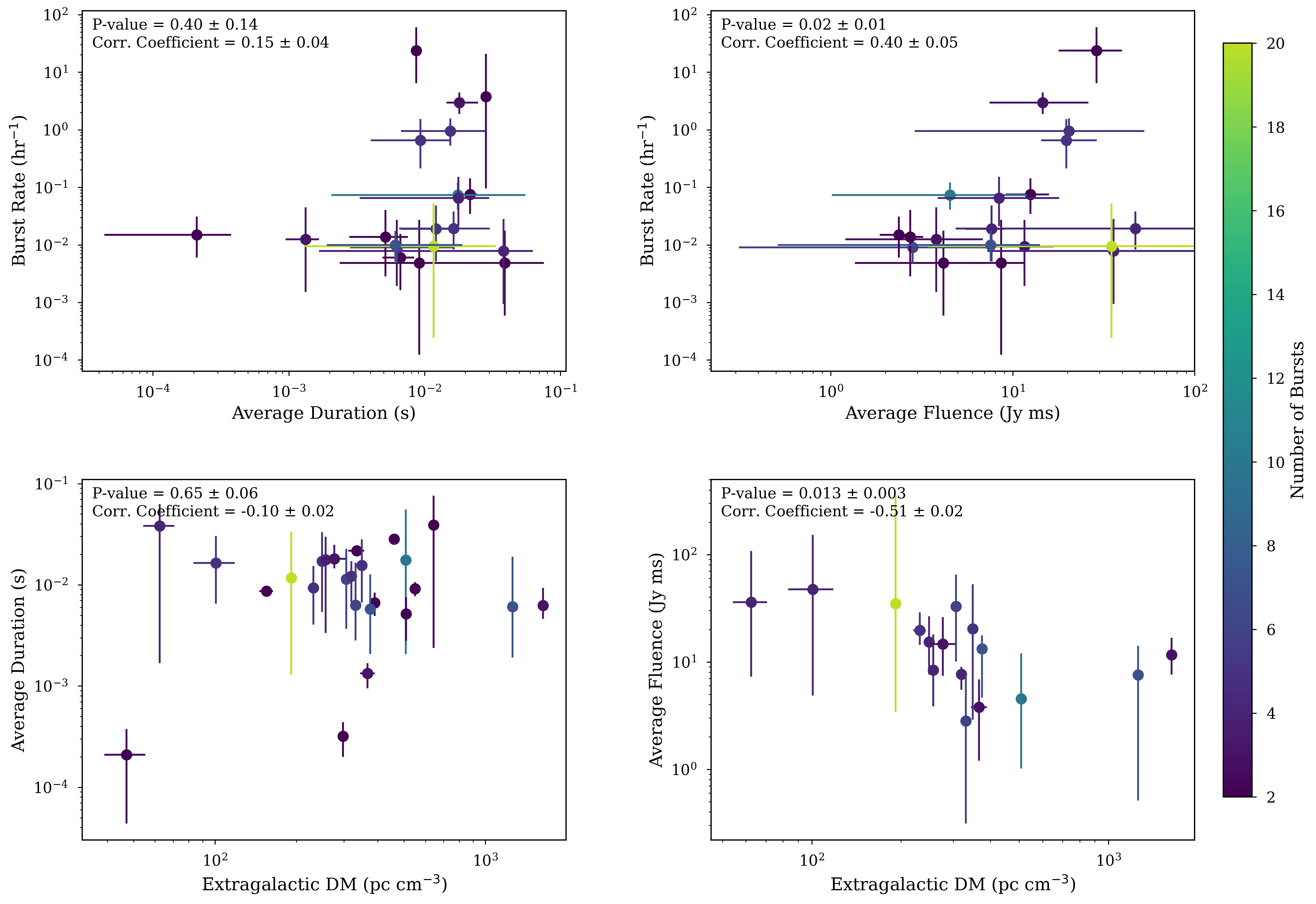}\hfill
    \caption{Correlations amongst burst rate, average burst duration, average burst fluence, and extragalactic (+halo) DM for the repeating FRBs in our sample. As discussed in Section \ref{ssec:Morphological Correlations}, we do not subtract a Galactic halo contribution from the total DM. We only include sources for which we have $\geq$ 2 bursts with the complex raw voltages recorded. The shown Spearman \textit{p}-values and correlation coefficients are the median value from 1000 MC iterations in which we iterate over the measured parameter uncertainties and calculate a new \textit{p}-value and correlation coefficient at each iteration. Uncertainties are calculated using the MADFM. }
    \label{Figure: Morphological Correlations}

\end{figure*}

In Figure \ref{Figure: Morphological Correlations}, we plot combinations of burst rate, average source duration, average source fluence, and extragalactic (+halo) DM in order to search for correlations. While we primarily focus on repeating FRBs for these correlations, we include non-repeaters from \citet{2024Sand} in some of the comparisons for which it makes sense to do so (e.g., correlations not including burst rate). For the average burst duration and fluence, the plotted error bars encapsulate the range of values for that source, e.g., the range of burst durations exhibited by that repeater over all of its bursts.

We define the extragalactic DM as the excess after subtracting the NE2001 Galactic component \citep{2002astro.ph..7156C} from the DM determined using \texttt{fitburst}. We assume a 20$\%$ uncertainty on the NE2001 DM contribution. While some models and constraints on the Milky Way halo predict significant latitude-dependent structure \citep[e.g.,][]{2020ApJ...888..105Y, 2021MNRAS.500..655D}, the variation between the median DM halo values from most commonly accepted models is of order that expected from the variations along individual lines-of-sight \citep{cbg+23}. Thus, we do not account for the halo contribution to the observed DMs. This has the same effect on the analysis as would subtracting a constant halo value from each line of sight. We refer to this DM as the extragalactic DM (+halo). We do not account for the Galactic scattering contribution, as it is not additive in the same way as the DM. However, all of our scattering values are significantly larger than the predicted Galactic scattering contribution \citep{ne2001}. For this specific analysis, we only include sources for which there are at least two bursts from that source that have the complex raw voltages recorded. This totals 25 different repeating sources.

To determine whether a given correlation is significant, we use the \citet{spearman1904} rank test as implemented in python using  \texttt{scipy.stats.spearmanr} \citep{scipySpearman}. For a correlation to be considered significant, we require the Spearman \textit{p}-value to be to be $\textrm{p} < 10^{-3}$. 

To account for uncertainties in our measured parameters, we follow the technique presented by \citet{RN3}. We first perform a Monte Carlo (MC) simulation in which we re-determine a given parameter (e.g., average duration of a source) 1000 times. We do this by sampling from the uncertainty region for each burst for that given parameter in order to determine a new average parameter. We calculate an associated Spearman \textit{p}-value and correlation coefficient for every iteration, and report the median \textit{p}-value and correlation coefficient from the MC along with the Median Absolute Deviation From the Median (MADFM). For repeaters for which the burst rate is zero\footnote{When calculating the burst rate, we only include bursts that occurred within the FWHM of a given CHIME/FRB beam at 600 MHz. Thus, it is possibly for a given repeater to have no bursts occur within these parameters, and thus have a burst rate of zero.} but the new, sampled rate will not be zero, we do not show them in the below plots but do include them when calculating correlation coefficients and \textit{p}-values.

\subsubsection{Burst Rate \& Average Duration} 
As seen in the top left panel of Figure \ref{Figure: Morphological Correlations}, we do not find a significant correlation between burst rate and average burst duration. \citet{2020WidthRateConnor} suggested that the burst rate and duration of FRBs may be correlated in order to explain the discrepancies in the widths of repeating and non-repeating FRBs observed in \citet{pleunismorph}. If the beaming angle of a source is directly related to the burst width, we would expect fewer bursts from sources emitting narrow bursts, as their emission would be highly beamed.

However, as seen in Figure \ref{Figure: Morphological Correlations}, many sources emit bursts spanning approximately an order of magnitude in burst duration, with FRB 20181030A emitting bursts over approximately two orders of magnitude in duration. If beaming angle is proportional to burst width, then this implies a significant change in the beaming angle for an individual source. It is also possible that the large range of burst durations seen from a given source obscures an underlying correlation. Therefore we also re-do the correlation study using only the maximum burst width from a given source (see Figure \ref{Figure:MaxWidthRate} in the Appendix). The median \textit{p}-value from 1000 MC samples for this correlation is 0.68, still not significant.

It is also possible that the large burst rate uncertainties (dominated by the number of bursts due to the assumed Poissonian nature) obscure an underlying correlation here. CHIME/FRB is also biased against detecting long bursts \citep{marcus_injection} and the sensitivity of the instrument to highly complex bursts (such as downward drifting bursts) is not yet quantified. Thus, it is possible that CHIME/FRB is missing the longest bursts from these sources, and hence we are not yet probing the widths needed to see a burst rate versus duration correlation. The burst rates and durations here are also determined in the observer's reference frame, while that predicted by \citet{2020WidthRateConnor} are in the burst reference frame.  

Nonetheless, our results may suggest that there is significant jitter in the relation between burst duration and beaming angle, or that burst duration is not correlated with the beaming angle for the radio emission. As discussed by \citet{2020WidthRateConnor}, emission models involving shocks far from a central compact object predict a direct scaling of pulse width with beaming angle. However, models involving emission close to the surface of a NS do not as clearly predict this correlation.

\subsubsection{Burst rate \& Average Fluence Correlation}
As seen in top right panel of Figure \ref{Figure: Morphological Correlations}, we do not find a significant correlation between burst rate and a source's average fluence. Theoretically, such a correlation might be expected if all FRBs share the same luminosity function and average luminosity, as nearby repeaters would be more likely to have a higher observed burst rate and higher average fluence. However, this correlation is likely heavily affected by our detection threshold, with our average fluences biased high due to our inability to detect low fluence events. Additionally, this comparison may be affected by our reduced ability to detect long duration bursts with high fluences \citep{marcus_injection}.  If indeed the lack of a correlation is astrophysical and is not impacted by our system biases, it suggests that repeating FRBs have a wide range of average intrinsic luminosities.

\subsubsection{Average Duration \& Extragalactic DM Correlation}
As seen in the bottom left panel of Figure \ref{ssec:Morphological Correlations}, we do not find a significant correlation between the average burst duration for our sources and their extragalactic (+halo) DM, as might be expected if higher-DM sources are on average farther away and undergo greater cosmological time dilation \citep{connor19}. We also re-do this analysis including the non-repeating bursts from \citet{2024Sand} as this analysis does not require a burst repetition rate. The \textit{p}-value for the correlation using this additional sample of bursts is 0.97, which is not significant (see Figure \ref{Figure:DurationDMRepeaterNonRepeater} in the Appendix). The p-value when just using the non-repeaters is 0.58 \citep[see also][]{2024Sand}.

Using the first CHIME/FRB catalog, \citet{2023Shin} predicted that 99.4\% of CHIME/FRB sources would be found at redshifts $\leq$ 2. A redshift of 2 corresponds to a cosmological expansion factor of 3. Given that the burst duration from a given source can vary by an order of magnitude, the effect of a cosmological expansion factor is likely too small at these redshifts to be discernible. Additionally, propagation effects such as multiple scattering screens may smear out such effects \citep{2024Sand}.

\subsubsection{Average Fluence \& Extragalactic DM Correlation}
As seen in the lower right panel of Figure \ref{Figure: Morphological Correlations}, we do not find a correlation between the average fluence of our sources and their extragalactic (+halo) DM, as might be expected if higher-DM sources are on average farther away and all repeating FRBs follow a universal luminosity function with the same average luminosity.
We re-do the correlation using the median fluence, but again find no correlation. We also re-do the correlation using the non-repeater sample from \citet{2024Sand}. The Spearman \textit{p}-value is 0.005 (See Figure \ref{Figure:FluenceDMRepeaterNonRepeater} in the Appendix), still larger than our significance threshold of $10^{-3}$ but smaller than that when examining the two samples separately. The p-value from the non-repeaters alone is 0.02. 

It is possible that a significant portion of the DM from our sources is coming from the host galaxies and hence this obscures the correlation. However, for the three repeaters in Figure \ref{Figure: Morphological Correlations} at DM $>700$ pc cm$^{-3}$, host DMs of order 500 pc cm$^{-3}$ would be needed to see the predicted negative DM-fluence correlation. While such a high host DM contribution has been seen for a handful of FRBs \citep{2022Ocker, 2023Caleb}, the average assumed DM host contribution is only $\sim 80^{+69}_{-49} - 90^{+29}_{-19}$ pc cm$^{-3}$ \citep{2023Shin, 2024FlimFlam}. One of these sources (FRB 20190417A) has a particularly large host RM \citep{mgm+23b}. However, the other two do not \citep[Ng et al. in prep.;][]{mgm+23b} Thus, large host DM contributions seems an unlikely explanation for all three sources. A more probable explanation is that repeating FRBs have a broad range of average intrinsic luminosities such that any correlation is undetectable.

\subsubsection{Extragalactic DM \& Scattering Correlation}
Galactic pulsars show a strong correlation between their DM and scattering timescale \citep{Bhat_2004}. A correlation between extragalactic DM and scattering for our FRBs could suggest that a significant portion of an FRB's DM originates from the host galaxy, as the intergalactic medium (IGM) is expected to contribute negligible scattering \citep{occ_2022}. 

In the left panel of Figure \ref{fig: Scattering vs. DM; repeaters vs. non-repeaters}, we show the extragalactic (+halo) DM versus total scattering timescale for our repeating FRB sample. Most of the bursts in our sample do not have measured scattering timescales, and instead we place an upper limit on the scattering timescale equal to the width of the narrowest sub-component. As we will show in Section \ref{ssec:Individual Sources}, we do not find evidence that the scattering time of a single repeating FRB changes with time. Hence, for each repeating source, we use (and show) either our scattering timescale with the smallest uncertainty or our most constraining upper limit. 

Only 9 of the repeating FRB sources in our sample have at least one burst with a measured scattering timescale. Thus, the sample for this correlation is extremely limited. Nonetheless, the Spearman \textit{p}-value using only the sources with measured scattering timescales is 0.63\footnote{Given the low number of sources, and the low significance, we do not perform the same MC employed for previous correlations.}. We also run a censored Kendall $\tau$ test using the \texttt{cenken} function in R's \texttt{NADA} package\citep{NADApackage, Rsoftware}. The censored Kendall $\tau$ test allows us to account for upper limits on our scattering timescales (also referred to as left censored data). The Kendall $\tau$ value is 0.1, suggesting little to no correlation. The lack of a correlation supports the notion that a significant portion of the extragalactic DM does not arise from the same location as that of the scattering. Our results are similar to those of \citet{2024Sand}.

\subsection{Repeating vs. Non-repeating FRBs Morphology Comparison} 
\label{ssec:RepeatersVsNonrepeaters}

We compare various morphological features between our sample of repeating FRBs and the thus-far non-repeating FRB sample presented by \citet{2024Sand}. A significant advantage to comparing these two populations rather than including FRBs from instruments other than CHIME/FRB is that both suffer from the same instrumental biases, aside from their detection thresholds. The difference in detection thresholds is the most relevant for comparisons of the spectral fluences (F$_\nu$) of the bursts. Thus, for all fluence comparisons, we only include bursts for which the real-time detection S/N\footnote{While the S/N threshold for non-repeaters is 12, we choose 20 to be well above the completeness level.} $>20$. For other tests, we place cuts on the time resolution at which the burst is studied. Thus, this is approximately equivalent to a S/N cut, as only bursts with sufficient S/N can be studied at high time resolutions. We compare the bandwidth, total burst duration, number of burst components, sub-burst widths, sub-burst waiting times, and fluences between the two samples in Figures \ref{fig: Scattering vs. DM; repeaters vs. non-repeaters} - \ref{fig: fluence repeaters nonrepeaters}. 

To assess the difference between the two populations, we use both the Anderson-Darling \citep[AD; ][]{ADTest} and Kolmogorov-Smirnov \citep[KS; ][]{KSTest} test statistics. Similar to in Section \ref{ssec:Morphological Correlations}, we consider two samples to be distinct if the AD and KS test \textit{p}-values are both $<10^{-3}$. Results from all of our comparisons are listed in Table \ref{Table:repeater_vs_non-repeater_ad_ks_test}. We include the total number of bursts from repeating and non-repeating FRBs used per comparison after applying various cuts in Table \ref{Table:repeater_vs_non-repeater_ad_ks_test}. We perform the comparisons using each individual burst from a given source, rather than the average of that quantity for the source. Thus, repeaters with a significant number of bursts will more heavily contribute to the given distributions.

\begin{deluxetable*}{llllcc}
\tablecaption{Comparisons between repeating and thus far non-repeating FRB parameters\label{Table:repeater_vs_non-repeater_ad_ks_test}}
\tablehead{
\colhead{Property} & \colhead{Cutoff} & \colhead{\# Repeat bursts} & \colhead{\# Non-repeat bursts} & \colhead{AD \textit{p}-value\tablenotemark{a}} & \colhead{KS \textit{p}-value\tablenotemark{b}}
}
\startdata
Bandwidth & N/A & 124 & 125 & $< 0.001$ & $2 \times 10^{-25}$ \\ \hline
Duration & N/A & 124 & 125 & $< 0.001$ & $2 \times 10^{-24}$ \\ \hline
Num. Components &  t$_{\textrm{res}} \leq 0.16$ ms & 20 & 79 & 0.3 & 0.7 \\ \hline
Sub-burst Width &  t$_{\textrm{res}} \leq 0.16$ ms & 89\tablenotemark{e} & 210\tablenotemark{e} & $< 0.001$ & $4 \times 10^{-19}$ \\
Sub-burst Width &  t$_{\textrm{res}} \leq 80 \upmu$s & 36\tablenotemark{e} & 190\tablenotemark{e} & $< 0.001$ & $9 \times 10^{-6}$ \\
Sub-burst Width &  t$_{\textrm{res}} \leq 40 \upmu$s & 18\tablenotemark{e} & 164\tablenotemark{e} & 0.008 & 0.007 \\ \hline
Duration-Normalized Sub-burst Width &  t$_{\textrm{res}} \leq 0.16$ ms & 89\tablenotemark{e} & 210\tablenotemark{e} & 0.3 & 0.4 \\
Duration-Normalized Sub-burst Width &  t$_{\textrm{res}} \leq 80 \upmu$s & 36\tablenotemark{e} & 190\tablenotemark{e} & 0.02 & 0.009 \\
Duration-Normalized Sub-burst Width &  t$_{\textrm{res}} \leq 40 \upmu$s & 18\tablenotemark{e} & 164\tablenotemark{e} & 0.008 & 0.007 \\ \hline
Wait time &  t$_{\textrm{res}} \leq 0.16$ ms & 48\tablenotemark{f} & 113\tablenotemark{f} & $< 0.001$ & $4 \times 10^{-20}$ \\
Wait time &  t$_{\textrm{res}} \leq 80 \upmu$s & 15\tablenotemark{f} & 108\tablenotemark{f} & $< 0.001$ & $7 \times 10^{-7}$ \\
Wait time &  t$_{\textrm{res}} \leq 40 \upmu$s & 6\tablenotemark{f} & 98\tablenotemark{f} & 0.02 & 0.02 \\ \hline
Duration-Normalized Wait time &  t$_{\textrm{res}} \leq 0.16$ ms & 48\tablenotemark{f} & 113\tablenotemark{f} & 0.25 & 0.98 \\
Duration-Normalized Wait time &  t$_{\textrm{res}} \leq 80 \upmu$s & 15\tablenotemark{f} & 108\tablenotemark{f} & 0.01 & 0.06 \\
Duration-Normalized Wait time &  t$_{\textrm{res}} \leq 40 \upmu$s & 6 & 98 & 0.005 & 0.02 \\ \hline
Fluence\tablenotemark{c} & S/N $>20$ & 35 & 50 & 0.04 & 0.12 \\
Band-limited Fluence\tablenotemark{d} & S/N $>20$ & 35 & 48 & 0.3 & 0.5 \\
\enddata
\tablenotetext{a}{Anderson-Darling test as implemented using \texttt{scipy.stats}. Note \textit{p}-values are floored at 0.001.}
\tablenotetext{b}{Kolmogorov-Smirnov test as implemented using \texttt{scipy.stats}.}
\tablenotetext{c}{Spectral fluence averaged over the entire CHIME/FRB 400-MHz band. We limit the sample bursts whose detection S/N is greater than 20.}
\tablenotetext{d}{Spectral fluence averaged over the spectral extent of the burst. We limit the sample to bursts whose detection S/N is greater than 20. }
\tablenotetext{e}{Number of individual sub-bursts rather than number of detected bursts.}
\tablenotetext{f}{The number of unique waiting times calculated as the differences between arrival times. Only applicable for multi-component bursts.}
\end{deluxetable*}

\subsubsection{Extragalactic DM and Scattering for Repeating vs. Non-repeating Sources}

\citet{RN3} found a significant difference between the DM distributions of repeating and non-repeating FRBs. This DM distribution difference persisted as the S/N threshold of their sample was lowered, as well as when extragalactic DM instead of total DM was considered. They found an extragalactic DM difference of 161 $\pm$ 55 pc cm$^{-3}$ between the mean DM of repeaters and non-repeaters detected above a S/N threshold of 12, with repeaters having on average lower extragalactic DMs. While our analysis has been able to refine the DM of repeating FRBs, the improvement has been on the order of $\sim$ 1 pc cm$^{-3}$ or less. Hence, we do not expect a significant change from the results of \citet{RN3} regarding the DM distribution. 

The difference in the DM distributions of the two samples could be due to either a) a difference in the host galaxy contributions or b) a difference in the distances to which we detect them (and hence the IGM contributions). However, unlike DM, the scattering contribution from the IGM is expected to be small \citep{2013ApJ...776..125M, occ_2022}. A difference in the scattering timescale distributions between the two samples would, in principle, provide direct evidence of different host environments for the two populations, or provide evidence that the IGM scattering contribution is significant. However, highly scattered, narrow bursts are harder to detect than highly scattered, wide bursts and so there is also a natural observational bias here.

\begin{figure*}[]
    \centering   \includegraphics[width=0.48\textwidth]{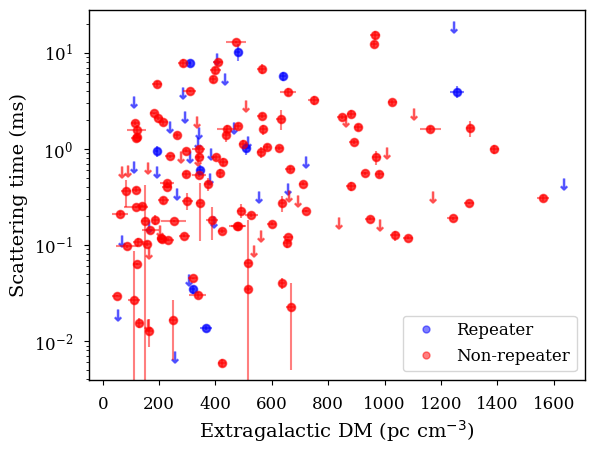}
    \includegraphics[width=0.48\textwidth]{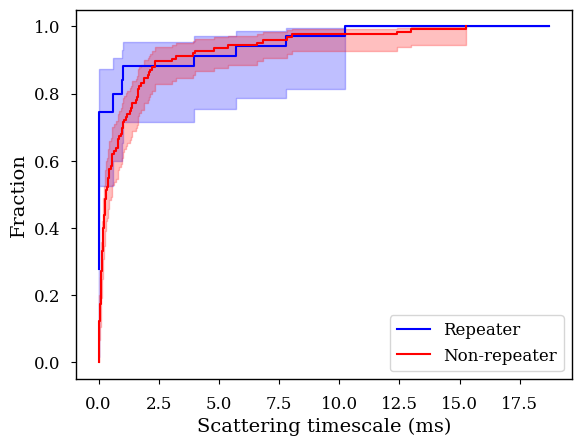}
    \caption{\textit{Left panel:} 
    Scattering versus extragalactic (+halo) DM for repeaters and non-repeaters. For DM, we subtract the Galactic component as determined using NE2001. Downward arrows represent upper limits on scattering timescales. For repeating FRBs, this is assumed to be the width of the narrowest sub-burst for that event. For non-repeating FRB parameter descriptions, see \citet{2024Sand}.
    \textit{Right panel:} Cumulative distribution function (CDF) for scattering for repeating and thus-far non-repeating FRBs using the Kaplan-Meier method for left-censored data. }
    \label{fig: Scattering vs. DM; repeaters vs. non-repeaters}
\end{figure*}

The majority of the repeating FRBs in our sample do not show visible signs of scattering. However, we can still compare the scattering timescales (including upper limits) of our sample with those of non-repeating FRBs. In the left panel of Figure \ref{fig: Scattering vs. DM; repeaters vs. non-repeaters}, we show the scattering timescale versus the extragalactic (+halo) DM  for both repeating and non-repeating FRBs. Upper limits on the scattering timescales are shown with downward arrows. 

To compare the two populations, we use the Kaplan-Meier method, a part of the python package \texttt{lifelines}, and the Peto-Peto test statistic \citep{PetoPeto}. These methods give significant advantages over the AD and KS tests as data can be labeled as either censored or uncensored. In the case of scattering timescales, the censored data are those for which we have only upper limits, while the uncensored data are the measured scattering timescales. In the right panel of Figure \ref{fig: Scattering vs. DM; repeaters vs. non-repeaters}, we show the cumulative distribution function (CDF) of the repeating and non-repeating scattering distributions using the Kaplan-Meier method. The Peto-Peto test statistic for the two distributions is $p_{\textrm{PP}} = 1.0$ (significantly larger than our significance criteria of $10^{-3}$). Thus, given the current sample sizes, there is no evidence for a difference in the scattering timescales of repeating and thus-far non-repeating FRBs. However, we caution that our repeating FRB sample is small, and those with measured scattering timescales even smaller. Additional repeating FRBs with measured scattering timescales could still reveal a difference between the two populations.

\subsubsection{Burst Duration and Bandwidth for Repeating vs. Non-repeating Sources}

\begin{figure}[]
    \centering
    \includegraphics[width=0.99\columnwidth]{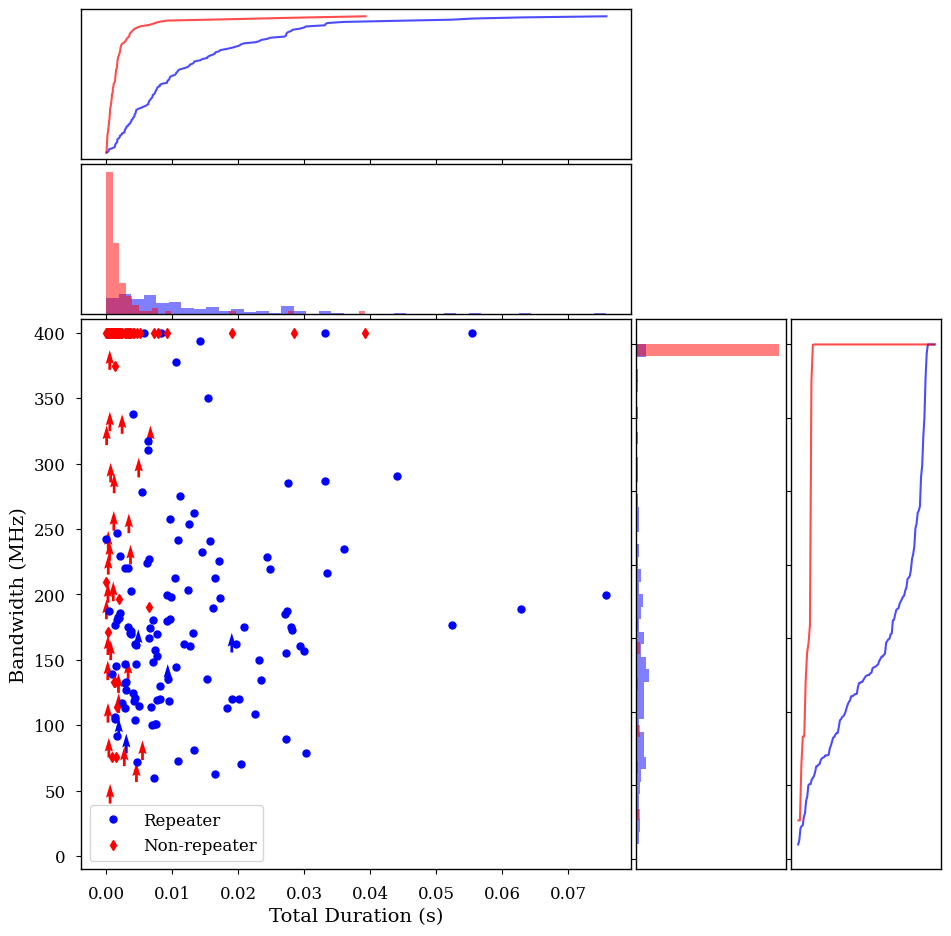}
    \caption{Bandwidth versus total duration for repeating and thus-far non-repeating FRBs. Bursts for which the full bandwidth was not available are indicated with arrows. Additionally, bursts for which the bandwidth is 400 MHz are again lower limits as this is solely set by the frequency range of CHIME. The normalized histograms and the cumulative distributions only include bursts for which the full bandwidth was recorded. }
    \label{fig: Bandwidth vs. width; repeaters nonrepeaters}
\end{figure}

\citet{pleunismorph} identified a difference in both the bandwidth and duration of repeating and non-repeating FRBs. Using ours and \citet{2024Sand}'s improved bandwidth and duration estimates, we compare the two in Figure \ref{fig: Bandwidth vs. width; repeaters nonrepeaters}. By eye, the two populations inhabit very different portions of the duration-bandwidth phase space, with repeating FRBs narrower in bandwidth and longer in duration than non-repeaters. We calculate the AD and KS test values for bandwidth and duration separately and list them in Table \ref{Table:repeater_vs_non-repeater_ad_ks_test}. The full band was not recorded for a significant fraction of the non-repeating FRBs \citep[see ][for a discussion of this]{2024Sand}. Thus, many of the bandwidths are lower limits. However, larger bandwidths for these sources would only increase the discrepancy seen between the two populations. The AD and KS test statistics provided in Table \ref{Table:repeater_vs_non-repeater_ad_ks_test} assume that the bandwidth for these sources is equivalent to the lower limit. If we instead remove sources for which the bandwidth is a lower limit, the AD and KS test statistic for BW are $p_{AD} < 0.001$ and $p_{KS} = 5 \times 10^{-39}$.  Thus, in both scenarios, this strongly indicates that the two populations are distinct in both BW and duration. 

\subsubsection{Sub-Burst Properties for Repeating vs. Non-repeating Sources}

In an effort to systematically search for other apparent differences in emission phenomenology between repeaters and non-repeaters, we compare the number of sub-bursts per event\footnote{Note this is not per source, but rather per individual burst from a given source.}, the width of individual sub-bursts, and the waiting time (defined as the difference in the time of arrival of individual sub-bursts) between the two populations. However, all of these comparisons can be biased by the time resolution at which we study the populations as additional components may be unresolved at lower resolutions. To try to address this bias, we limit the comparison to bursts (both repeating and non-repeating) studied at a time resolution $\leq$ 0.16 ms. However, a bias may still remain as many of the non-repeating FRBs are bright enough to be analyzed at time resolutions $\leq$ 20.48 $\upmu$s (e.g., 51 out of 125 bursts) while only 7 bursts from our repeating sample are studied at time resolutions $\leq$ 20.48 $\upmu$s. Additionally, we note that more active repeaters will influence the distributions more than less active sources, as we use every burst from each source with the required data for our comparisons.

\begin{figure*}[t]
    \centering
    \includegraphics[width=0.48\textwidth]{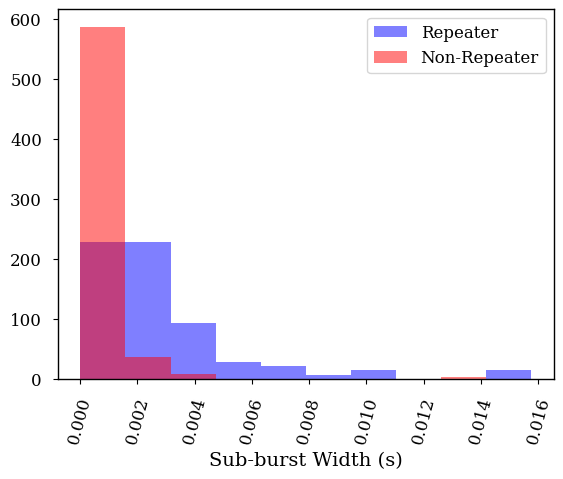}
    \includegraphics[width=0.48\textwidth]{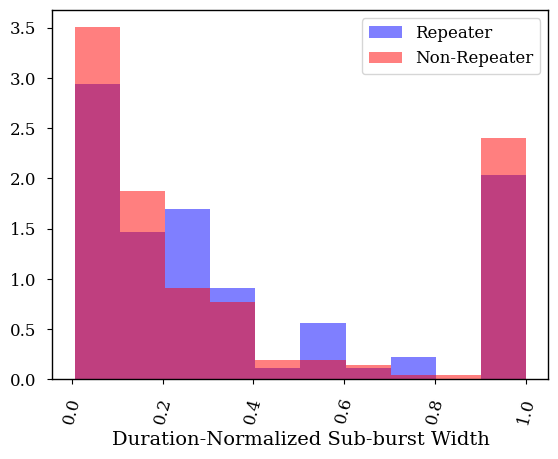}
    \includegraphics[width=0.455\textwidth]{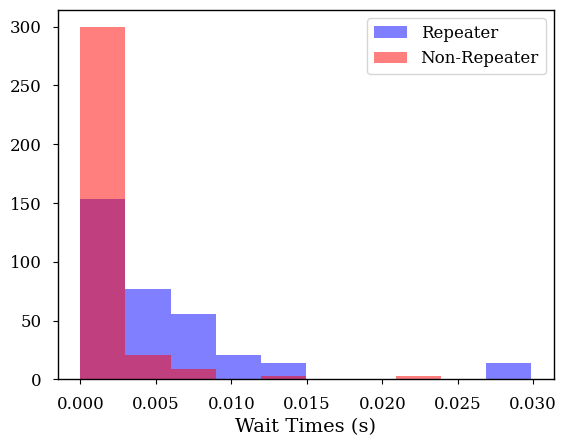}
    \includegraphics[width=0.455\textwidth]{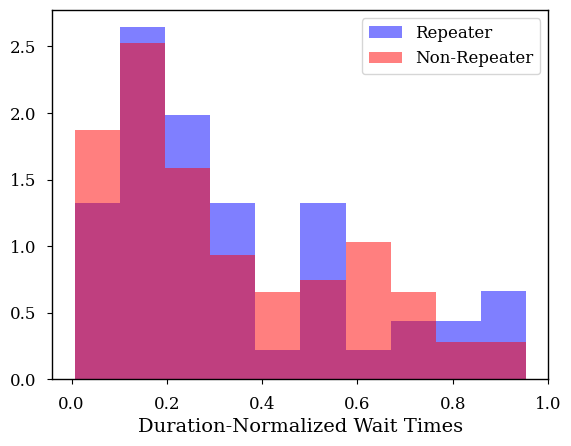}
    \includegraphics[width=0.48\textwidth]{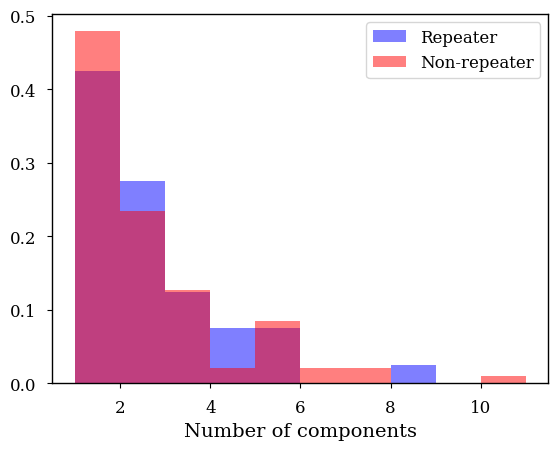}
    \caption{\textit{Left top panel: } Sub-burst width distribution for repeating and thus-far non-repeating FRBs. We limit the two samples to only include bursts studied a time resolution $\leq$ 0.16 ms. 
    \textit{Right top panel: } Same as left panel except for the sub-burst width normalized by the total burst duration.
    \textit{Left middle panel:} Wait time distribution for repeating and thus far non-repeating FRBs. The wait time is defined as the difference in arrival time between subsequent sub-bursts for a given FRB. The sample is limited to bursts which were analyzed at time resolutions equal to or less than 0.16 ms. 
    \textit{Right middle panel: } Same as the left panel except the duration-normalized wait time. Each wait time is normalized by the total duration of that given burst. 
    \textit{Bottom panel:} Number of sub-burst components for repeating and thus-far non-repeating FRBs. We limit the two samples to only include bursts studied a time resolution $\leq$ 0.16 ms.}
    \label{fig:sub-burst properties; repeaters and nonrepeaters}
\end{figure*}

Nonetheless, there are 94 bursts (210 sub-bursts) from non-repeating FRBs and 40 bursts (89 sub-bursts) from repeating FRBs that we analyze at a time resolution $\leq$ 0.16 ms.  In Figure \ref{fig:sub-burst properties; repeaters and nonrepeaters} we compare the number of sub-burst components, the sub-burst widths, and the waiting time of sub-bursts between the two populations. The AD and KS test statistics for each comparison is listed in Table \ref{Table:repeater_vs_non-repeater_ad_ks_test}.

While the number of sub-bursts is consistent between the two populations, the sub-bursts widths and waiting times are not (Figure \ref{fig:sub-burst properties; repeaters and nonrepeaters}). The average width of an individual sub-burst for repeaters is 2.7 ms while that for non-repeaters is 0.57 ms. Additionally, the average wait time for repeaters is 5.2 ms while that for non-repeaters is 1.2 ms. We re-do our analysis only using bursts studied at even higher time resolutions (e.g., $\leq$81.92 $\upmu$s and $\leq$ 40.96 $\upmu$s) The \textit{p}-values for the smaller samples are larger, suggesting that the sub-burst width and waiting time differences may not be inherent to the two populations and instead are solely due to missed components at lower time resolutions (see Figure \ref{fig:sub-burst properties; repeaters and nonrepeaters; higher time res} in the Appendix). However, we note the sample of repeating FRBs at resolutions $\leq$ 40.96 $\upmu$s is very small, and may simply be insufficient to reject the null hypothesis that the two samples originate from the same distributions. 

Given that repeating FRBs have longer durations than non-repeaters, we explore whether the duration-normalized sub-burst and waiting times between the two populations are consistent. In Figure \ref{fig:sub-burst properties; repeaters and nonrepeaters}, we show the duration-normalized sub-burst width and waiting times for the larger burst sample (e.g., bursts studied at time resolutions $\leq$ 0.16 ms). Interestingly, the AD and KS test-statistics for both duration-normalized sub-burst width and duration-normalized waiting times are not significant. We also re-do the comparison excluding single component bursts for which the duration-normalized individual sub-burst width would be 1. This new sample consists of 165 non-repeater sub-bursts and 72 repeater-sub-bursts. The \textit{p}-values are slightly smaller but still significantly above $10^{-3}$.

We also re-do our analysis with even higher time resolution limits (e.g., 81.92 $\upmu$s and 40.96 $\upmu$s). The \textit{p}-values for the duration-normalized wait times and widths are now slightly smaller, but still greater than $10^{-3}$ (see Figure \ref{fig:sub-burst properties; repeaters and nonrepeaters; higher time res} in the Appendix). 

In Figure \ref{fig: Duration vs. sub-burst width; repeater non-repeater}, we show sub-burst width versus total duration for all repeating and non-repeating bursts in our sample studied at time resolutions $\leq$ 0.16 ms. We exclude bursts for which there is only one component, as these would cluster along a 1:1 line. There is a positive correlation between sub-burst width and total burst duration. This suggests that there may be a shared physical scaling for repeating and non-repeating FRBs relating to their total widths and sub-burst widths.  Additional repeating FRBs studied at the highest time resolutions will provide further insight into this new finding.

\begin{figure}[]
    \centering
    \includegraphics[width=0.99\columnwidth]{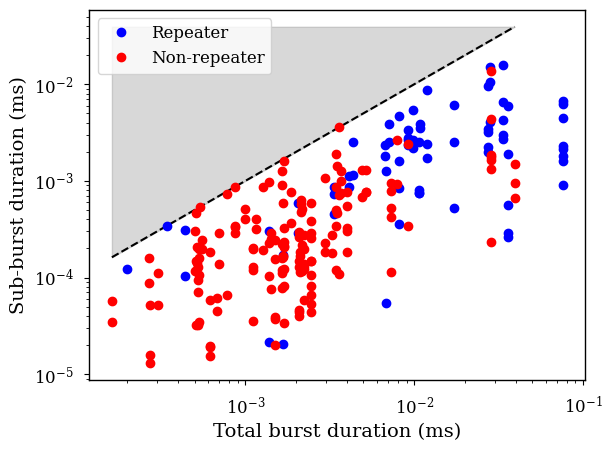}
    \caption{Sub-burst width versus total burst duration for all repeating and non-repeating bursts studied at time resolutions $\leq$ 0.16 ms. We do not include bursts for which there is only one component. The grey shaded region indicates an unphysical phase space where the sub-burst width would be larger than the total burst duration. }
    \label{fig: Duration vs. sub-burst width; repeater non-repeater}
\end{figure}

\subsubsection{Spectral Fluences for Repeating vs. Non-repeating Sources}

There is an inherent bias in a comparison of the spectral fluences\footnote{Unless specified otherwise, the spectral fluences referred to throughout this work will be those averaged over the full 400 to 800 MHz CHIME/FRB band. } of the two populations since the triggering threshold at CHIME/FRB for the complex raw voltages is lower for repeaters than for non-repeaters. Thus, we only include bursts in our sample for which the initial detection S/N was $> 20$, as this is sufficiently above the triggering threshold for both repeating and non-repeating sources. Additionally, missing channels due to RFI masking could also affect this comparison. However, given we are only comparing bursts detected by CHIME, we do not expect drastic changes in the total fraction of the band that is masked due to RFI. We remove bursts for which the top available channel was less than 800 MHz, though (e.g., those for which DM $> 1000$ pc cm$^{-3}$) as the fluence may biased high if we assume a negative spectral index.

\begin{figure*}[t]
    \centering
    \includegraphics[width=0.45\textwidth]{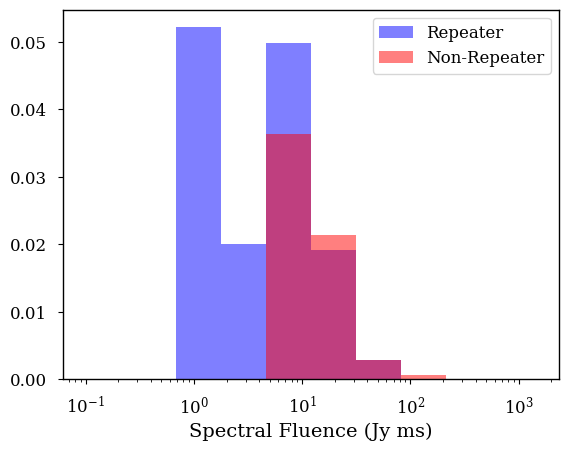}
    \includegraphics[width=0.45\textwidth]{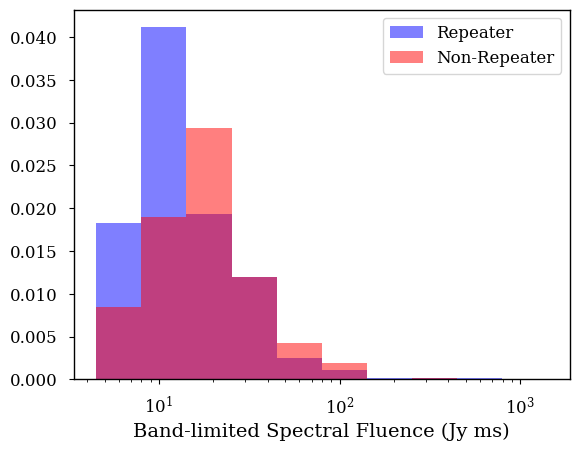}
    \caption{\textit{Left panel: } Band-averaged spectral fluence distribution for repeating and thus-far non-repeating FRBs over the entire 400- MHz CHIME/FRB bandwidth. As repeating FRBs have a different threshold for triggering the complex raw voltages than non-repeaters at CHIME/FRB, we limit the sample to bursts (repeating \& non-repeating) whose detection S/N is greater than 20. We also only include bursts for which the full CHIME/FRB band was recorded. \textit{Right panel: } Same as left panel except a band-limited spectral fluence calculation in which the spectral fluence is only determined over the frequency channels that contain emission from the source.}
    \label{fig: fluence repeaters nonrepeaters}
\end{figure*}

With the S/N cut of 20, the AD and KS test statistics do not suggest the two are drawn from different distributions. However, repeaters have on-average lower spectral fluences than non-repeaters (17 Jy ms versus 91 Jy ms, respectively;  see left panel of Figure \ref{fig: fluence repeaters nonrepeaters}). This is expected as repeating FRBs are on average more narrowband than non-repeating FRBs, and thus their spectral fluences over the entire CHIME/FRB bandwidth will be lower than those of a broadband non-repeater as more noise is averaged in. To alleviate this, we re-calculate each burst's spectral fluence solely over the spectral extent of that burst. We compare this new, band-limited spectral fluence for repeating and non-repeating FRBs in the right panel of Figure \ref{fig: fluence repeaters nonrepeaters}. We again limit our sample to bursts detected with a real-time S/N $>20.$ The two are consistent with being drawn from the same distribution, with the AD and KS test values listed in Table \ref{Table:repeater_vs_non-repeater_ad_ks_test}. If we again re-do this analysis but with a real-time S/N limit of $>15$, the two remain consistent with being drawn from the same distribution. 

CHIME/FRB's detection algorithm uses a boxcar matched filter that effectively integrates over the burst in frequency and time. Thus, a non-repeating FRB with a larger bandwidth yet the same band-limited fluence as a repeating FRB should have a higher detection S/N. We note, however, that there are a few instances in which this does not occur. Significant frequency-dependent noise weighting or RFI flagging can boost a narrowband signal's S/N. Additionally, CHIME/FRB's detection algorithm searches over two spectral indices, such that a narrowband burst at the top or bottom of the band can again be boosted in S/N. 

Nonetheless, in most cases, a broadband burst with the same specific fluence should have a larger S/N than a narrowband burst. Thus, we should be able to detect broadband non-repeating FRBs out to larger distances than narrowband repeaters. This is consistent with the differences seen in the DM distribution for repeating and thus-far one-off FRBs \citep{RN3}. However, for the spectral fluence over the spectral extent of the bursts to be consistent between the two populations, this would imply that non-repeaters have, on-average, intrinsically higher specific energies than repeating FRBs.

To test the hypothesis that an FRB with the same band limited spectral fluence should have a higher detection S/N for a larger bandwidth, we investigate the detection S/N versus bandwidth for our repeating sources (see Figure \ref{Figure: BWvs.SN} in the Appendix). However, we cannot draw any significant conclusions due to the large range of spectral fluences exhibited by a single source along with the small bandwidth ranges.
Future injections of bursts to better understand CHIME/FRB's detection sensitivity should lend significant insights into this.

\subsection{Spectral Fluence, Bandwidth, DM, and Scattering Time Variations} 
\label{ssec:Individual Sources}

If FRBs are young NS embedded in turbulent environments, we might expect the local environment (probed through the DM, rotation measure, scattering timescale, and scintillation features) of these sources to change with time \citep{pg18}.\footnote{We note, however, that changes in the DM, rotation measure, etc. could also be caused by the relative motion of the source with respect to us and thus a changing DM does not directly imply a changing environment.} \citet{michilli2018extreme} found extreme changes in the rotation measure (RM) of FRB 20121102A \citep[which may be embedded in a young nebula e.g., see][]{Marcote_2017} while \citet{Hessels_2019} found evidence for DM changes of $\sim$1-4 pc cm$^{-3}$ over a four year period for this source. \citet{Mckinven2023b} and Ng et al. (in prep.) also both find that most repeating FRBs show subtle RM changes of order a few rad m$^{-2}$, while some show significant RM variations. However, \citet{Mckinven2023b} did not perform detailed studies of the DMs of the sources in their analysis. \citet{bsm+22} also found changes in the scintillation bandwidth for FRB 20180916B with time. 

Here, we explore whether any sources in our sample show changes in their DM and scattering timescale as a function of time. In particular, we focus on the repeating sources FRB 20191106C, FRB 20200929C, FRB 20181119A, FRB 20190208A, FRB 20190303A, and FRB 20190417A. These sources are chosen as they each have six or more bursts with the complex raw voltages recorded\footnote{A minimum of six bursts was chosen in order to have a sufficiently large sample of burst property measurements while also enabling examination of multiple sources from our sample.}. In addition to DM and scattering timescales, we also explore whether the spectral fluence of a burst impacts (or is impacted by) any other burst property and whether there is a systematic change in the emitting bandwidth of these sources \citep[largely motivated by FRB 20180916B, whose activity window is frequency dependent e.g., see ][]{pcl+21,pmb+21, bsm+22}.  To quantify changes in a given parameter, we perform a weighted least-squares fit assuming a linear function. We do not include bursts for which we either calculate an upper limit on the scattering timescale, or for which we do not report a DM uncertainty.

\begin{figure}[]
    \centering    
    \includegraphics[width=0.99\columnwidth]{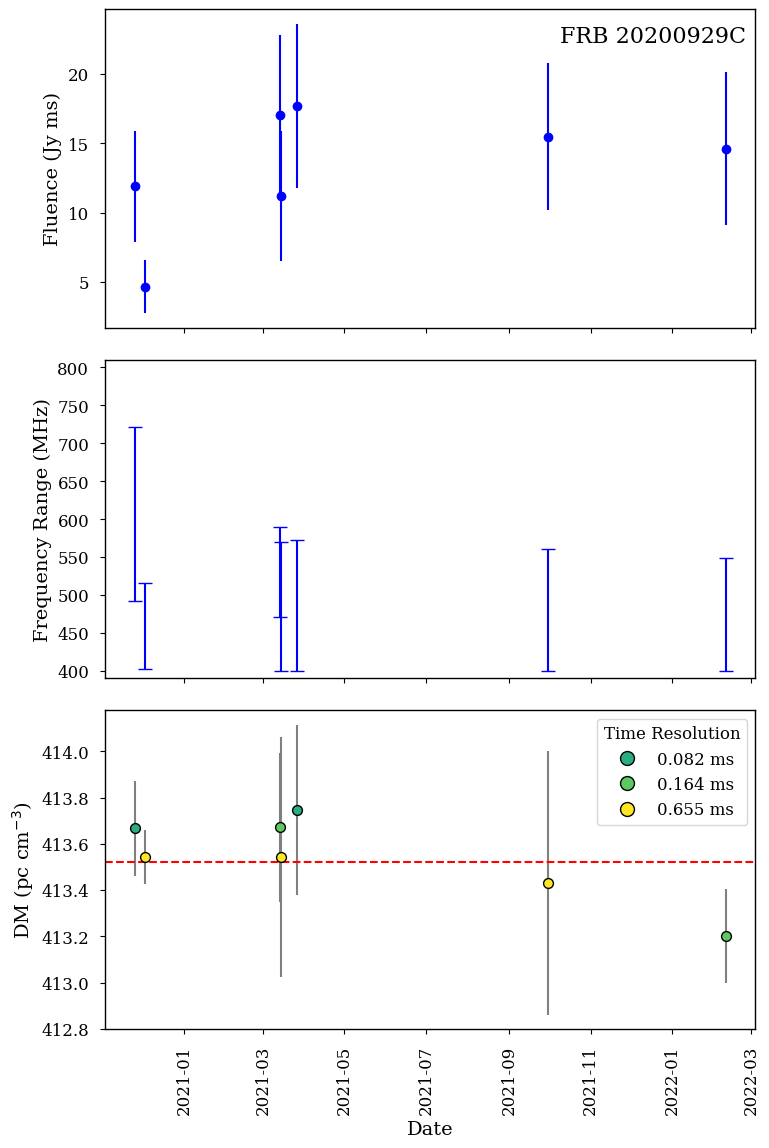}
    \caption{Same as Figure \ref{fig: R6, FRB 20181119A time dependency} except for FRB 20200929C.}
    \label{fig: R68; FRB 20200929C time dependency.}
\end{figure}

In Figures \ref{fig: R6, FRB 20181119A time dependency} - \ref{fig: R68; FRB 20200929C time dependency.}, we show the fluence, DM, scattering timescale (if applicable\footnote{Only shown if we can measure a scattering timescale for at least one of the bursts from the source.}), and frequency range for FRBs 20181119A, 20190208A, 20190303A, 20190417A,
20191106C, and 20200929C, respectively. The best-fit slopes for spectral fluence, DM, and scattering timescale (along with their 1$\sigma$ uncertainties) are listed in Table \ref{Table:individual source variation w time}. Except for the temporal variation of the spectral fluence of FRB 20181119A, all other derived slopes are either not statistically significant or are due to underestimated measurement uncertainties. We do not perform the same comparison for bandwidth, as the central frequency is highly dependent on the CHIME/FRB beam response which we have not corrected for in this work. Thus, for bandwidth, only by-eye trends are reported\footnote{The actual bandwidths are still calculated using the \texttt{fitburst} model assuming a running power law.}.

\subsubsection{Spectral Fluence Variations over Time}
FRB 20181119A is the only source for which there is a significant (slope greater than the 3$\sigma$ uncertainties) change in the spectral fluence (over the entire 400-MHz CHIME/FRB band) with time (see Figure \ref{fig: R6, FRB 20181119A time dependency}). The best fit slope is $-1.2 \pm 0.2$ Jy ms yr$^{-1}$.  However, the reduced $\chi$-squared for this fit is $\sim$25, implying a poor fit to the data.

\begin{deluxetable*}{lccc}
\tablecaption{Individual Source Parameter Variations with Time\label{Table:individual source variation w time}}
\tablehead{
\colhead{Source} & \colhead{Spectral Fluence} & \colhead{DM} & \colhead{Scat. Time} \\
\colhead{} & \colhead{(Jy ms yr$^{-1}$)} & \colhead{(pc cm$^{-3}$ yr$^{-1}$)} & \colhead{(ms yr$^{-1}$)}
}
\startdata
FRB 20181119A & $-1.2 \pm 0.2$ & $1.3 \pm 0.3$ & --\\
FRB 20190208A & $-0.001 \pm 0.9$ & $0.1 \pm 0.07$ & -- \\
FRB 20190303A & $-0.7 \pm 0.4$ & $-0.04 \pm 0.02$ & $-0.2 \pm 0.07$\\
FRB 20190417A & $-1.8 \pm 2.5$ & $-0.09 \pm 0.6$ & $0.1 \pm 0.2$\\
FRB 20191106C & $14 \pm 6$ & $1 \pm 2$ & $-1.5 +/- 0.6$\\
FRB 20200929C & $9 \pm 5$ & $-0.3 \pm 0.06$ & -- \\
\enddata
\end{deluxetable*}

\subsubsection{Bandwidth Variations over Time}
No sources in our sample show any systematic, significant trend of frequency range with time. \citet{2024MNRAS.52710425S} found a decrease in both the central frequency and bandwidth of FRB 20220912A over an approximately two month observational period. However, there was significant scatter in their measurements, and given our small sample size per source, it is possible that such a correlation is not yet obvious. Additionally, as we have not corrected for the primary beam effects of CHIME for individual source spectra, we do not perform an analysis of the central frequency as a function of time, as these measurements would be highly biased by beam effects.

\subsubsection{DM Variations over Time}
Determining the DM for a given burst using \texttt{fitburst} is particularly difficult for repeating FRBs due to both the complex frequency and time structure displayed and the (on average) low S/N of these bursts. Thus, for our DM comparisons, we instead use the DM determined using the \texttt{DM phase} routine \citep{dm_phase} as we found that it visually performed better for our sample. Unlike \texttt{fitburst} which fits a set of modified Gaussians to the data, \texttt{DM phase} maximizes the structure of the burst and is significantly more indifferent to the intrinsic shape of the data than \texttt{fitburst}. Thus, for highly complex bursts, the \texttt{DM phase} package often better visually aligns a burst. 

Other than FRB 20181119A (Figure \ref{fig: R6, FRB 20181119A time dependency}), none of the sources in our sample have a significant secular increase or decrease in DM as a function of time. While the trend for FRB 20181119A is significant, it is based on only four data points, and does not include the final two measurements that were estimated by-eye. Additionally, the reduced chi-squared for this fit is $\sim$ 60, suggesting it is a very poor fit to the data. We also manually de-disperse all bursts from this source to the weighted-average DM and all bursts still appear fully de-dispersed. Thus, the variation for this source is likely not astrophysical and is instead due to difficulty in determining the DMs of the bursts as many have low S/Ns without obvious sub-structure. 

FRB 20190208A also shows significant variation about the mean DM. However, we de-disperse the three outlier bursts to the average DM and all appear fully de-dispersed. Thus, similar to FRB 20181119A, the variation is likely not astrophysical but solely due to our inability to determine precise DMs for certain bursts. The DMs for the remaining sources (FRB 20190303A, FRB 20190417A, FRB 20191106C, and FRB 20200929C) show small variations, but most are still consistent with the respective inverse-variance weighted DM within uncertainties.

\subsection{Scattering Timescale Variations over Time}
\begin{figure}[]
    \centering    \includegraphics[width=0.99\columnwidth]{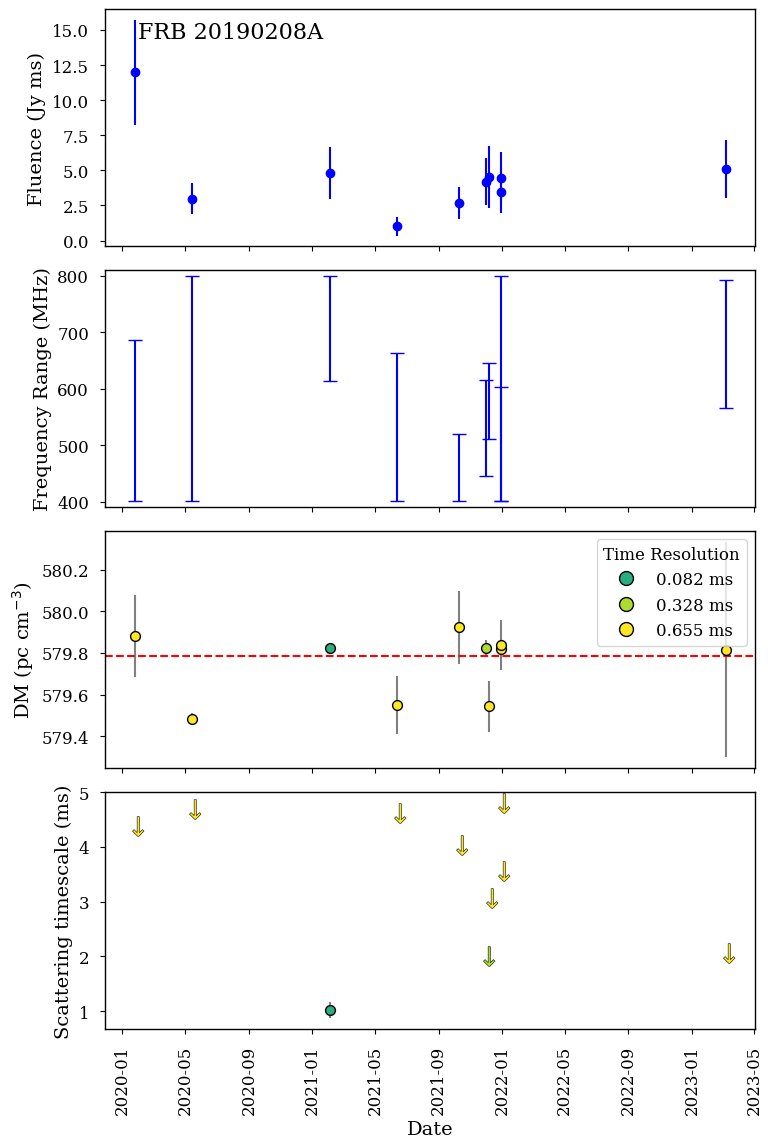}
    \caption{Same as Figure \ref{fig: R6, FRB 20181119A time dependency} except for FRB 20190208A. As FRB 20190208A has at least one burst with a measured scattering timescale, we also include the scattering timescale. For bursts for which there was no by-eye visible scattering timescale, we determine an upper limit on the scattering timescale as equivalent to the reported width of the narrowest sub-burst.}
    \label{fig: R12; FRN 20190208A time dependency}
\end{figure}

\begin{figure}[]
    \centering    \includegraphics[width=0.99\columnwidth]{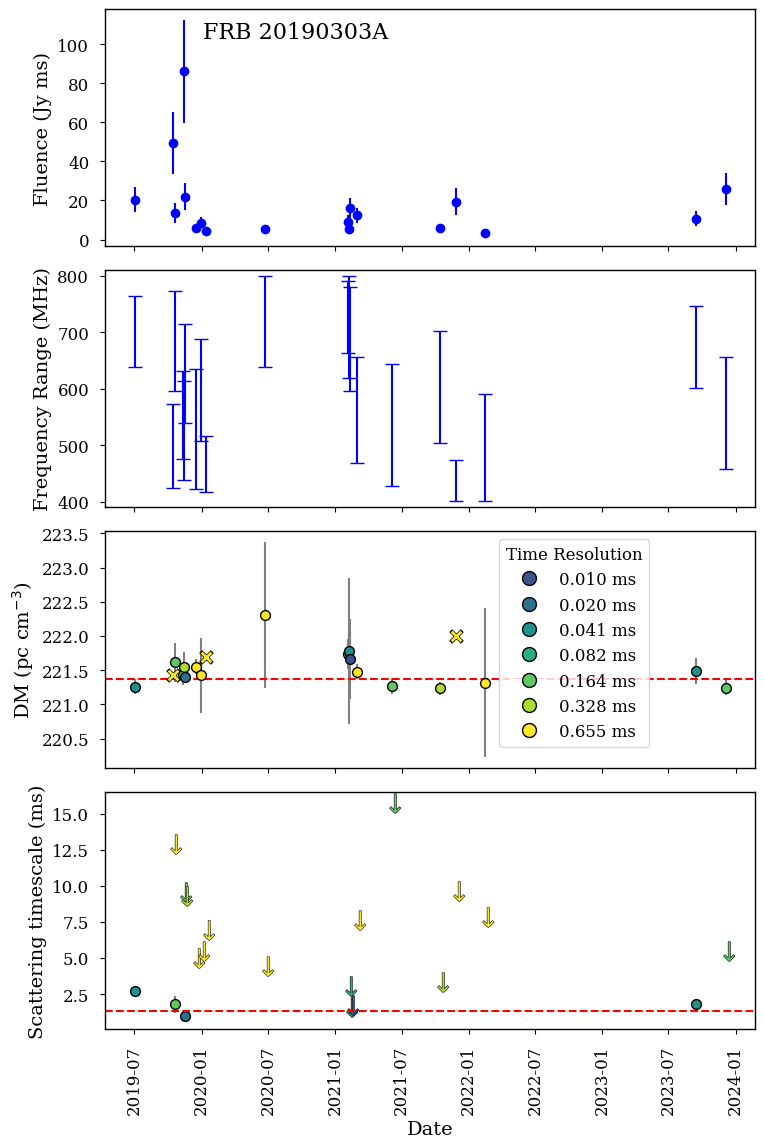}
    \caption{Same as Figure \ref{fig: R12; FRN 20190208A time dependency} except for FRB 20190303A. We do not include an extreme outlier burst for which the fluence was 351 Jy ms on 2021-06-06.}
    \label{fig: R17; FRB 20190303A time dependency}
\end{figure}

\begin{figure}[]
    \centering    \includegraphics[width=0.99\columnwidth]{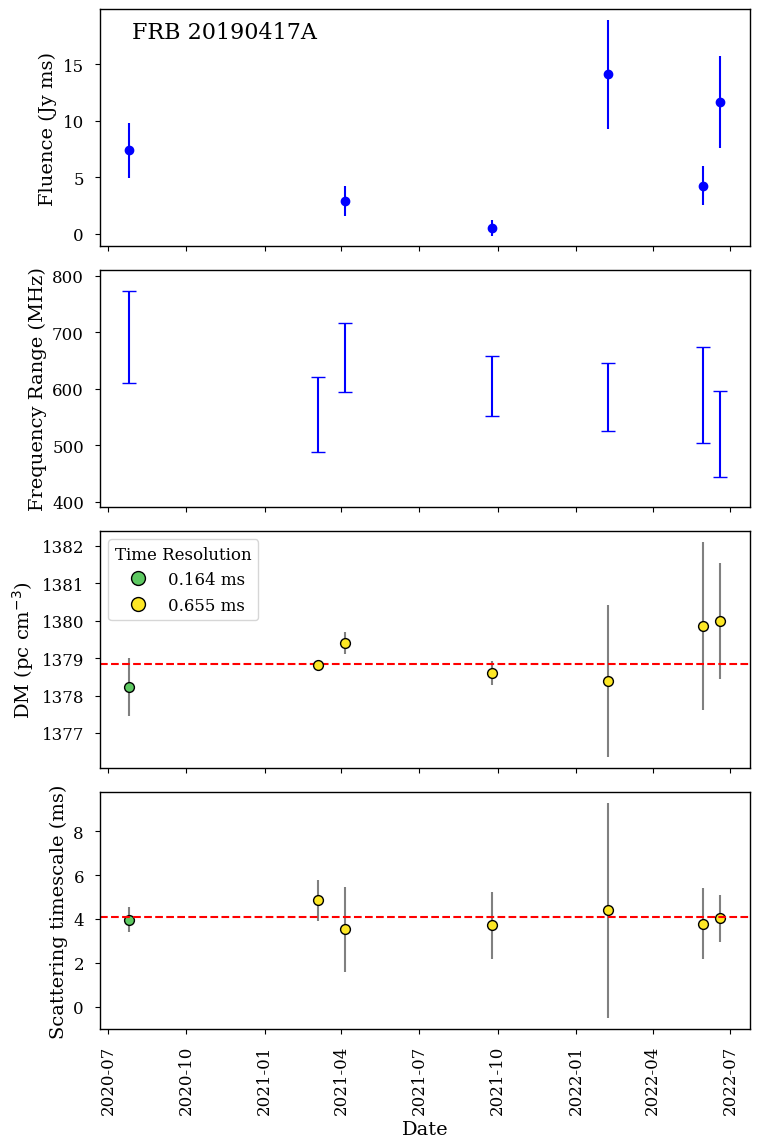}
    \caption{Same as Figure \ref{fig: R12; FRN 20190208A time dependency} except for FRB 20190417A}
    \label{fig: R18; FRB 20190417A time dependency}
\end{figure}

\begin{figure}[]
    \centering    \includegraphics[width=0.99\columnwidth]{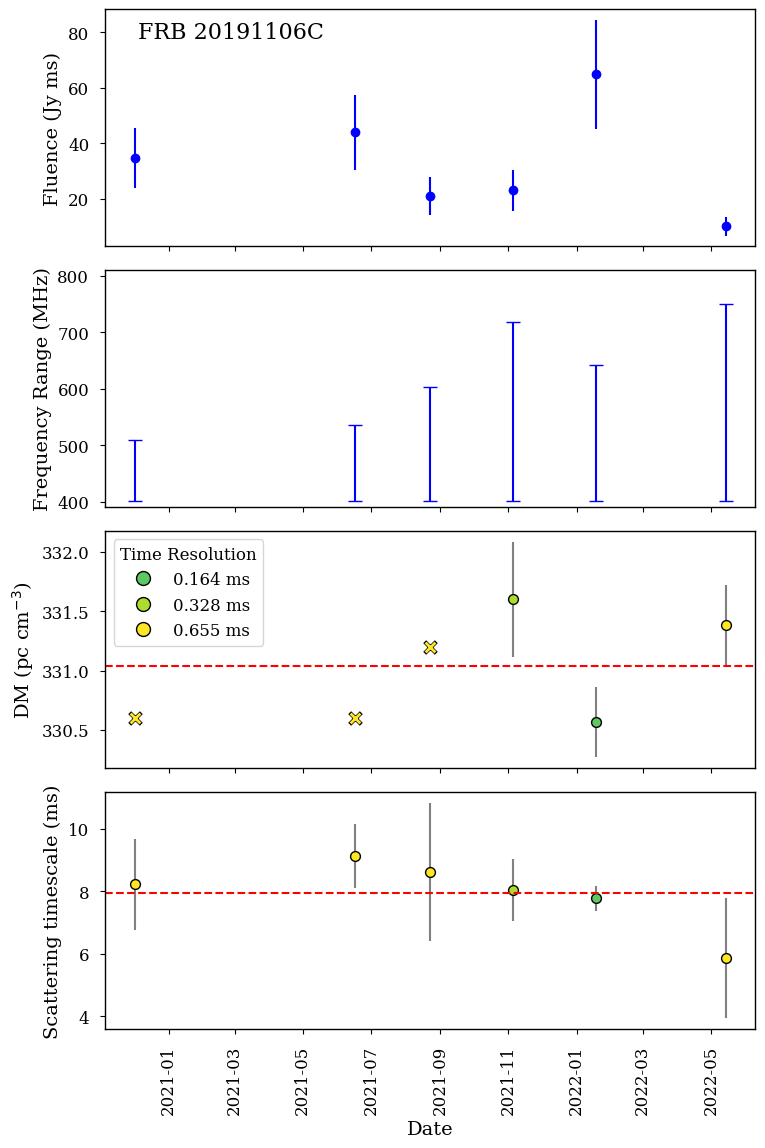}
    \caption{Same as Figure \ref{fig: R12; FRN 20190208A time dependency} except for FRB 20191106C.}
    \label{fig: R34; FRB 20191106C time dependency}
\end{figure}

\noindent The measured scattering timescale per burst is determined using the 2D spectro-temporal fit from \texttt{fitburst}. Due to the degeneracy between scattering and downward drifting, we only report scattering timescales for bursts for which we can see by-eye evidence for scattering. We do not find evidence for a secular change in the scattering timescales of FRB 20190303A (Figure \ref{fig: R17; FRB 20190303A time dependency}), FRB 20190417A (Figure \ref{fig: R18; FRB 20190417A time dependency}), or FRB 20191106C (Figure \ref{fig: R34; FRB 20191106C time dependency}). All of our measured scattering timescales for the two FRBs with scattering upper limits, FRB 20190208A (Figure \ref{fig: R12; FRN 20190208A time dependency}) and FRB 20190303A (Figure \ref{fig: R17; FRB 20190303A time dependency}), are consistent with the scattering times that we report for the sources. 

\begin{figure}[]
    \centering    \includegraphics[width=0.99\columnwidth]{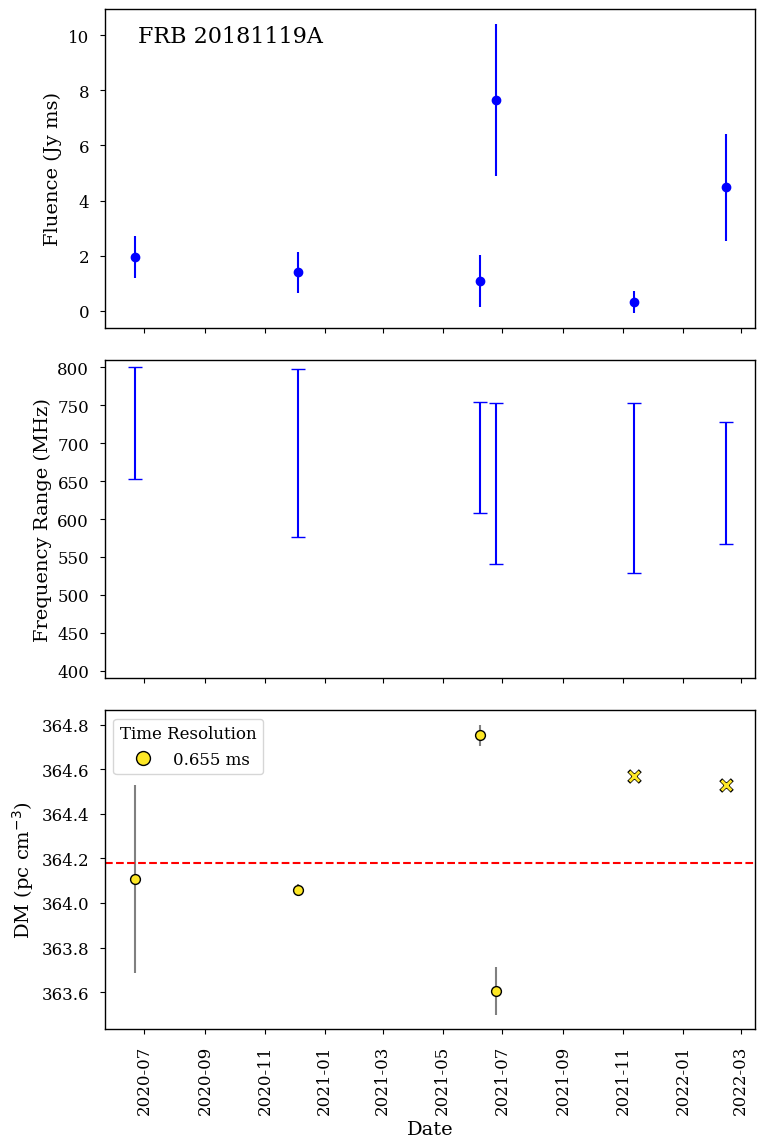}
    \caption{Band-averaged fluence (calculated over the entire CHIME/FRB bandwidth), CHIME/FRB frequency range, and DM for FRB 20181119A as a function of time. Error bars for fluence and DM are shown at the 99\% confidence level. The time resolution at which the burst is analyzed is shown by the color of the points. The lowest time resolution is 0.655 ms while the highest is 2.56 $\upmu \textrm{s}$.  For this source, all bursts were analyzed at a time resolution of 0.655 ms, so no variation is seen in the color. We note that the uncertainties for bursts studied at a time resolution of 0.655 ms may be underestimated due to the inability to discern structure at such a low time resolution. For the DM, we show the inverse-variance weighted average as a red line. Bursts for which we could not calculate an optimal structure-maximized DM, and hence for which a by-eye DM was estimated, are shown as crosses. By-eye DMs were only estimated to within $\sim$0.5 pc cm$^{-3}$}
    \label{fig: R6, FRB 20181119A time dependency}
\end{figure}

\section{Discussion}
\label{sec: discussion}

\subsection{Local Environments}
For the six repeating FRBs presented in Section \ref{ssec:Individual Sources}, we do not find evidence for significant changes in scattering or DM variations with time. Long-term monitoring of other highly active sources such as FRB 20180916B \citep{Sand2023R3}, FRB 20201124A \citep{lac+22}, and FRB 20220912A \citep{2024Nancay} have also not yet shown any significant DM variations with time, although  they found low-level
($\sim$ 1 pc cm$^{-3}$), short timescale DM variations. While this might suggest that most FRBs do not lie in environments similar to that of FRB 20121102A \citep{michilli2018extreme,hss+19}, the total sample of repeaters for which possible DM variations have been studied remains small. Additionally, possible changes in the DM due to a putative expanding supernova shell are highly dependent on the age of the remnant, with certain periods having little to no change in the DM \citep{pg18}.

\citet{pkj+13} searched for systematic DM variations in 168 Galactic radio pulsars, and found that only four showed significant variations of order 0.005 to 0.2 pc cm$^{-3}$ yr$^{-1}$. They attributed these DM variations to either an associated pulsar wind nebulae/supernova remnant or a turbulent ISM. However, these variations are significantly below the constraints placed here. \citet{dep+18} also searched for a systematic DM variation in the Galactic centre magnetar, PSR J1745$-$2900. They found a decline of $\sim$10 pc cm$^{-3}$ over a period of four years. However, their results were consistent with no secular change within 2$\sigma$.

As discussed in Section \ref{ssec:RepeatersVsNonrepeaters}, we do not find a statistical difference in the scattering timescales between repeating and non-repeating FRBs. \citet{2024BasecatPolarization} found no significant difference in the linear polarization fraction or the extragalactic RM when comparing these two populations. They did not, however, find any thus-far non-repeaters with extreme RMs, in contrast with the extreme RMs seen in the repeating FRBs 20121102 and 20190520B \citep{michilli2018extreme, Anna_Thomas_2023}. This could be explained by intra-channel depolarization in the CHIME/FRB band as CHIME/FRB would have an approximately 50\% decrease in its sensitivity to bursts with RMs $\approx$ 5000 rad m$^{-2}$ at 600 MHz due to depolarization \citep{mmm+21b}.

Nonetheless, taken together, these findings suggest that repeating and non-repeating FRBs share similar local environments. This supports the notion that the differences observed in the DM distributions by \citet{RN3} are unlikely to be driven by local environmental factors. Nonetheless, the small number of known repeating FRBs compared to non-repeaters suggests that these conclusions should be viewed cautiously, as additional data may reveal more subtle differences.

\subsection{Luminosity}
Similar to the results of \citet{2024Sand}, we do not find a correlation between extragalactic (+halo) DM and fluence for our repeating FRB sources.  As seen in Figure \ref{Figure: Morphological Correlations}, a single source can emit bursts with fluences ranging from $10^{-2}$ to 10 Jy ms. This is similar to FRB 20201124A, which exhibits fluence values spanning five orders of magnitude \citep{xnc+22, koh+24} and to SGR 1935+2154, which also emits bursts spanning $\sim$7 orders of magnitude \citep{2021NatAs...5..414K}. Together, this suggests that FRBs have broad luminosity distributions. 

\subsection{Multiple FRB Classes?}
Our work further supports the results of \citet{pleunismorph} and \citet{2024Sand} who find clear morphological distinctions between repeating and thus-far non-repeating FRBs. Repeating FRBs tend to be narrower in bandwidth and longer in total duration than non-repeating FRBs (see Figure \ref{fig: Bandwidth vs. width; repeaters nonrepeaters}). We also find differences between the band-averaged fluences, sub-burst durations, and sub-burst waiting times of the two populations. Interestingly, when we normalize the sub-burst durations and waiting times of the two populations by the respective burst's total duration, the distributions are consistent with being drawn from the same population (see Figure \ref{fig:sub-burst properties; repeaters and nonrepeaters}). This could hint at a similar physical mechanism for the scaling of sub-burst width to the common envelope. However, we caution that for the repeating FRB population, we may be missing additional sub-bursts as most bursts were not studied at as high of a time resolution as the non-repeating sample. A larger number of repeating sources        studied at microsecond time resolutions should provide further insight into this.

The band-averaged fluences over the burst spectral extents for the two populations are consistent with being drawn from the same distribution. Given that \citet{RN3} found a significant difference between the DM distributions of the two samples, along with the larger bandwidths of non-repeaters, this could suggest a difference in their intrinsic specific energies. One possible interpretation is that repeaters and non-repeaters are produced by the same populations of physical object with non-repeaters the brightest bursts from a given repeater, primarily due to their larger bandwidths but also due to higher specific energies.

\citet{koh+24} found that the energy-distribution for the high-energy bursts from FRB 20201124A was consistent with that of non-repeaters, similarly suggesting that non-repeaters might be the brightest burst from a repeater. Most recently, \citet{2024NorthernCrossProject} and \citet{ouldboukattine2024probemaximumenergeticsfast} both found similar results for FRB 20220912A. However, \citet{koh+24} did not find any significant morphological or spectral differences between the high-energy bursts from FRB 20201124A and the lower-energy bursts. This could be in-part due to their comparatively smaller bandwidth. We also did not find any distinct morphological differences between the brightest bursts from our sources and the other bursts from the same source. A larger population of bright, repeating sources detected over large bandwidths is needed to explore this in further detail.

\section{Summary and Future Work}
\label{sec: summary}

In this work, we present for the first time a morphological analysis of the raw voltage data for 124 bursts from 35 unique repeating sources detected by CHIME/FRB. We present spectro-temporal fits for each burst at timescales down to microseconds, along with improved flux and fluence measurements for each burst. We search for correlations amongst the burst fluences, DMs, burst rate, and burst duration. We do not find any significant correlations amongst these parameters, even with improved fluences, burst rates, and burst duration estimates. The lack of a correlation between fluence and extragalactic DM (+MW halo) for our sources suggests that repeating FRBs have intrinsically large luminosity ranges. Additionally, the lack of a correlation between burst rate and total duration may suggest that that burst widths are not correlated with their beaming angle. We also search for an extragalactic DM (+MW halo) and scattering correlation, but again do not find one. A lack of a correlation suggests that the medium responsible for the majority of the extragalactic DM is not the same which primarily contributes to the sources scattering timescales. 

We perform the first large-scale comparison of the morphology of the repeating and non-repeating FRB populations at timescales down to microseconds. Using the non-repeating FRB sample presented in \citet{2024Sand}, we confirm the results of \citet{pleunismorph} that non-repeating FRBs have, on average, larger bandwidths and shorter temporal durations than repeating sources. Additionally, with increased time resolution, we compare the sub-burst properties of the two populations. Our results suggest that their may be a common duration normalized sub-burst duration amongst the two populations. Additionally, we find that the spectral fluences of the two populations over their spectral extents are consistent with being draw from the same distribution, despite non-repeating FRBs having larger bandwidths and, on average, larger DMs. This suggests that non-repeating FRBs may have intrinsically higher specific energies than repeating FRBs. 

We also do not find a significant difference in the scattering timescales of the two populations. In combination with the work of \citet{pandhi2024polarizationproperties128nonrepeating}, this suggests that repeating and non-repeating FRBs do not generally have different local environments. However, the repeating FRB sample size for this comparison is low, and hence we caution strong interpretations of this result.

Lastly, we examine the temporal variations of spectral fluence, bandwidth, DM, and scattering for six of the FRB sources in our sample. Only one source, FRB 20181119A, shows a significant trend in its fluence as a function of time, with a rate of $-1.2 \pm 0.2$ Jy ms yr$^{-1}$. None of our sources show secular increases/decreases in the DM and scattering timescales, constrained to $\lessapprox 1$ pc cm$^{-3}$ yr$^{-1}$ and $\lessapprox 2$ ms yr$^{-1}$, respectively.

CHIME/FRB, with its daily view of the sky north of $\delta = - 11^{\circ}$, excels at finding repeating FRBs.  The prospects for increasing the size of the known repeater population in the coming years is therefore excellent and will enable future similar studies with greater statistics. Additionally, with the upcoming CHIME/FRB Outrigger telescopes \citep{lanman_kko}, sub-arcsecond localizations will be available for most events, enabling host galaxy and redshift determinations, which will greatly aid many property comparisons and may eventually allow important correlations to emerge.

\section*{acknowledgments}

We acknowledge that CHIME is located on the traditional, ancestral, and unceded territory of the Syilx/Okanagan people. We are grateful to the staff of the Dominion Radio Astrophysical Observatory, which is operated by the National Research Council of Canada.  CHIME is funded by a grant from the Canada Foundation for Innovation (CFI) 2012 Leading Edge Fund (Project 31170) and by contributions from the provinces of British Columbia, Qu\'{e}bec and Ontario. The CHIME/FRB Project, which enabled development in common with the CHIME/Pulsar instrument, is funded by a grant from the CFI 2015 Innovation Fund (Project 33213) and by contributions from the provinces of British Columbia and Qu\'{e}bec, and by the Dunlap Institute for Astronomy and Astrophysics at the University of Toronto. Additional support was provided by the Canadian Institute for Advanced Research (CIFAR), McGill University and the McGill Space Institute thanks to the Trottier Family Foundation, and the University of British Columbia. The CHIME/Pulsar instrument hardware was funded by NSERC RTI-1 grant EQPEQ 458893-2014.

\allacks
\appendix

Below, in Figure \ref{fig:burst_wfalls}, we include the waterfall plots for a sub-sample of the bursts analyzed in this work. In Table \ref{Table: Burst Parameters}, we include the measured burst parameters for each burst. In Figure \ref{Figure:MaxWidthRate}, we show the maximum burst width from a given source versus its burst rate. In Figure \ref{Figure:DurationDMRepeaterNonRepeater}, we show the burst duration versus extragalactic DM (+halo) correlation including non-repeating FRBs from \citet{2024Sand}. In Figure \ref{Figure:FluenceDMRepeaterNonRepeater}, we show the fluence versus extragalactic DM (+halo) correlation including non-repeating FRBs from \citet{2024Sand}. In Figure \ref{fig:sub-burst properties; repeaters and nonrepeaters; higher time res}, we show comparisons of the repeating and non-repeating sample, but limit the sample to bursts studied at time resolutions $\leq 40 \; \upmu$s. In Figure \ref{Figure: BWvs.SN} we show the bandwidth versus detection S/N for all sources in our sample for which there are more than five bursts. Lastly, in Table \ref{Table:Source rates}, we list updated burst rates for the repeaters in our sample.

\setcounter{table}{0}
\renewcommand{\thetable}{A\arabic{table}}

\setcounter{figure}{0}
\renewcommand{\thefigure}{A\arabic{figure}}
\newpage

\begin{figure*}
\centering
\gridline{
          \fig{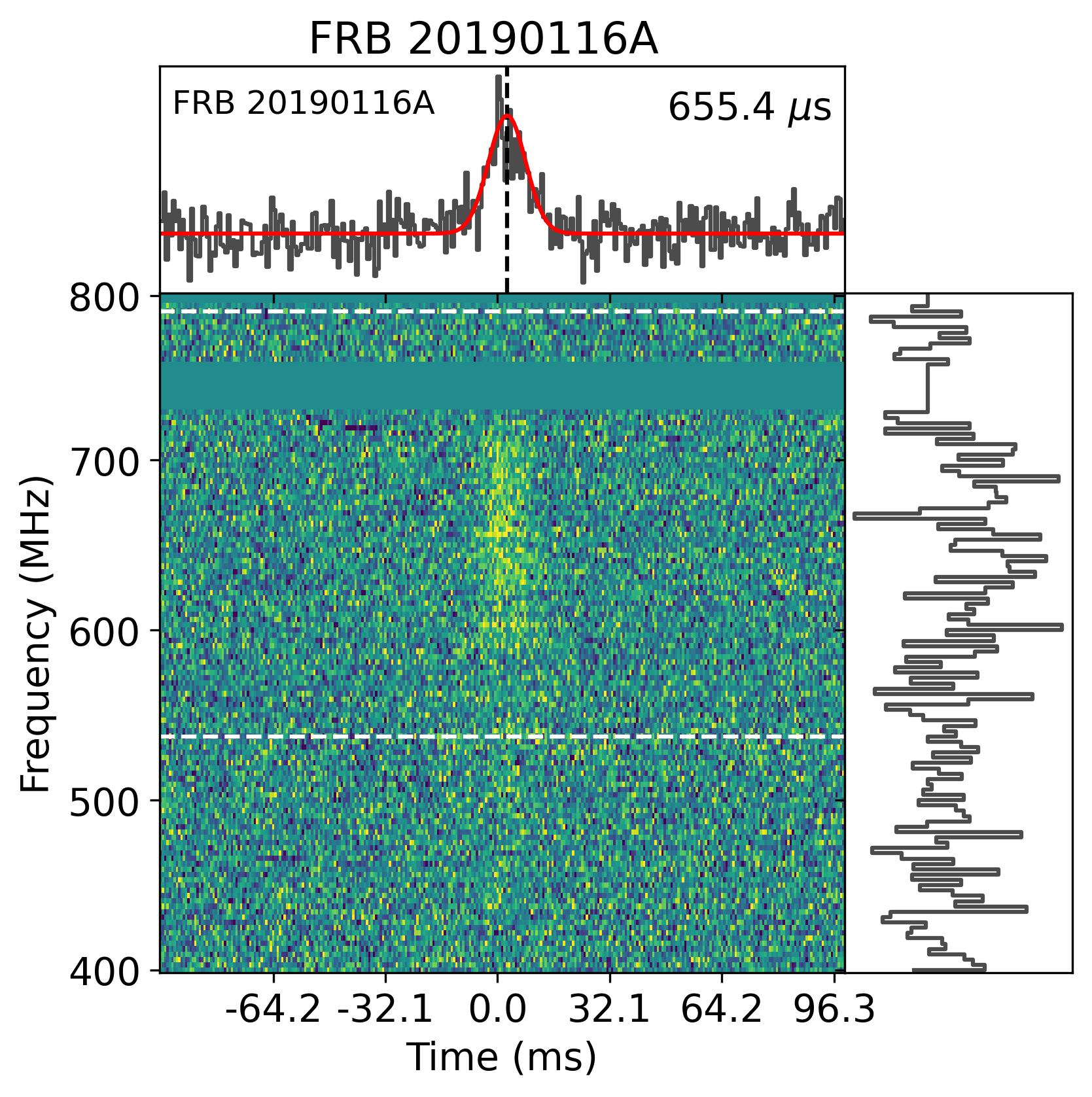}{0.24\textwidth}{}
          \fig{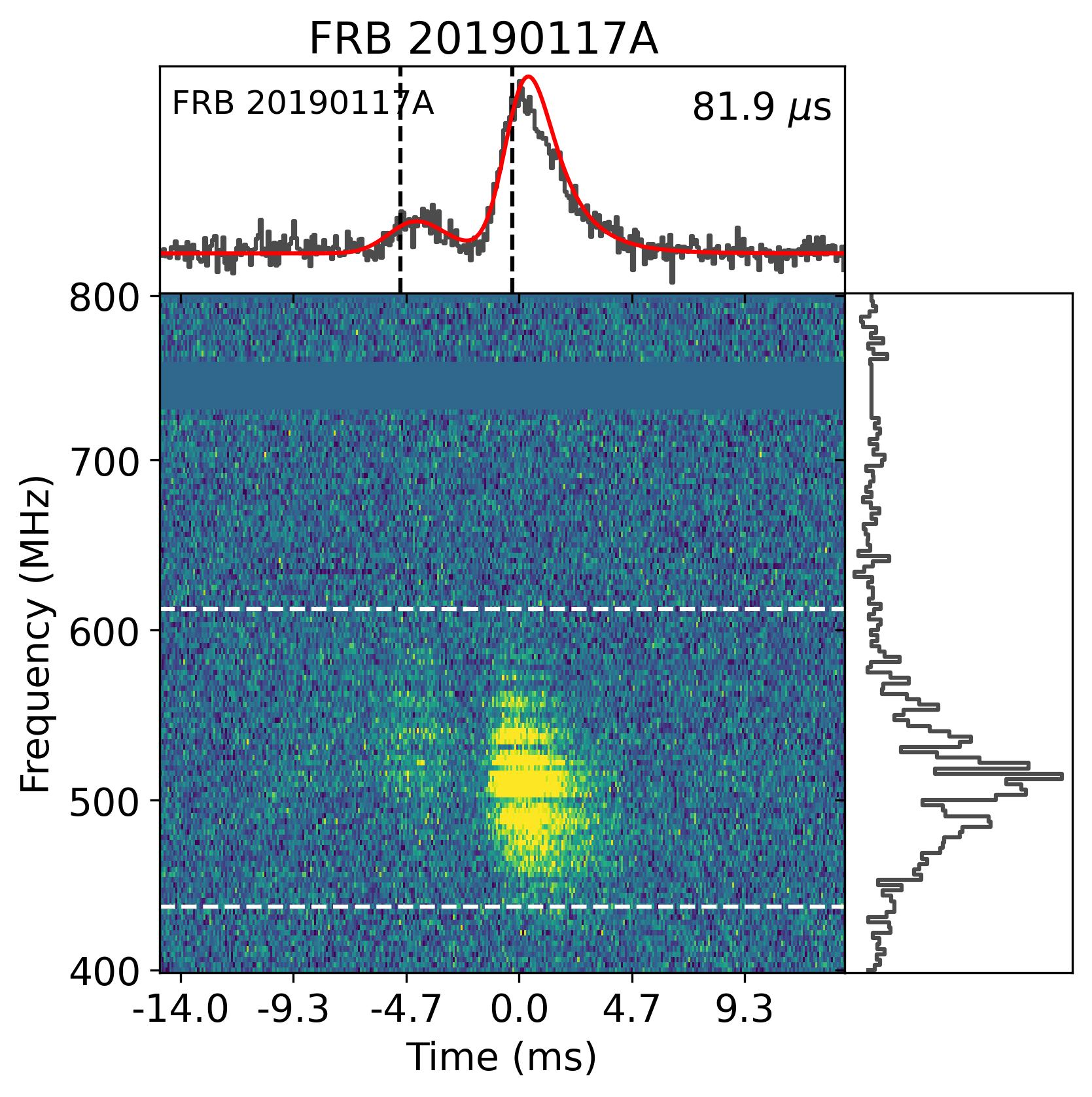}{0.24\textwidth}{}
          \fig{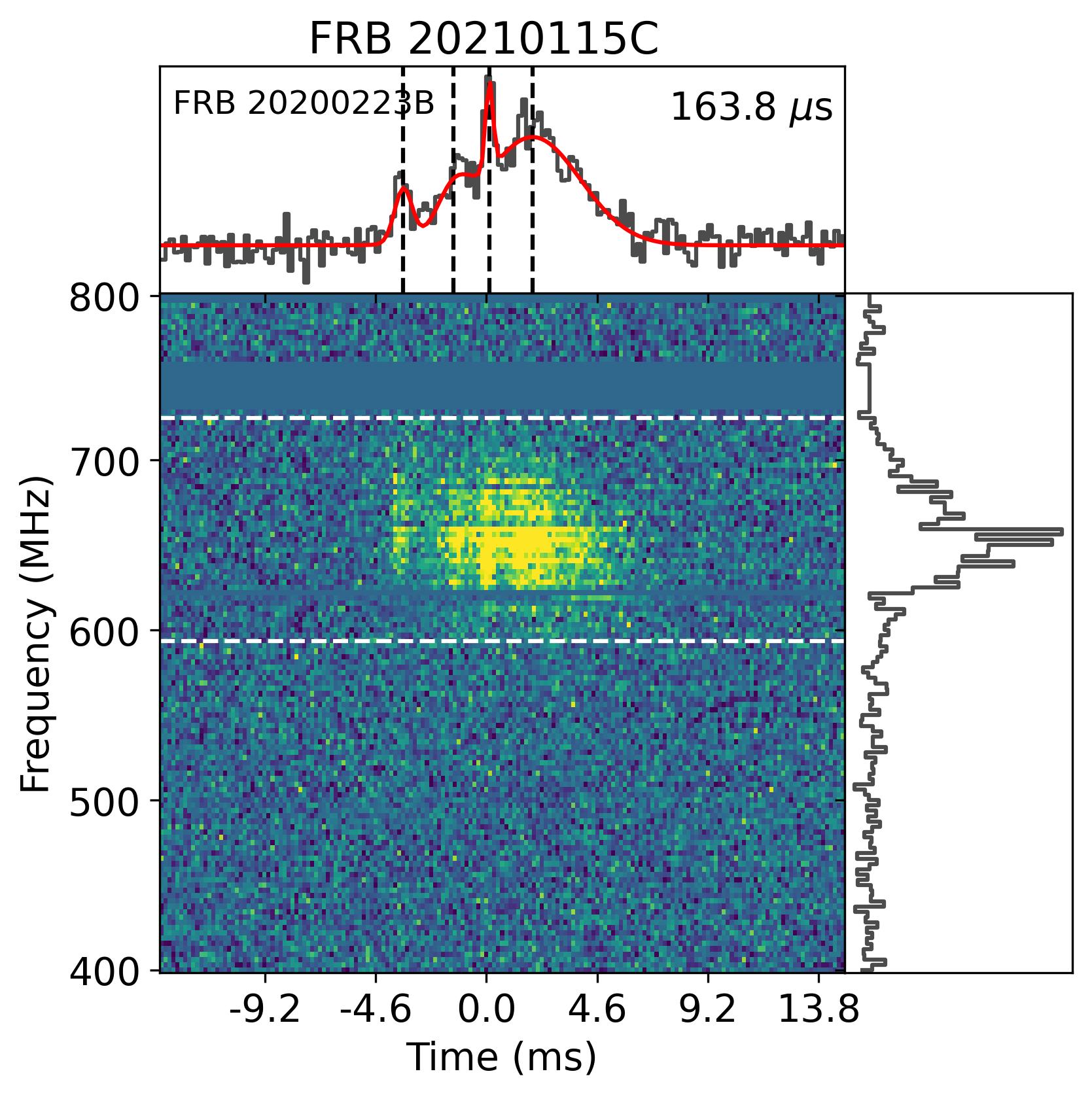}{0.24\textwidth}{}
          \fig{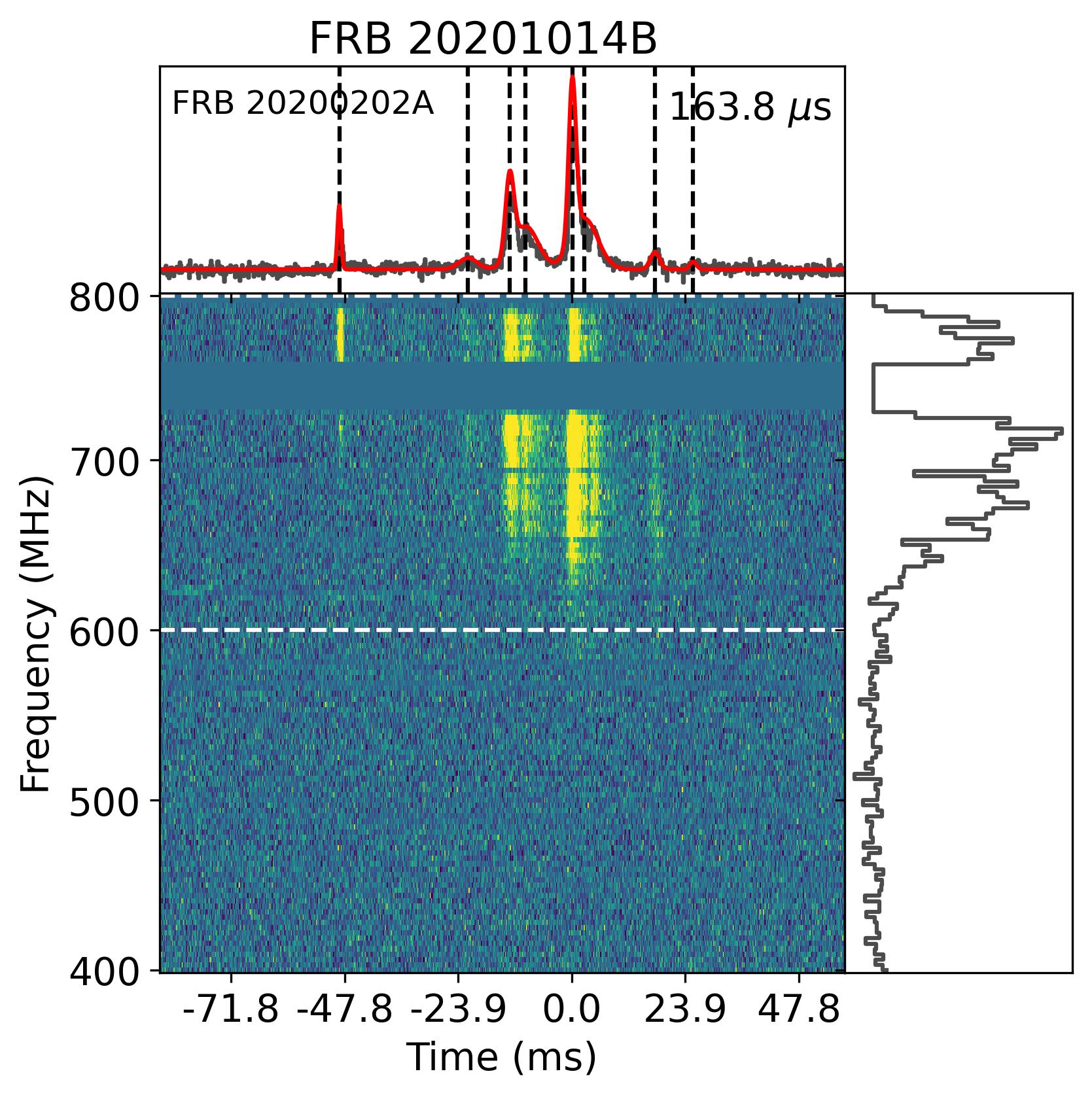}{0.24\textwidth}{}
          }
\gridline{\fig{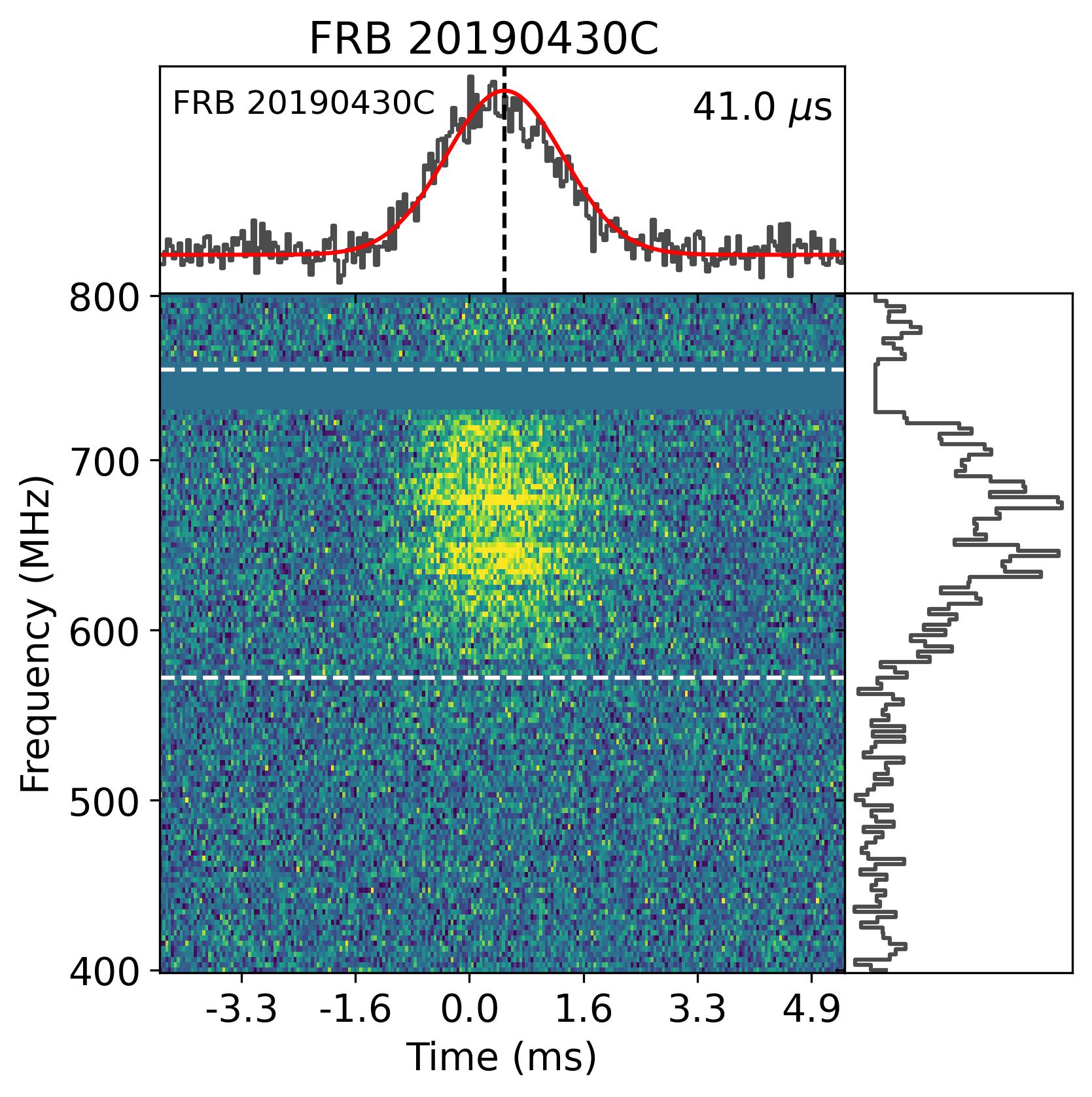}{0.24\textwidth}{}
          \fig{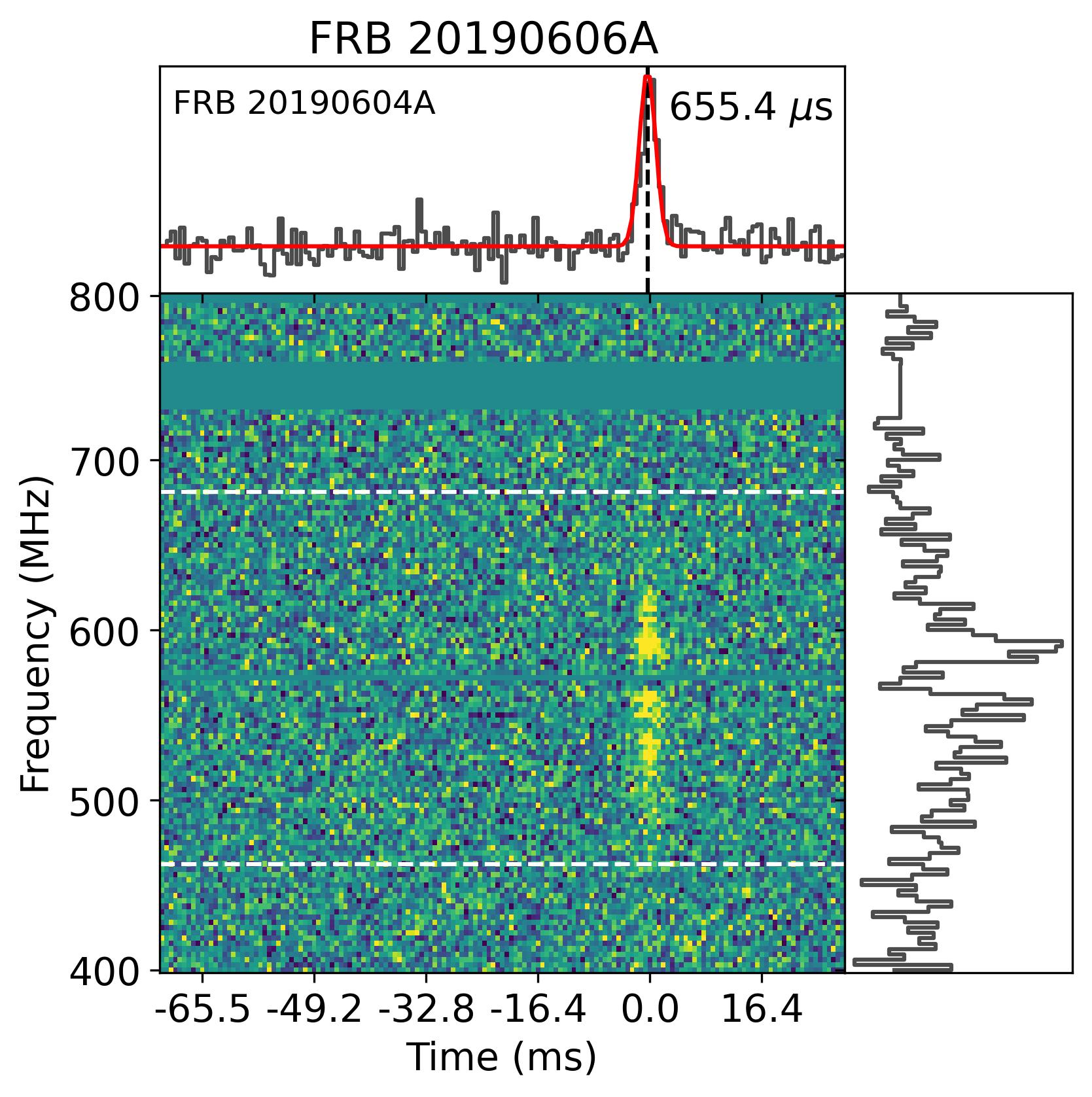}{0.24\textwidth}{}
          \fig{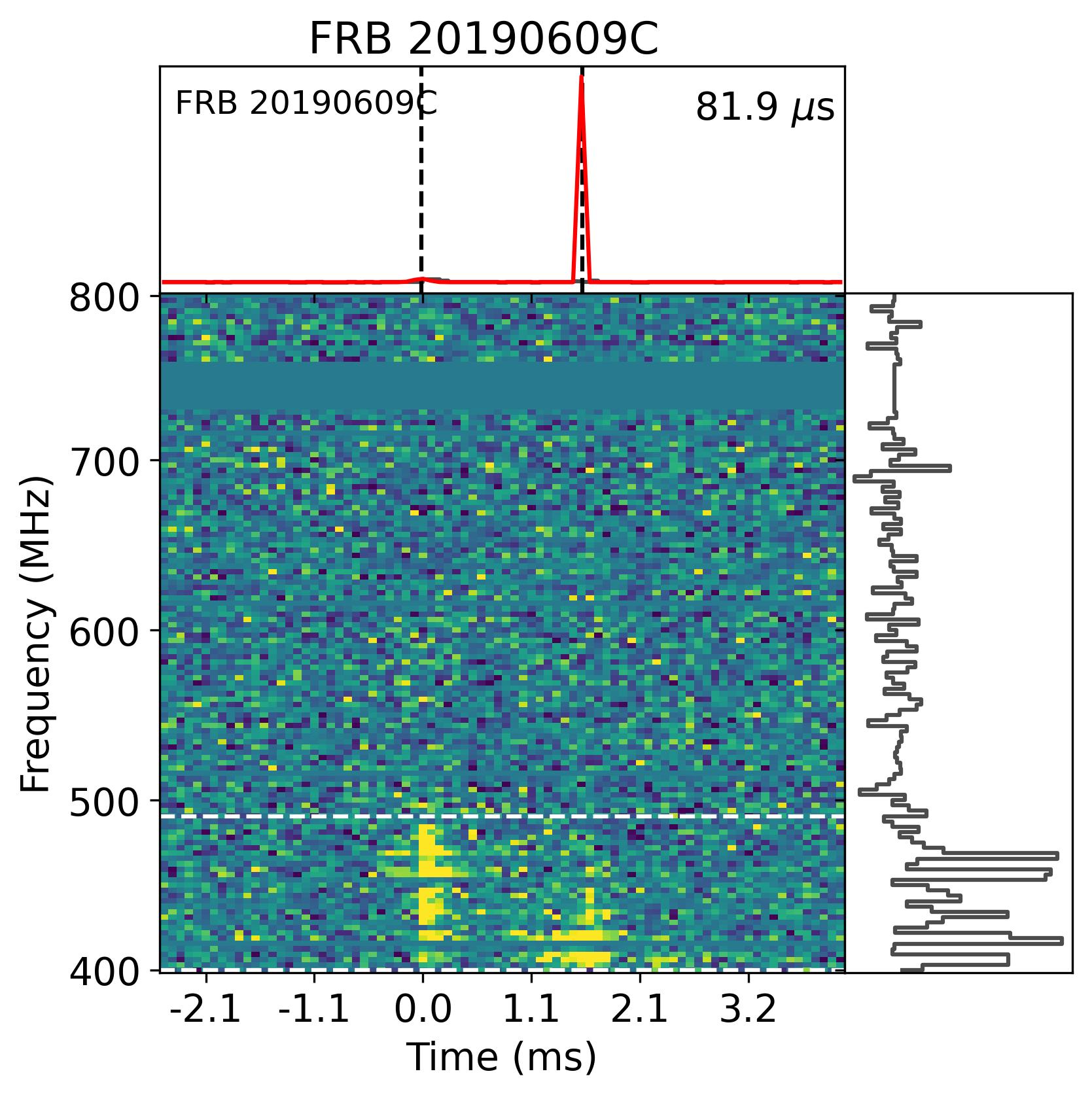}{0.24\textwidth}{}
          \fig{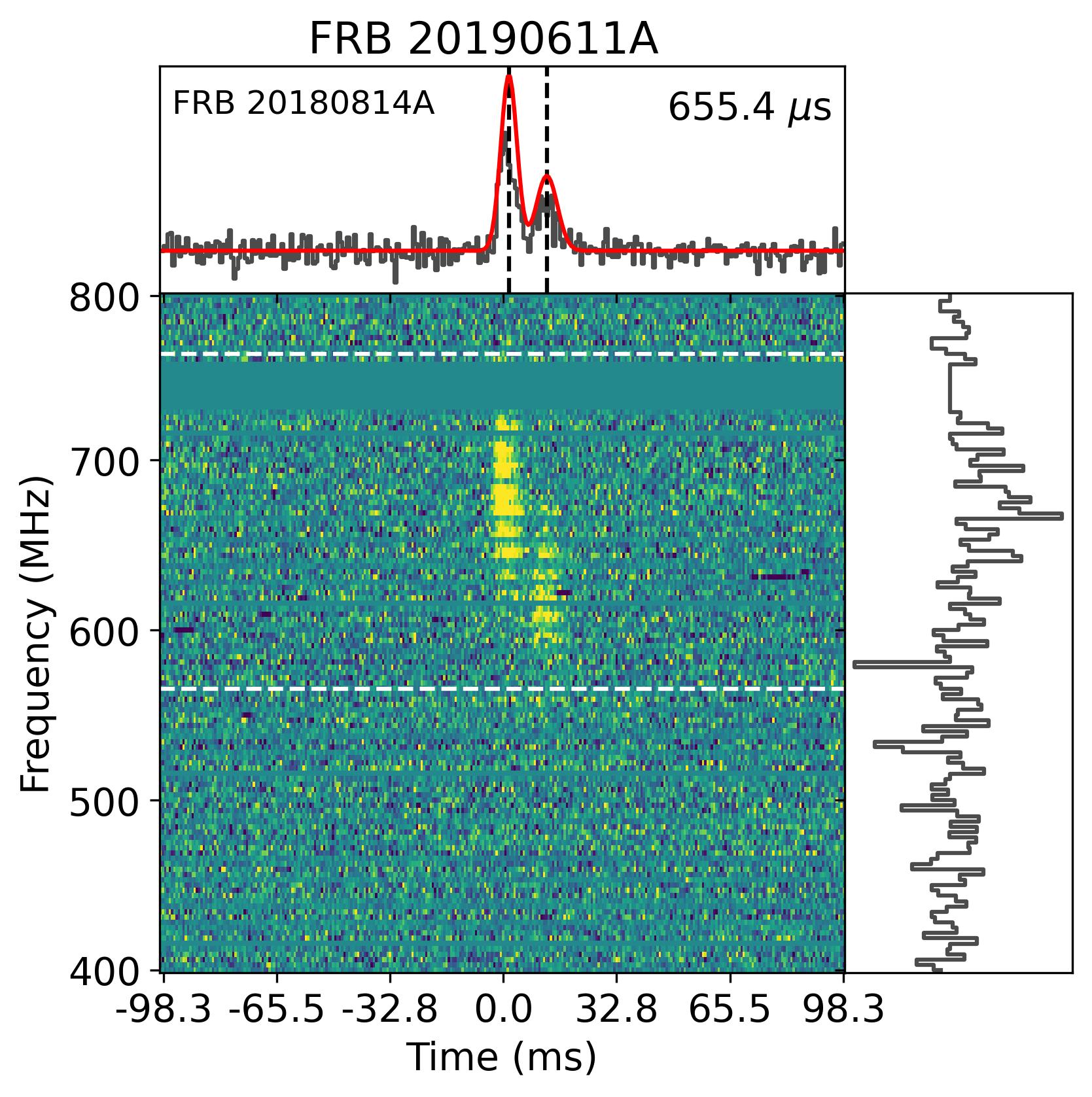}{0.24\textwidth}{}
          }
\gridline{\fig{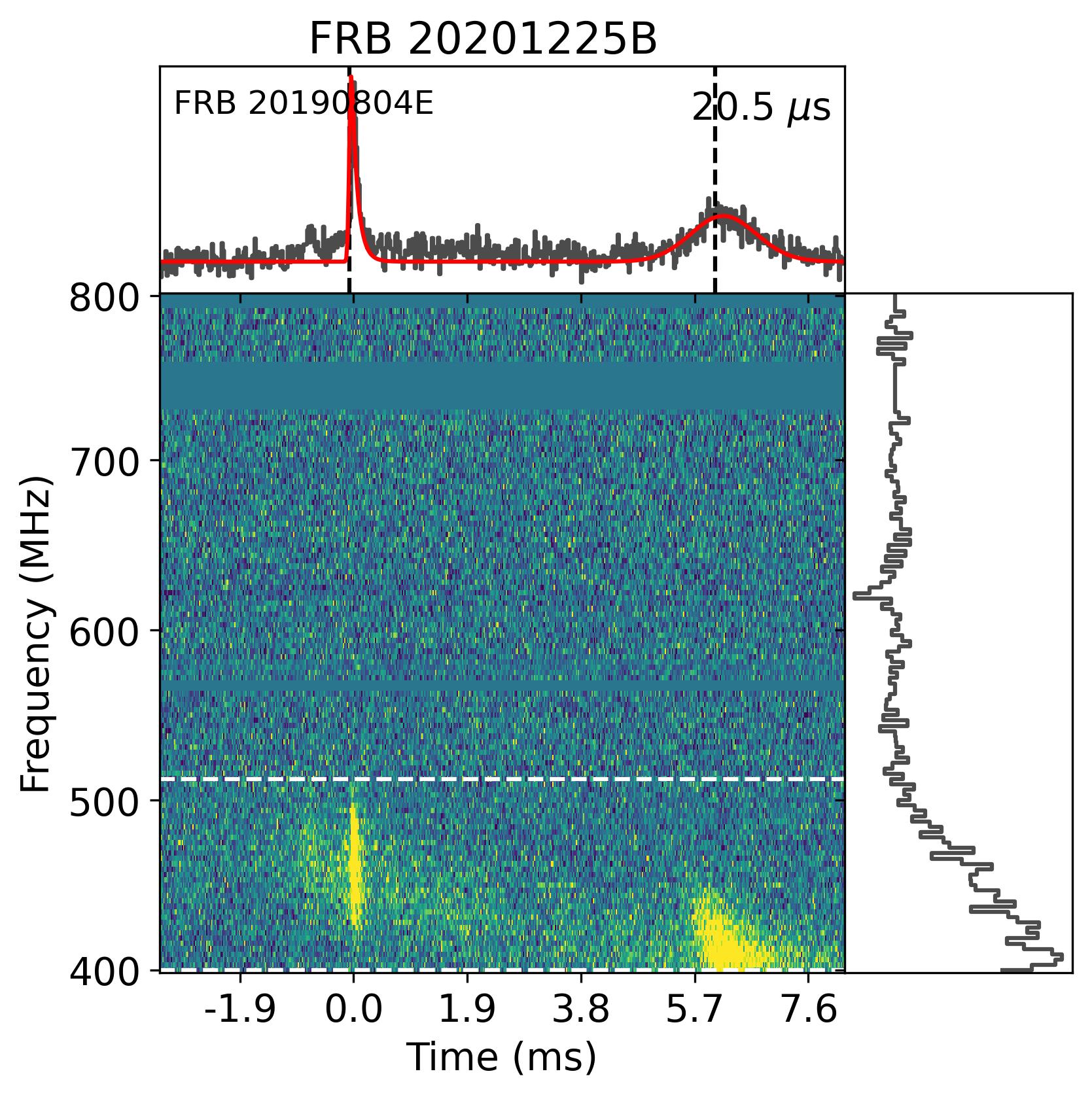}{0.24\textwidth}{}
          \fig{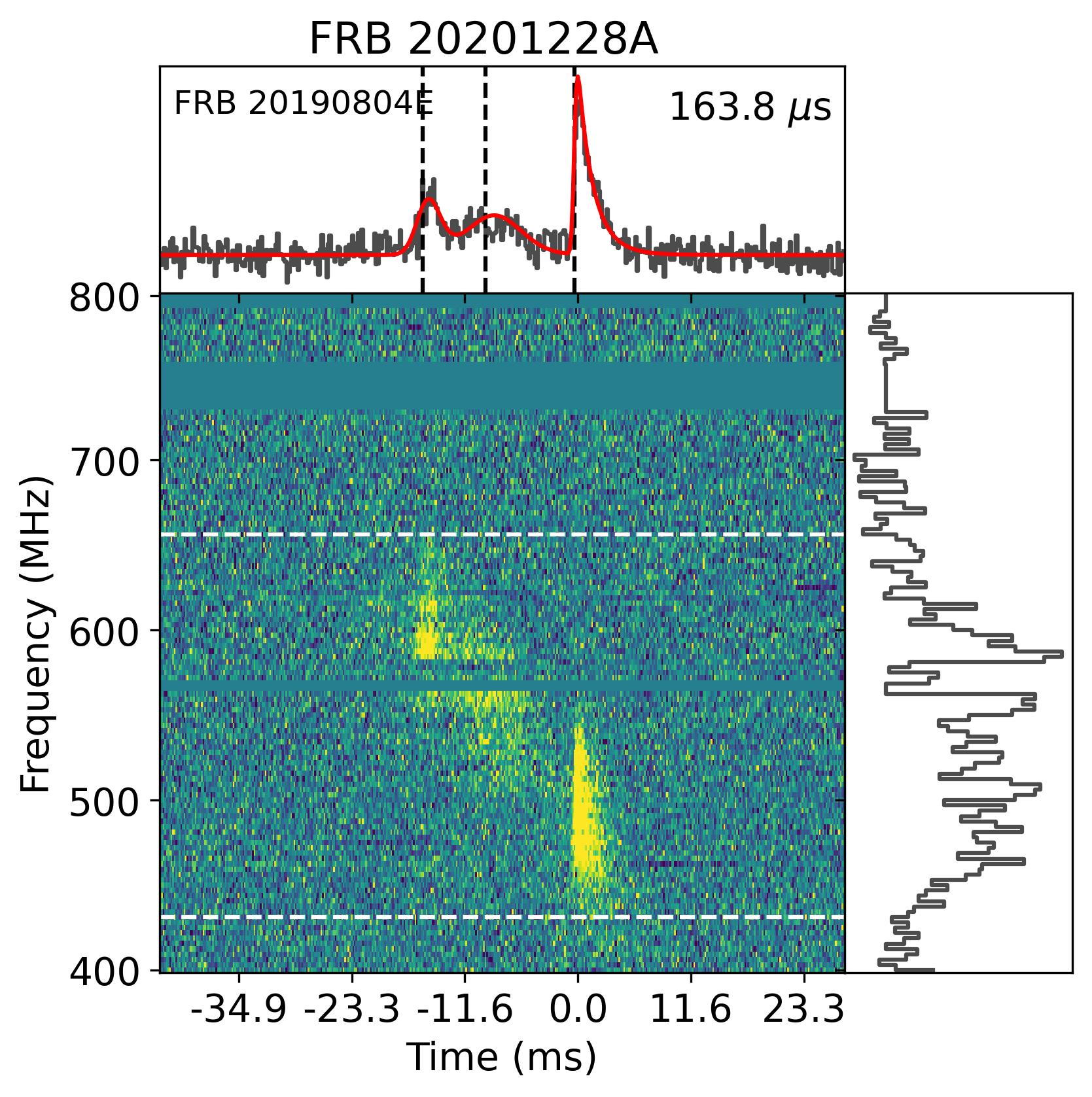}{0.24\textwidth}{}
          \fig{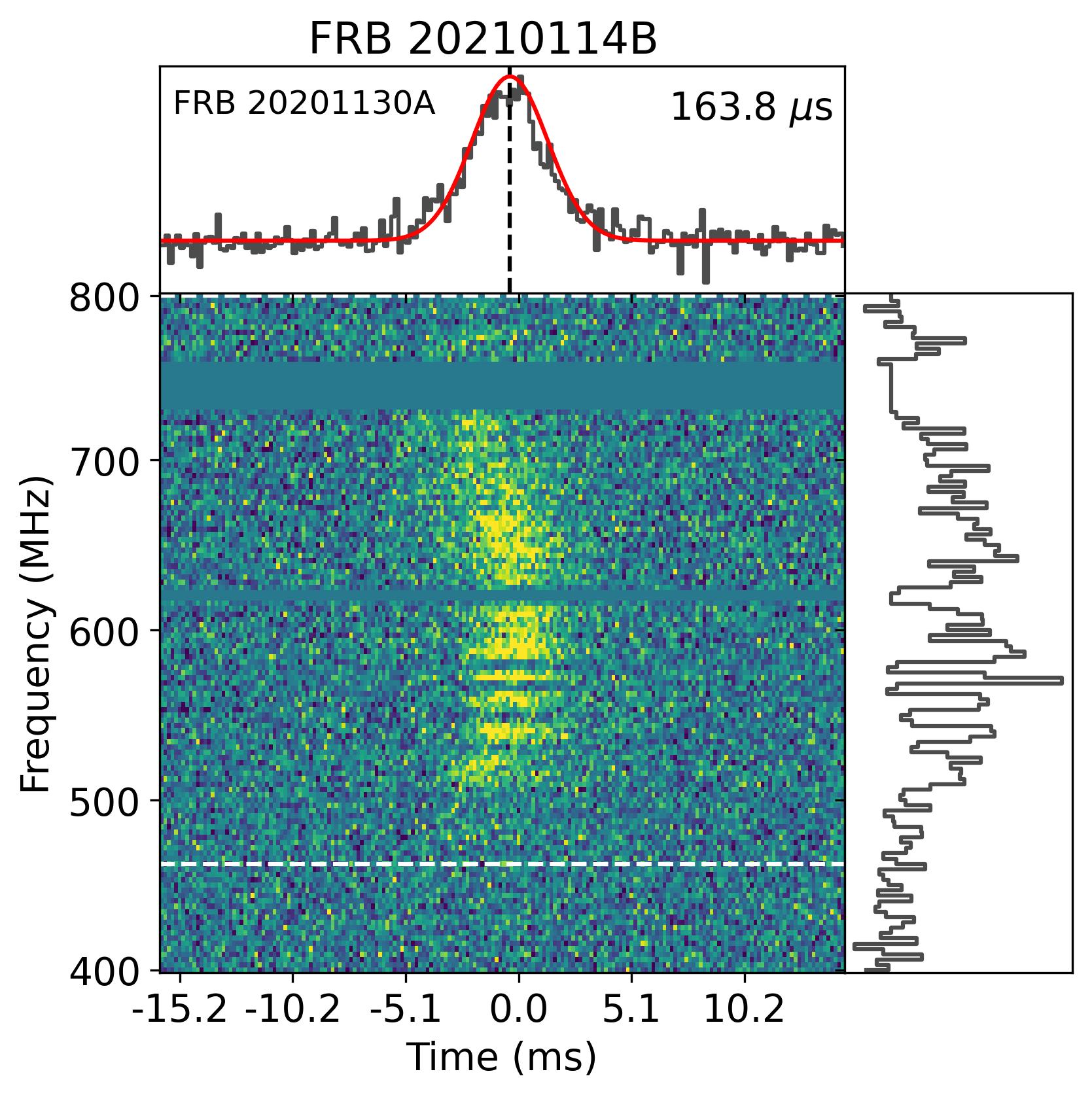}{0.24\textwidth}{}
          \fig{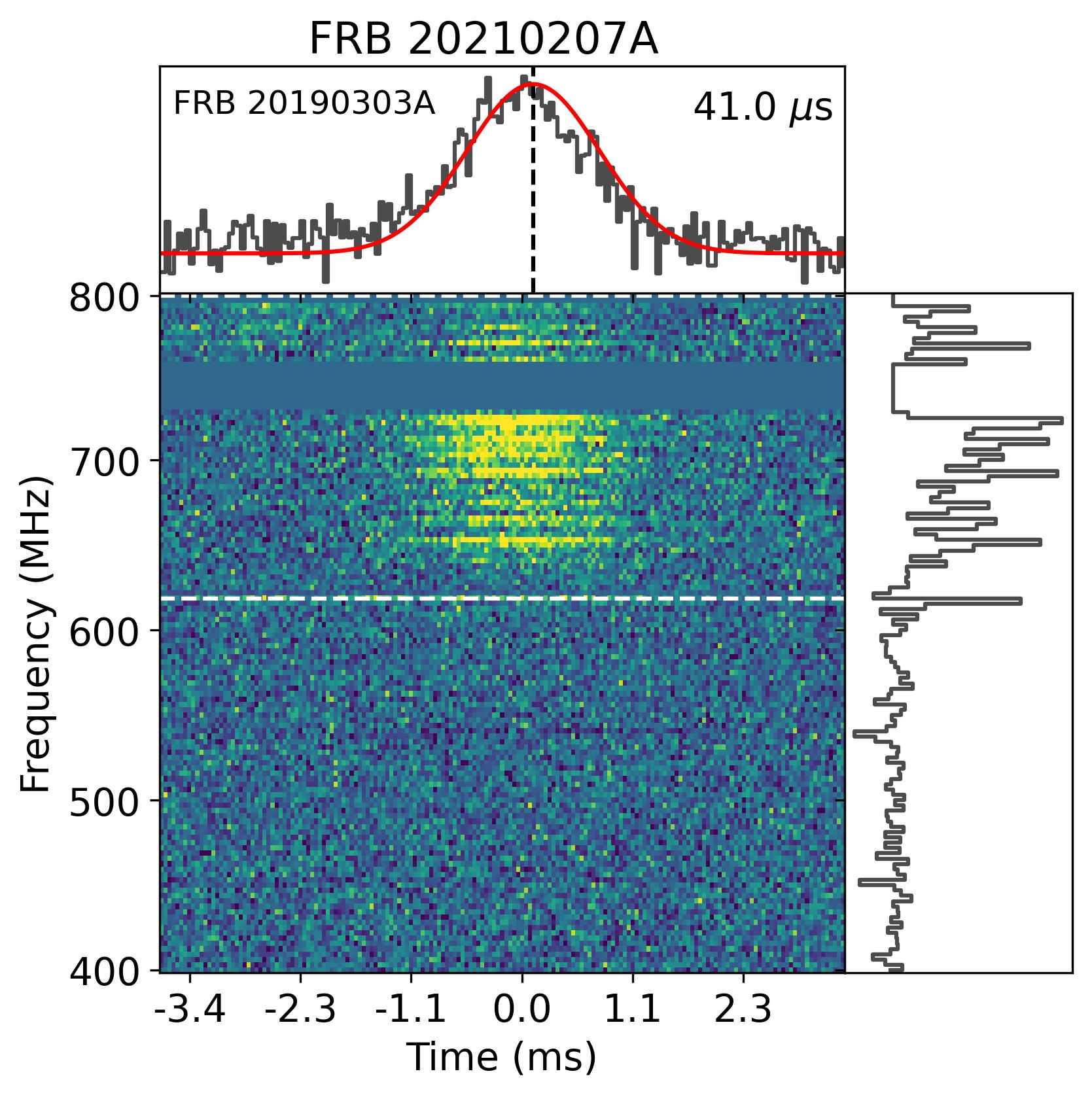}{0.24\textwidth}{}
          }
\gridline{
          \fig{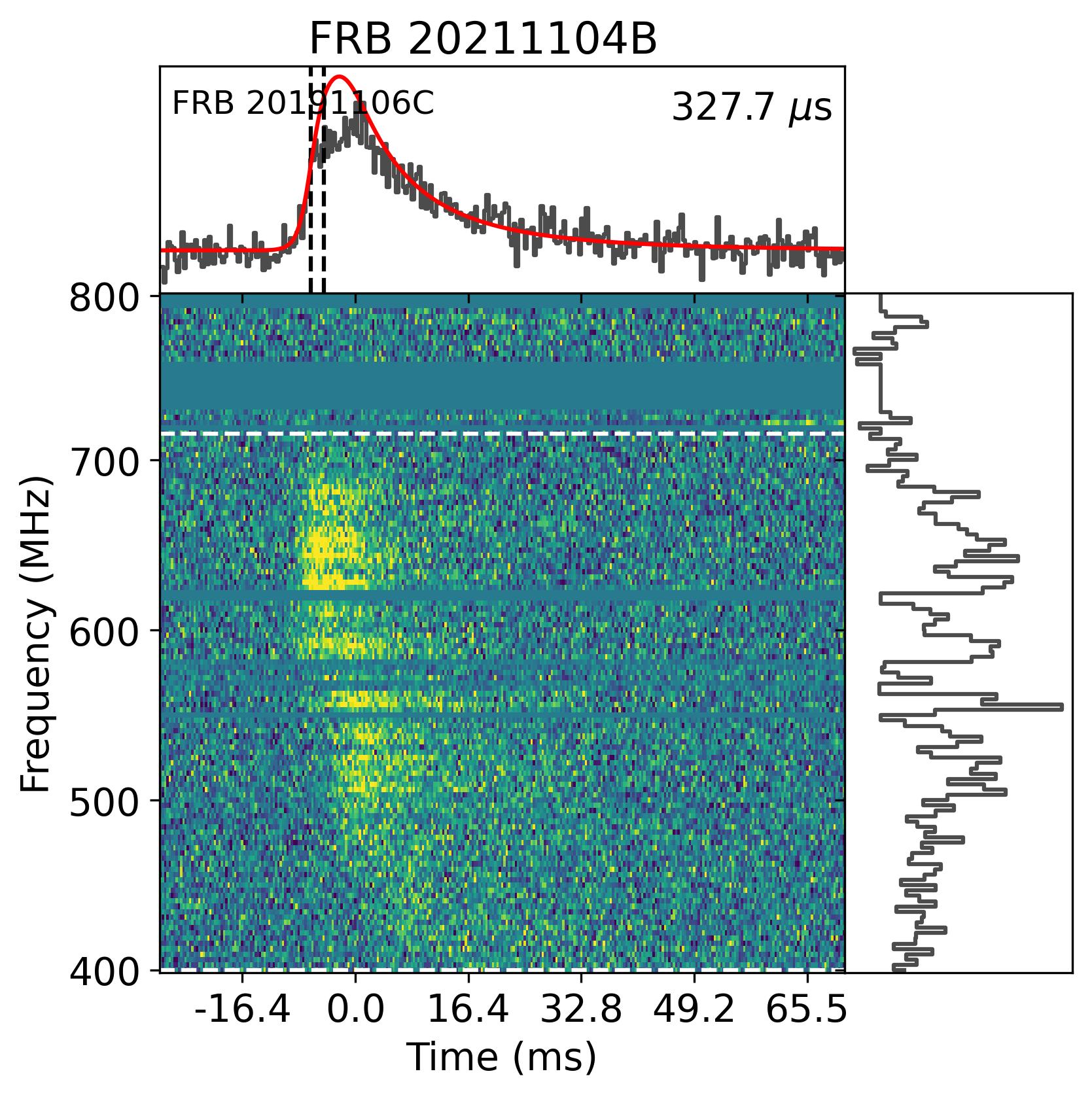}{0.24\textwidth}{}
          \fig{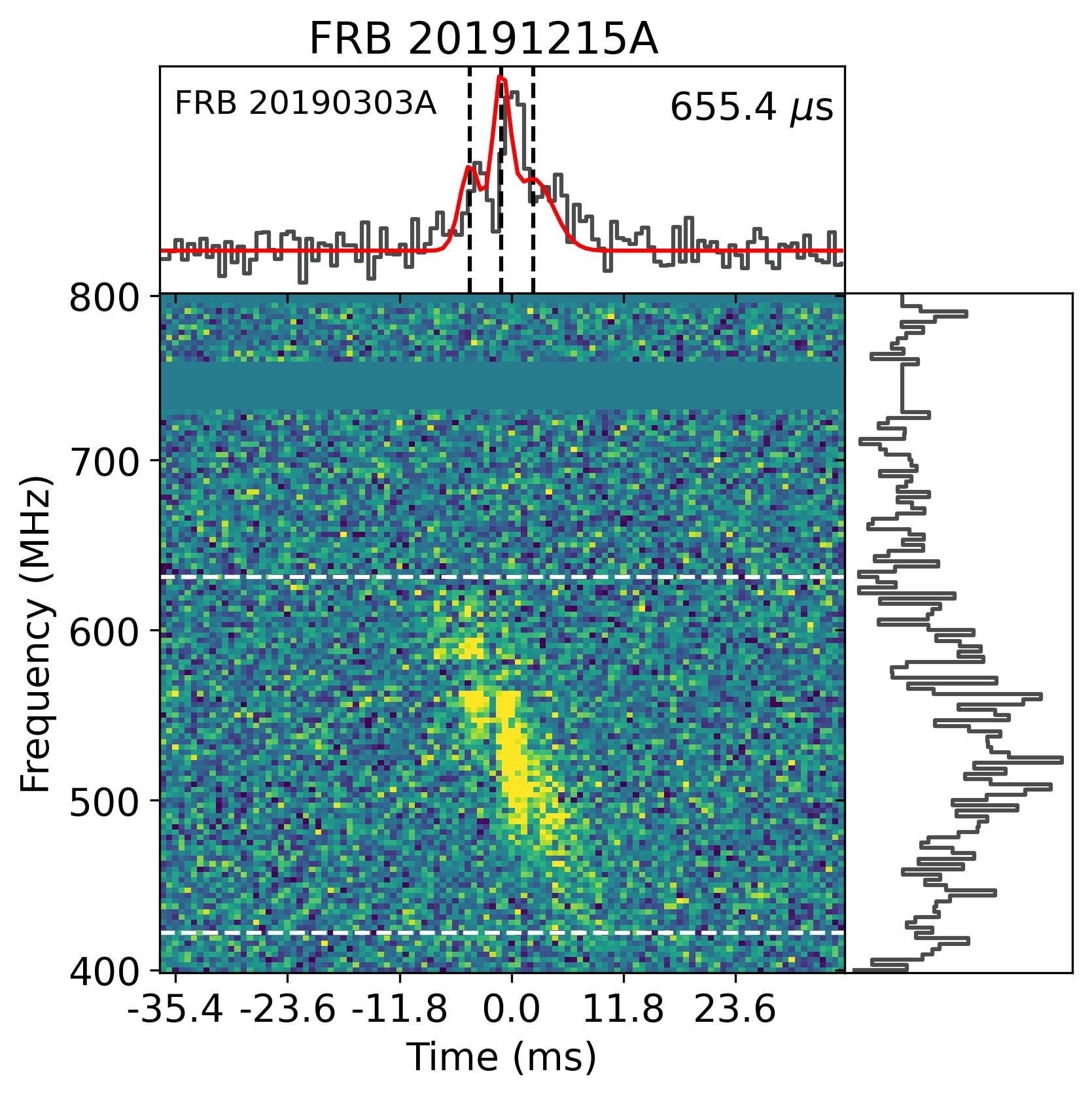}{0.24\textwidth}{}
          \fig{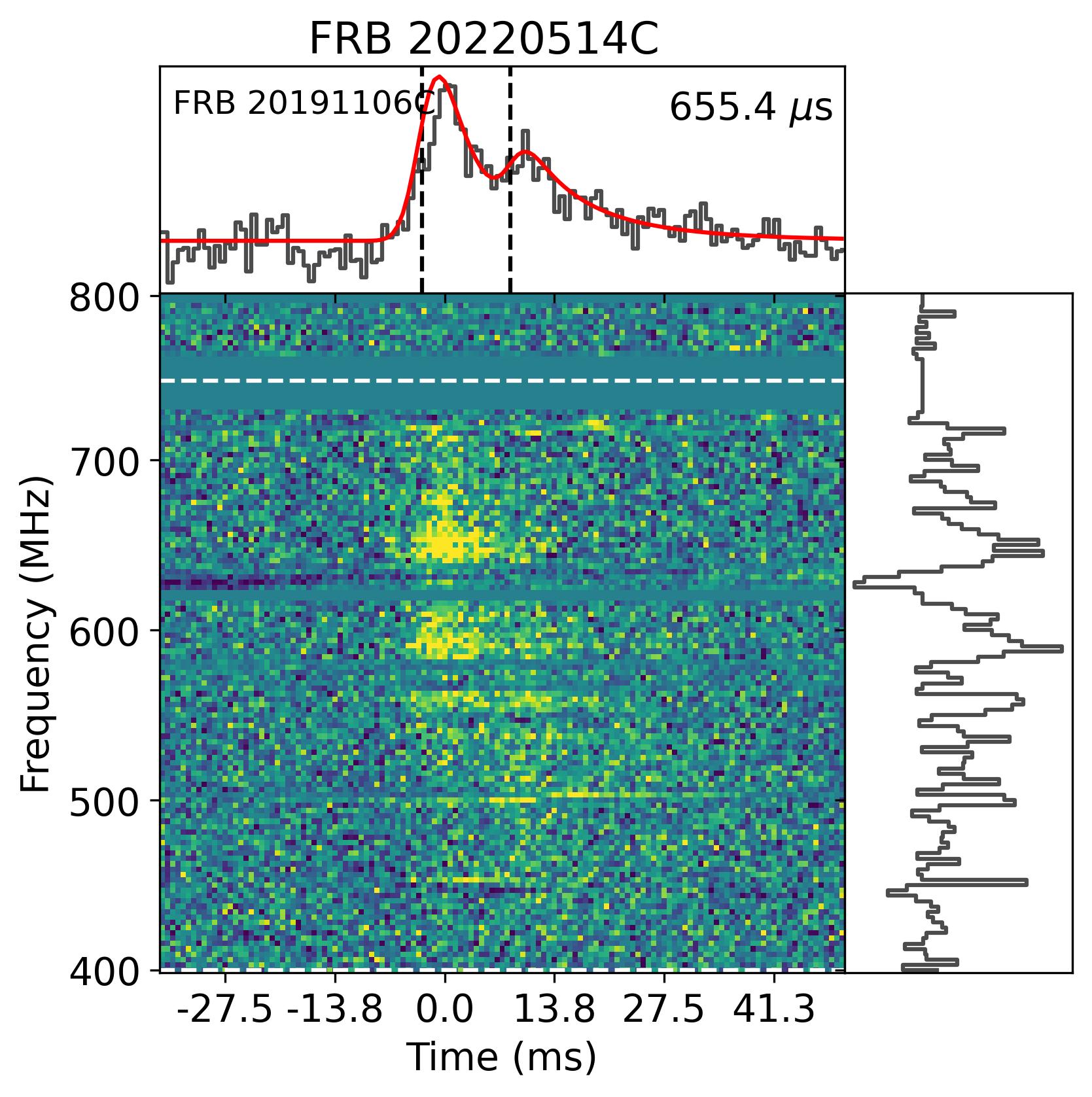}{0.24\textwidth}{}
          \fig{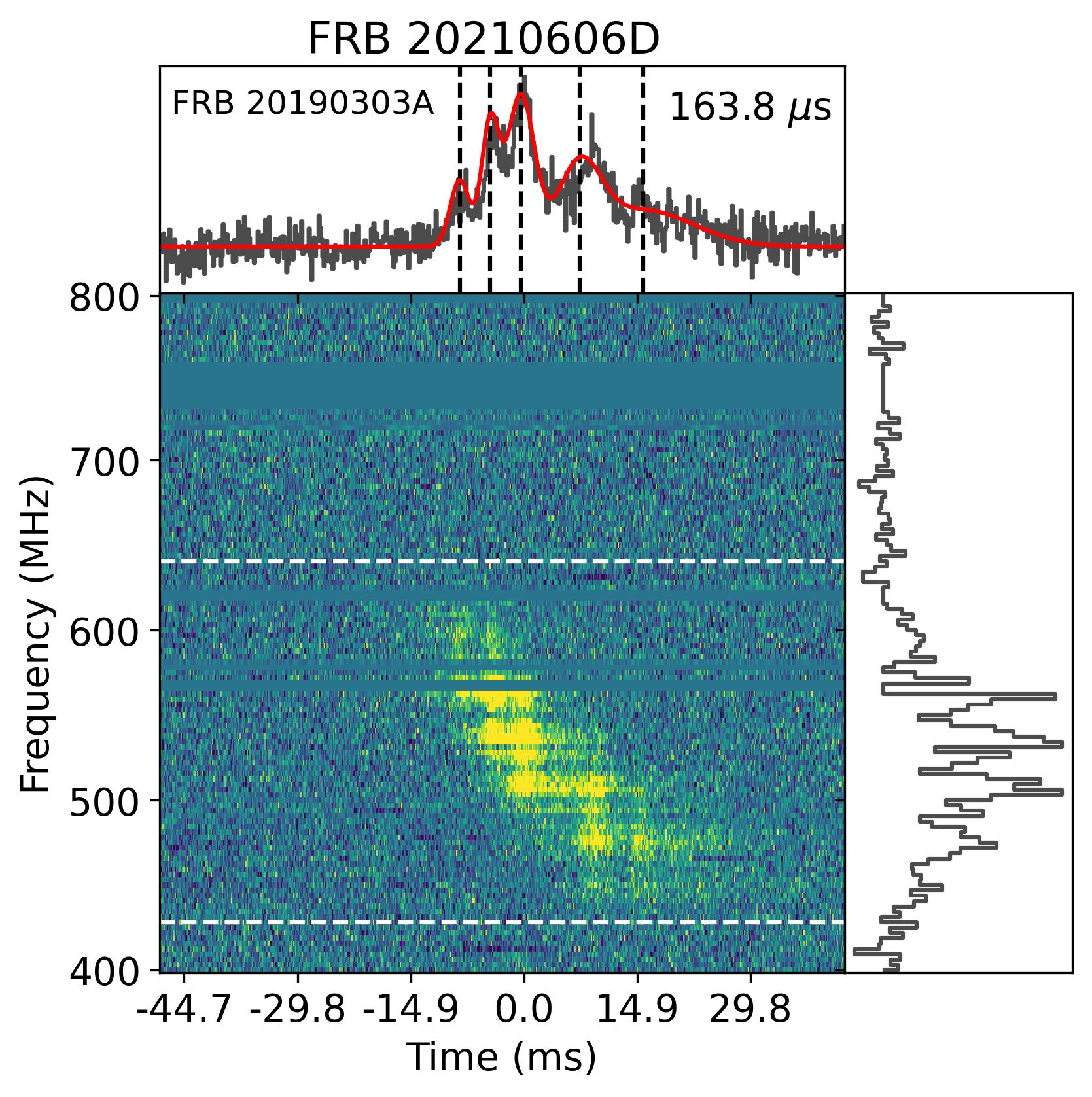}{0.24\textwidth}{}
          }
\caption{A sub-set of the burst waterfalls from the repeating sources in our sample. Each burst is de-dispersed to its structure maximizing DM, with the relevant \texttt{fitburst} fit shown on top of the burst. The time resolution at which the burst is displayed is listed in the rop right corner. The dashed, white lines indicated the derived bandwidth range for the source. Channels which were masked due to RFI are shown as dark green/teal. The repeater name is shown in the upper left corner, while the individual burst name is listed on top. }\label{fig:burst_wfalls}
\end{figure*}

\begin{figure}[]
    \centering
    \includegraphics[width=0.5\textwidth]{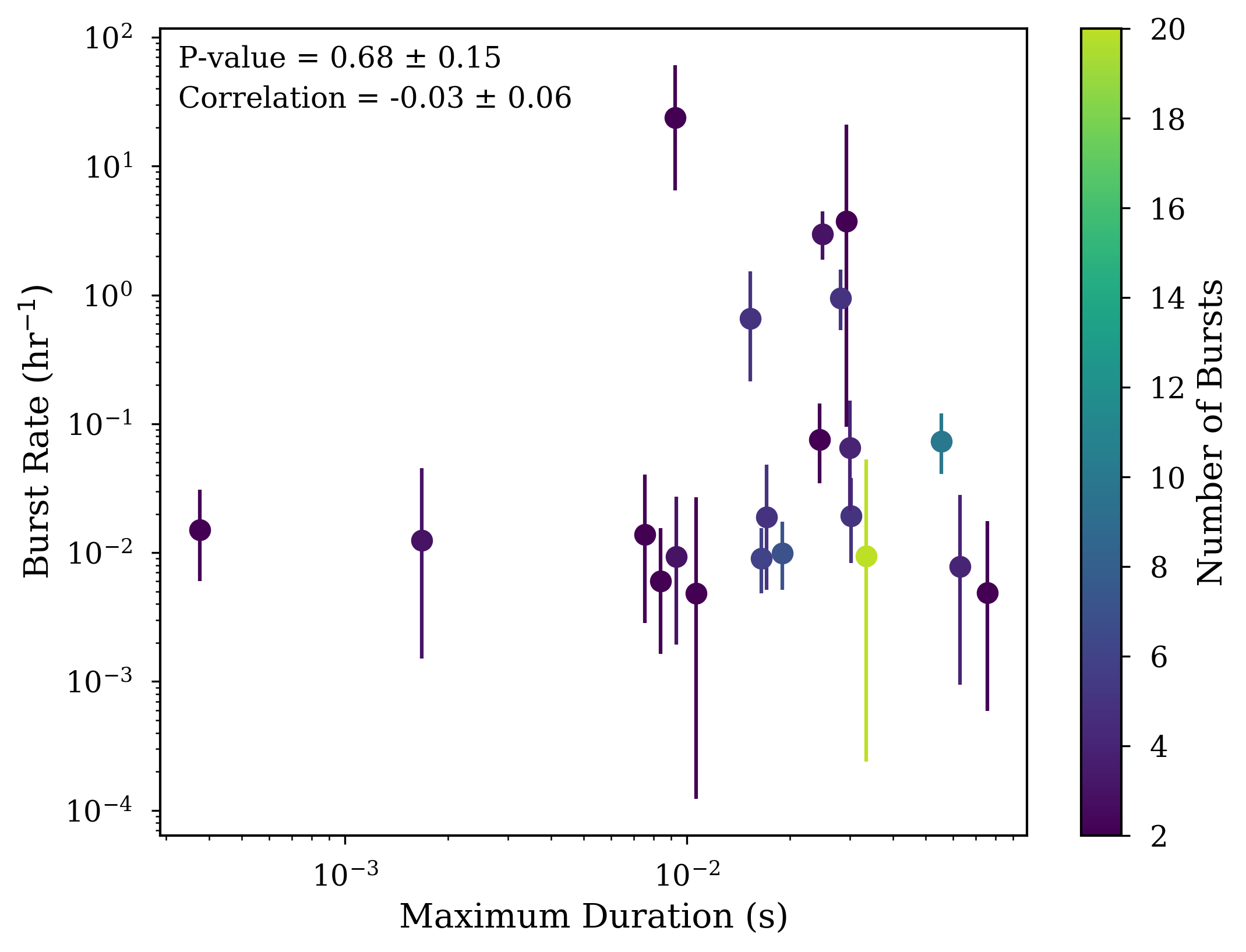}\hfill
    \caption{Same as top left panel of Figure \ref{Figure: Morphological Correlations} except using the maximum burst duration per repeating FRB source. }
    \label{Figure:MaxWidthRate}
\end{figure}

\begin{figure}[]
    \centering
    \includegraphics[width=0.5\textwidth]{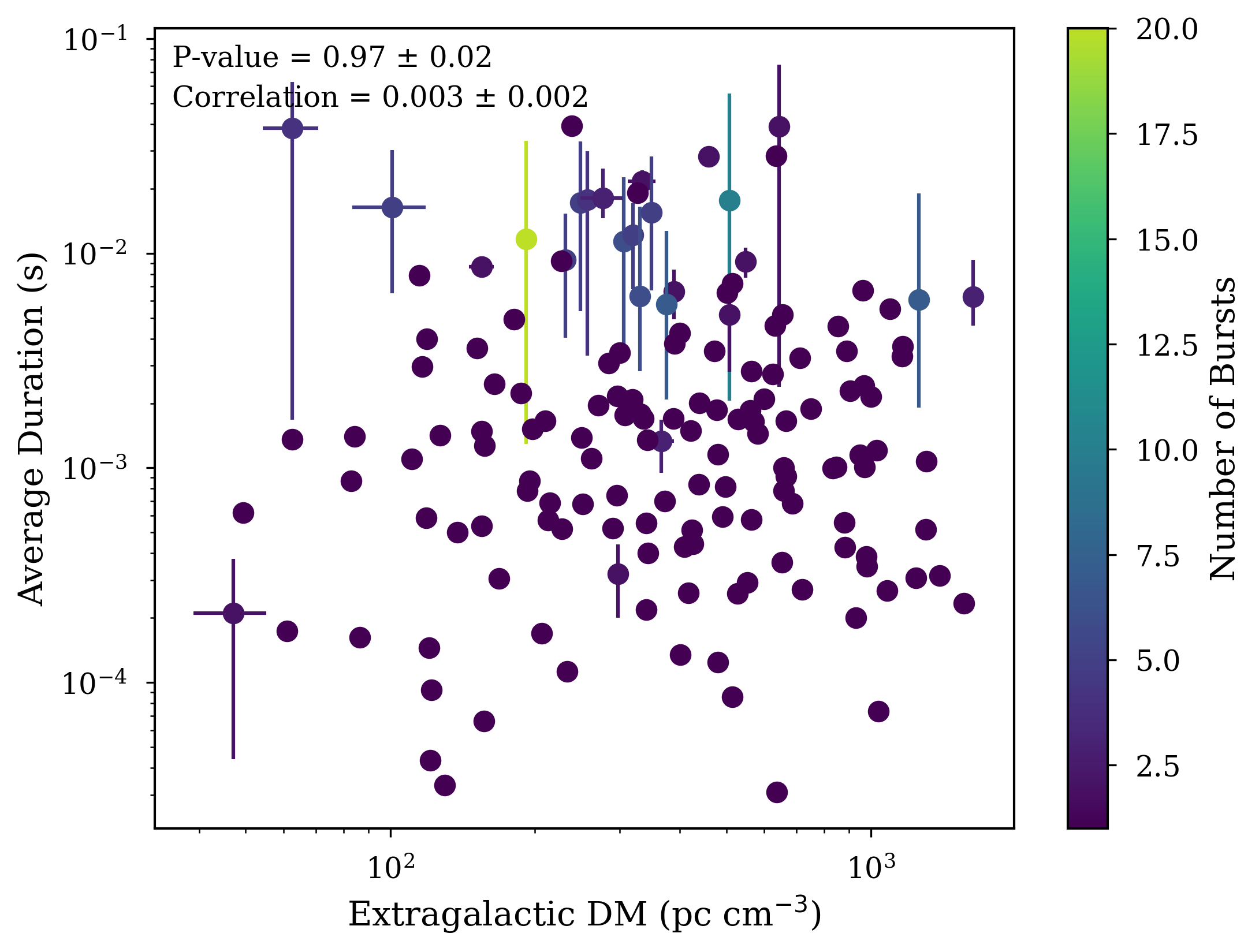}\hfill
    \caption{Same as bottom left panel of Figure \ref{Figure: Morphological Correlations} except including non-repeating FRBs from \citet{2024Sand}. }
    \label{Figure:DurationDMRepeaterNonRepeater}
\end{figure}

\begin{figure}[]
    \centering
    \includegraphics[width=0.5\textwidth]{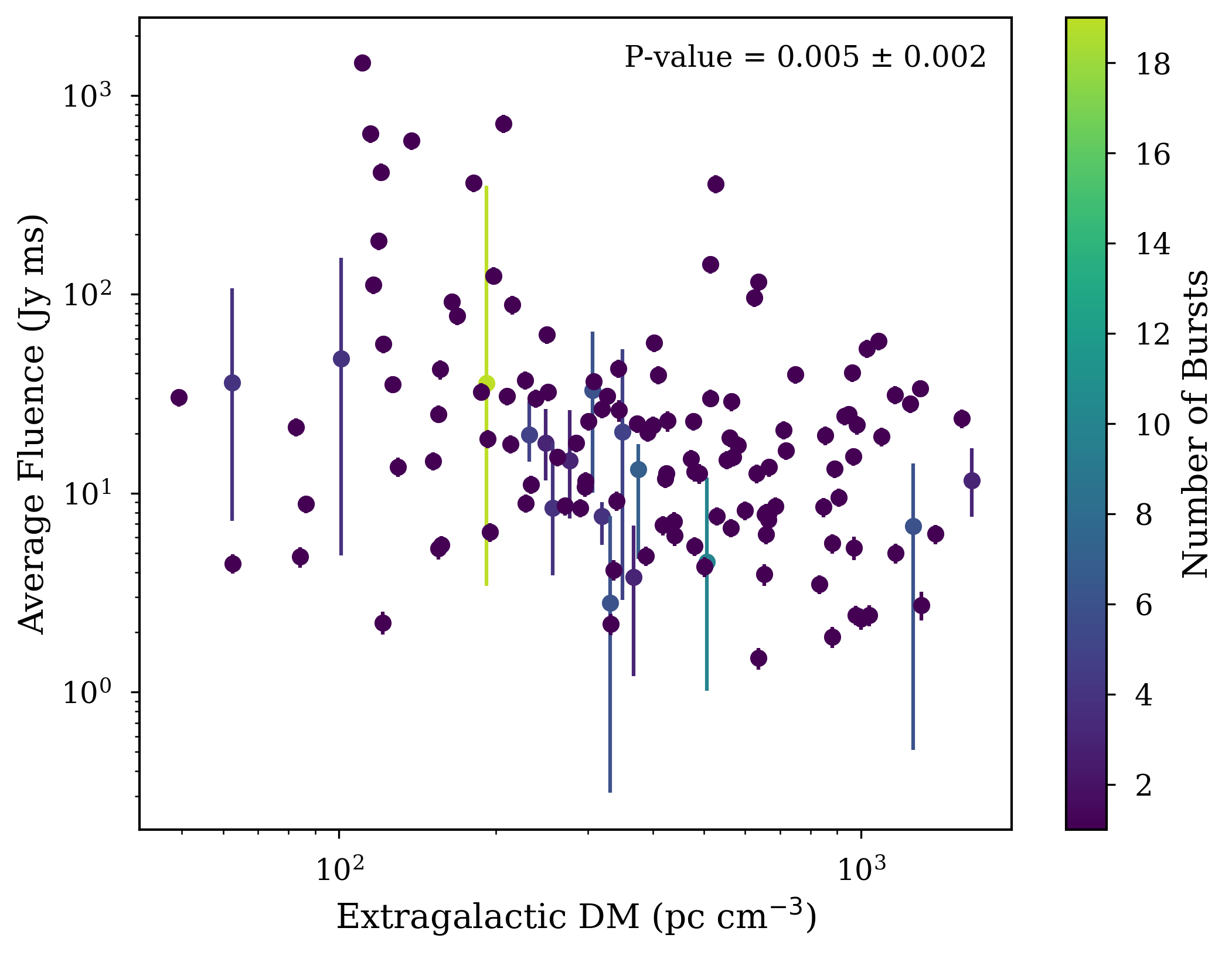}\hfill
    \caption{Same as bottom right panel of Figure \ref{Figure: Morphological Correlations} except including non-repeating FRBs from \citet{2024Sand}. }
    \label{Figure:FluenceDMRepeaterNonRepeater}
\end{figure}

\begin{figure*}[t]
    \centering
    \includegraphics[width=0.48\textwidth]{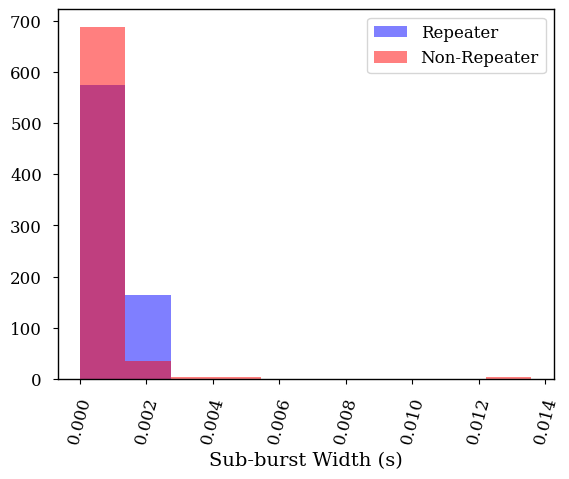}
    \includegraphics[width=0.48\textwidth]{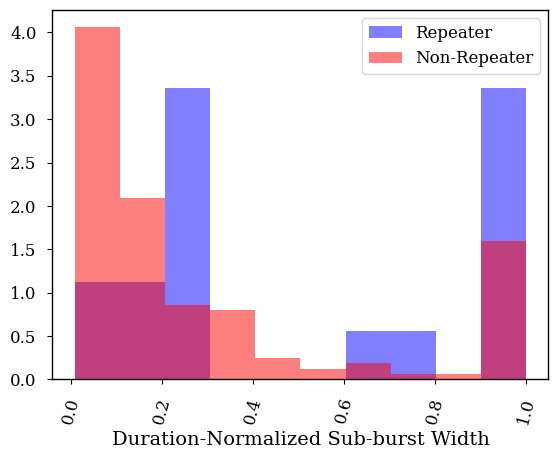}
    \includegraphics[width=0.455\textwidth]{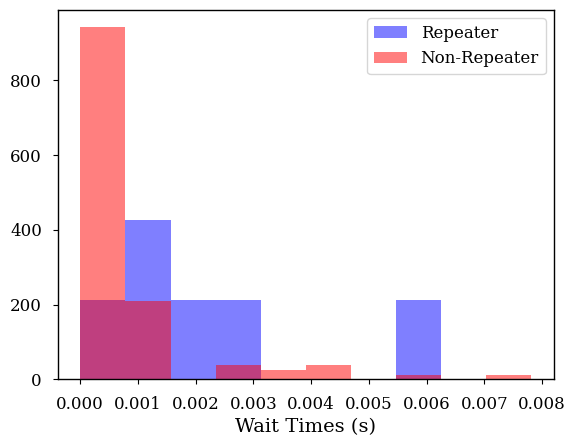}
    \includegraphics[width=0.455\textwidth]{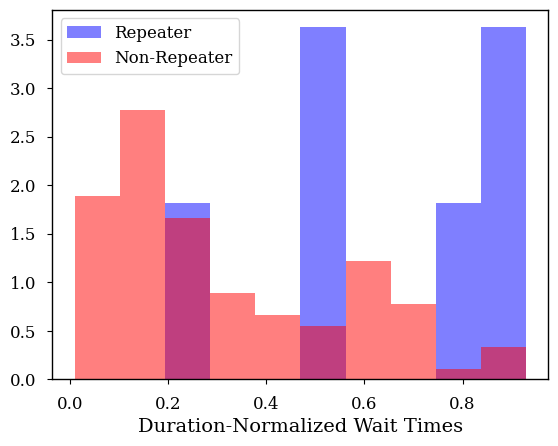}
    \includegraphics[width=0.48\textwidth]{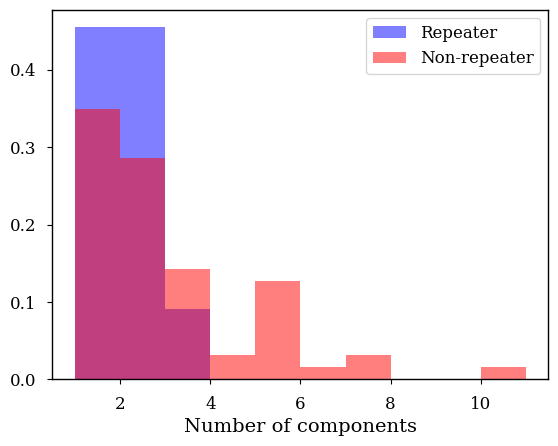}
    \caption{Same as Figure \ref{fig:sub-burst properties; repeaters and nonrepeaters} except limiting the burst to those studied at time resolutions $\leq$ 40 $\upmu$s}
    \label{fig:sub-burst properties; repeaters and nonrepeaters; higher time res}
\end{figure*}

\begin{figure*}[t]
    \centering
    \includegraphics[width=0.95\textwidth]{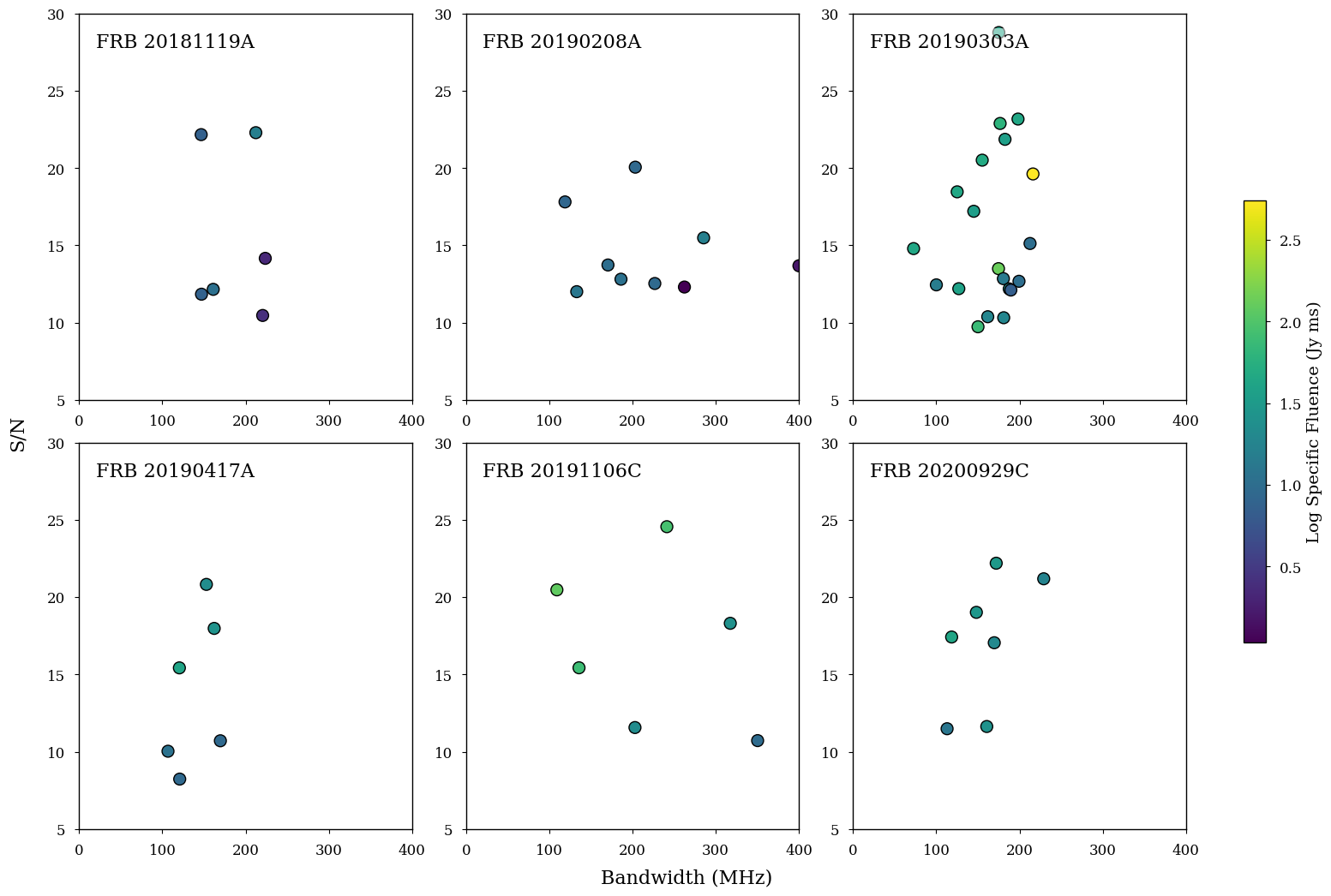}\hfill
    \caption{Bandwidth versus detection S/N for all sources in our sample for which there are more than 5 bursts with the complex raw voltages recorded. The color scale indicates the log of the spectral fluence over the spectral extent of the burst.}
    \label{Figure: BWvs.SN}

\end{figure*}

\clearpage

\setlength{\tabcolsep}{4pt}
\startlongtable
\begin{deluxetable*}{ccccccccccc}
\tabletypesize{\scriptsize}
\tablecaption{Measured morphological properties for 124 repeating FRBs detected by CHIME/FRB.\label{Table: Burst Parameters}}
\tablehead{\colhead{TNS Name} & \colhead{Repeater of} & \colhead{t$_{\textrm{res}}$\tablenotemark{a}} & \colhead{\# Comp.\tablenotemark{b}} & 
\colhead{DM$_{\textrm{struct}}$\tablenotemark{c}} & \colhead{DM$_{\textrm{fit}}$\tablenotemark{d}} & \colhead{Scat.\tablenotemark{e}} & \colhead{Duration\tablenotemark{f}} & \colhead{BW} & \colhead{Fluence\tablenotemark{g}} & \colhead{Flux} \\ 
\colhead{} & \colhead{} & \colhead{(ms)} & \colhead{} & \colhead{(pc cm$^{-3}$)} & 
\colhead{(pc cm$^{-3}$)} & \colhead{(ms)} & \colhead{(ms)} & \colhead{(MHz)} & \colhead{(Jy ms)}& \colhead{(Jy)}} 
\startdata
FRB 20190611A & FRB 20180814A & 0.66 & 2 & 189.7(5) & 190.20(11) & $<$2.3 & 17.3 & 197.07 & 22.0(2.3) & 2.11(24)  \\
FRB 20190625E & FRB 20180814A & 0.66 & 3 & 188.53(4) & 188.57(4) & $<$1.1 & 30.26 & 78.59 & 152(16) & 15.7(2.2)  \\
FRB 20190626A & FRB 20180814A & 0.66 & 1 & 191.38(3) & 192.24(24) & $<$8.7 & 20.38 & 70.77 & -- & --  \\
FRB 20191029A & FRB 20180814A & 0.66 & 2 & 189.129(21) & 189.10(10) & $<$0.66 & 7.43 & 157.97 & 10.2(1.1) & 2.01(26)  \\
FRB 20191111C & FRB 20180814A & 0.66 & 2 & 188.01(--) & 188.41(8) & $<$0.81 & 6.52 & 166.96 & 4.9(6) & 0.85(13)  \\
FRB 20190621A & FRB 20180908B & 0.66 & 1 & 195.81(--) & 196.04(15) & $<$3.1 & 7.29 & 59.82 & 3.9(5) & 1.01(24)  \\
FRB 20200621D & FRB 20180910A & 0.16 & 1 & 696.37(16) & 696.297(14) & 5.73(22) & 2.35 & 400.0 & 40(4) & 5.9(6)  \\
FRB 20200122A & FRB 20181030A & 0.33 & 2 & 103.542(12) & 103.538(4) & $<$1.2 & 62.88 & 188.86 & 9.9(1.1) & 2.6(3)  \\
FRB 20200122B & FRB 20181030A & 0.16 & 5 & 103.5(--) & 103.501 7(1 4) & $<$0.11 & 35.96 & 234.6 & 7.3(8) & 3.1(3)  \\
FRB 20200122E & FRB 20181030A & 0.66 & 2 & 103.44(--) & 103.55(3) & $<$0.39 & 52.35 & 176.34 & 107(12) & 23(3)  \\
FRB 20200122G & FRB 20181030A & 0.16 & 1 & 103.53(--) & 103.561(15) & $<$0.71 & 1.68 & 246.73 & 19.0(2.4) & 11.6(1.4)  \\
FRB 20200621C & FRB 20181119A & 0.66 & 1 & 364.11(14) & 364.25(11) & $<$1.2 & 2.83 & 147.02 & 1.95(26) & 0.76(13)  \\
FRB 20201204D & FRB 20181119A & 0.66 & 1 & 364.061(8) & 363.93(14) & $<$1.4 & 3.37 & 220.53 & 1.40(25) & 0.88(24)  \\
FRB 20210608B & FRB 20181119A & 0.66 & 1 & 364.752(16) & 364.45(18) & $<$1.9 & 4.51 & 146.63 & 1.1(3) & 0.96(26)  \\
FRB 20210624B & FRB 20181119A & 0.66 & 3 & 363.61(4) & 364.13(20) & $<$1.2 & 16.49 & 212.32 & 7.7(9) & 1.10(20)  \\
FRB 20211111A & FRB 20181119A & 0.66 & 1 & 364.57(--) & 364.86(23) & $<$2.6 & 6.11 & 223.66 & 0.31(14) & 1.02(26)  \\
FRB 20220213A & FRB 20181119A & 0.66 & 1 & 364.53(--) & 364.54(16) & $<$1.9 & 4.49 & 161.09 & 4.5(6) & 2.4(5)  \\
FRB 20201215C & FRB 20181128A & 0.66 & 3 & 446.976(15) & 447.151(28) & $<$1.5 & 24.39 & 228.35 & 9.2(1.1) & 1.60(26)  \\
FRB 20231006C & FRB 20181128A & 0.66 & 2 & 445.7(4) & 446.24(20) & $<$2.9 & 19.03 & 163.05 & 15.8(1.9) & 2.0(4)  \\
FRB 20190110C & FRB 20190110C & 0.33 & 1 & 222.03(4) & 222.051(9) & $<$0.57 & 1.35 & 106.35 & 5.7(7) & 3.5(4)  \\
FRB 20190116A & FRB 20190116A & 0.66 & 1 & 445.17(10) & 445.52(25) & $<$5.3 & 12.57 & 254.15 & 7.6(8) & 1.00(15)  \\
FRB 20190117A & FRB 20190117A & 0.08 & 2 & 393.018(16) & 392.990(5) & 0.609(26) & 6.71 & 174.39 & 24.9(2.5) & 6.8(7)  \\
FRB 20211114A & FRB 20190117A & 0.16 & 3 & 395.89(5) & 396.105(20) & $<$1.7 & 28.2 & 173.22 & 53(5) & 4.8(5)  \\
FRB 20230106C & FRB 20190117A & 0.66 & 2 & 397.21(8) & 397.36(7) & 0.97(15) & 19.6 & 162.27 & 8.2(1.2) & 2.2(6)  \\
FRB 20230504C & FRB 20190117A & 0.66 & 2 & 397.1(4) & 397.56(13) & $<$1.6 & 15.67 & 240.47 & 12.8(1.6) & 1.6(3)  \\
FRB 20231210C & FRB 20190117A & 0.66 & 1 & 396.96(--) & 397.40(16) & $<$3.1 & 7.41 & 100.88 & 2.9(4) & 1.7(4)  \\
FRB 20200124A & FRB 20190208A & 0.66 & 3 & 579.88(7) & 580.031(25) & $<$0.79 & 27.55 & 285.43 & 12.0(1.2) & 1.56(18)  \\
FRB 20200513B & FRB 20190208A & 0.66 & 2 & 579.485(9) & 579.59(6) & $<$0.85 & 33.13 & 400.0 & 3.0(4) & 1.09(17)  \\
FRB 20210203B & FRB 20190208A & 0.08 & 2 & 579.824(4) & 579.869(7) & 1.02(5) & 2.06 & 186.12 & 4.8(6) & 3.5(4)  \\
FRB 20210612B & FRB 20190208A & 0.66 & 2 & 579.55(5) & 580.01(9) & $<$1.7 & 13.3 & 262.37 & 1.02(22) & 1.07(28)  \\
FRB 20211129B & FRB 20190208A & 0.33 & 2 & 579.823(14) & 579.785(16) & $<$0.78 & 13.09 & 170.48 & 4.2(6) & 3.1(4)  \\
FRB 20211205A & FRB 20190208A & 0.66 & 1 & 579.54(4) & 579.83(5) & $<$1.3 & 3.07 & 132.94 & 4.6(7) & 2.1(5)  \\
FRB 20230306D & FRB 20190208A & 0.66 & 2 & 579.82(17) & 579.85(5) & $<$0.76 & 6.47 & 226.78 & 5.1(7) & 2.2(4)  \\
FRB 20211009B & FRB 20190208A & 0.66 & 2 & 579.92(6) & 579.99(4) & $<$1.1 & 9.55 & 118.87 & 2.7(4) & 1.08(24)  \\
FRB 20211228A & FRB 20190208A & 0.66 & 3 & 579.819(7) & 579.759(24) & $<$0.49 & 12.34 & 203.32 & 4.4(6) & 1.41(28)  \\
FRB 20211229A & FRB 20190208A & 0.66 & 6 & 579.84(4) & 579.803(25) & $<$0.49 & 55.47 & 400.0 & 3.4(5) & 1.18(25)  \\
FRB 20191217A & FRB 20190212A & 0.66 & 4 & 301.6(--) & 301.61(5) & $<$0.99 & 33.22 & 287.0 & 20(2) & 2.1(3)  \\
FRB 20200725B & FRB 20190212A & 0.66 & 2 & 301.56(--) & 301.577(7) & $<$1.3 & 8.12 & 120.43 & 11.6(1.4) & 2.5(4)  \\
FRB 20210216B & FRB 20190212A & 0.66 & 2 & 301.335(16) & 301.326(6) & $<$0.0068 & 18.23 & 113.0 & 15.3(2.2) & 9.9(1.8)  \\
FRB 20210410C & FRB 20190212A & 0.66 & 2 & 301.81(--) & 301.795(29) & $<$0.72 & 5.37 & 278.01 & 7.6(1.2) & 4.0(8)  \\
FRB 20210805A & FRB 20190212A & 0.66 & 2 & 301.37(--) & 301.371(8) & $<$0.47 & 20.9 & 175.17 & 27(5) & 17(4)  \\
FRB 20220322C & FRB 20190216A & 0.66 & 1 & 1279(--) & 1 281.83(19) & $<$19.0 & 44.12 & 290.13 & 1.8(4) & 1.2(3)  \\
FRB 20190301A & FRB 20190222A & 0.16 & 2 & 459.74(--) & 459.83(4) & $<$0.49 & 4.31 & 104.01 & 0.67(12) & 0.53(16)  \\
FRB 20190702B & FRB 20190303A & 0.04 & 2 & 221.26(4) & 221.43(3) & 2.74(10) & 4.08 & 125.12 & 20.4(2.1) & 2.78(29)  \\
FRB 20191013A & FRB 20190303A & 0.66 & 3 & 221.43(--) & 221.72(6) & $<$0.94 & 23.21 & 150.15 & 49(5) & 3.6(5)  \\
FRB 20191020A & FRB 20190303A & 0.16 & 1 & 221.61(10) & 221.85(5) & 1.79(20) & 1.29 & 176.74 & 13.4(1.7) & 5.9(8)  \\
FRB 20191110A & FRB 20190303A & 0.16 & 5 & 221.43(5) & 221.426(10) & $<$0.85 & 27.25 & 155.23 & 21.3(2.2) & 2.05(21)  \\
FRB 20191113A & FRB 20190303A & 0.33 & 5 & 221.55(7) & 221.65(6) & $<$0.74 & 28.06 & 174.78 & 86(9) & 7.0(8)  \\
FRB 20191116A & FRB 20190303A & 0.02 & 3 & 221.400(27) & 221.452(10) & 0.95(4) & 3.35 & 175.17 & 22.0(2.2) & 8.8(9)  \\
FRB 20191215A & FRB 20190303A & 0.66 & 3 & 221.55(4) & 221.63(3) & $<$0.89 & 10.41 & 212.71 & 6.1(6) & 0.99(12)  \\
FRB 20191231A & FRB 20190303A & 0.66 & 2 & 221.42(18) & 221.87(9) & $<$1.2 & 9.65 & 181.04 & 8.6(9) & 1.66(22)  \\
FRB 20200112A & FRB 20190303A & 0.66 & 1 & 221.69(--) & 221.53(8) & $<$3.0 & 6.97 & 100.1 & 4.5(5) & 0.63(9)  \\
FRB 20200622A & FRB 20190303A & 0.66 & 1 & 222.3(4) & 222.96(8) & $<$1.9 & 4.44 & 161.88 & 5.3(6) & 1.42(19)  \\
FRB 20210203C & FRB 20190303A & 0.08 & 1 & 221.73(7) & 222.03(6) & $<$1.3 & 3.05 & 127.08 & 9.0(1.2) & 3.7(5)  \\
FRB 20210207A & FRB 20190303A & 0.04 & 1 & 221.8(4) & 221.831(19) & $<$0.69 & 1.62 & 180.65 & 5.6(7) & 3.0(4)  \\
FRB 20210209B & FRB 20190303A & 0.01 & 1 & 221.66(20) & 221.829(8) & $<$0.73 & 1.72 & 182.6 & 16.3(1.7) & 8.4(9)  \\
FRB 20210302C & FRB 20190303A & 0.66 & 3 & 221.48(4) & 221.80(8) & $<$2.5 & 27.35 & 187.68 & 12.4(1.3) & 1.10(15)  \\
FRB 20211014B & FRB 20190303A & 0.33 & 3 & 221.23(3) & 221.43(5) & $<$0.77 & 9.23 & 199.41 & 5.8(8) & 1.60(27)  \\
FRB 20211125A & FRB 20190303A & 0.66 & 2 & 222(--) & 222.42(15) & $<$1.1 & 10.88 & 72.73 & 19.4(2.3) & 4.0(9)  \\
FRB 20210606D & FRB 20190303A & 0.16 & 5 & 221.26(4) & 221.47(3) & $<$1.1 & 33.46 & 216.23 & 35(4)0 & 30(3)  \\
FRB 20220214D & FRB 20190303A & 0.66 & 2 & 221.3(4) & 221.86(10) & $<$2.2 & 16.19 & 189.64 & 3.4(5) & 0.94(22)  \\
FRB 20230913E & FRB 20190303A & 0.04 & 1 & 221.49(7) & 221.54(4) & 1.84(10) & 1.45 & 145.06 & 10.7(1.3) & 4.5(6)  \\
FRB 20231204A & FRB 20190303A & 0.16 & 3 & 221.25(4) & 221.341(21) & $<$0.92 & 9.89 & 198.24 & 26.0(2.7) & 4.4(5)  \\
FRB 20200726D & FRB 20190417A & 0.16 & 1 & 1 378.23(26) & 1 378.58(6) & 3.96(19) & 1.91 & 162.27 & 7.4(8) & 1.90(23)  \\
FRB 20210304B & FRB 20190417A & 0.66 & 1 & 1 378.81(4) & 1 378.59(8) & 4.9(3) & 2.91 & 132.55 & 12.1(1.4) & 1.9(3)  \\
FRB 20210404E & FRB 20190417A & 0.66 & 1 & 1 379.41(10) & 1 379.10(19) & 3.5(6) & 4.3 & 120.82 & 2.9(4) & 1.2(3)  \\
FRB 20210924B & FRB 20190417A & 0.66 & 1 & 1 378.60(11) & 1 378.6(3) & 3.7(5) & 3.04 & 106.74 & 0.51(22) & 2.2(4)  \\
FRB 20220207B & FRB 20190417A & 0.66 & 1 & 1 378.4(7) & 1 378.2(4) & 4.4(1.6) & 18.99 & 120.43 & 14.1(1.6) & 1.27(26)  \\
FRB 20220530B & FRB 20190417A & 0.66 & 1 & 1 379.9(8) & 1 379.67(15) & 3.8(5) & 3.83 & 169.7 & 4.3(6) & 2.2(5)  \\
FRB 20220619A & FRB 20190417A & 0.66 & 1 & 1 380.0(5) & 1 380.54(11) & 4.0(4) & 7.63 & 152.88 & 11.6(1.4) & 1.45(27)  \\
FRB 20190430C & FRB 20190430C & 0.04 & 1 & 400.36(17) & 400.400(15) & $<$0.83 & 1.96 & 181.82 & 8.8(9) & 4.1(4)  \\
FRB 20190606A & FRB 20190604A & 0.66 & 1 & 552.52(5) & 552.60(3) & $<$1.2 & 2.8 & 220.14 & 3.2(4) & 1.22(18)  \\
FRB 20210329A & FRB 20190604A & 0.66 & 1 & 552.5(5) & 552.69(9) & $<$3.2 & 7.54 & 100.88 & 2.3(3) & 0.62(14)  \\
FRB 20190609C & FRB 20190609C & 0.08 & 2 & 479.86(--) & 479.869(4) & $<$0.0087 & 1.68 & 92.28 & 3.3(4) & 4.7(5)  \\
FRB 20201030B & FRB 20190609C & 0.02 & 2 & 479.796 0(7) & 479.793 36(8) & 0.01360(22) & 1.37 & 105.18 & 6.9(8) & 35(4)  \\
FRB 20210113D & FRB 20190609C & 0.66 & 1 & 225.43(10) & 479.793(10) & $<$0.4 & 0.95 & 139.59 & 1.2(3) & 2.7(5)  \\
FRB 20200629C & FRB 20190804E & 0.66 & 2 & 317.75(14) & 363.474(29) & $<$1.5 & 7.03 & 180.25 & 9.0(1.0) & 1.96(23)  \\
FRB 20200709C & FRB 20190804E & 0.66 & 2 & 362.35(--) & 362.94(11) & $<$2.6 & 13.37 & 80.94 & 5.5(6) & 0.51(8)  \\
FRB 20201225B & FRB 20190804E & 0.02 & 2 & 362.724(6) & 362.728 4(7) & 0.0345(1 1) & 6.78 & 114.17 & 8.0(1.0) & 6.6(7)  \\
FRB 20201228A & FRB 20190804E & 0.16 & 3 & 362.969(27) & 362.963(11) & 0.736(26) & 17.11 & 225.22 & 8.1(1.0) & 2.4(3)  \\
FRB 20220203A & FRB 20190804E & 0.66 & 2 & 362.58(--) & 362.90(10) & $<$1.9 & 16.55 & 62.95 & -- & --  \\
FRB 20200729A & FRB 20190907A & 0.33 & 4 & 309.12(--) & 309.243(20) & $<$0.81 & 14.27 & 393.74 & 18.0(1.9) & 2.30(26)  \\
FRB 20220313B & FRB 20190907A & 0.33 & 2 & 309.58(13) & 309.61(4) & $<$0.62 & 3.65 & 170.48 & 3.9(5) & 2.0(3)  \\
FRB 20220525A & FRB 20190907A & 0.66 & 4 & 309.31(4) & 309.650(19) & $<$0.34 & 29.97 & 156.79 & 5.6(7) & 1.23(21)  \\
FRB 20221207E & FRB 20190907A & 0.66 & 3 & 309.54(5) & 309.76(4) & $<$1.2 & 23.49 & 134.51 & 6.2(7) & 1.53(23)  \\
FRB 20200214B & FRB 20190915D & 0.66 & 1 & 487.6(3) & 489.30(9) & $<$8.6 & 20.18 & 120.43 & 16.0(1.6) & 0.81(10)  \\
FRB 20200515A & FRB 20191013D & 0.66 & 1 & 522.7(4) & 522.43(5) & 10.3(7) & 5.75 & 400.0 & 47(5) & 2.8(3)  \\
FRB 20201201A & FRB 20191106C & 0.66 & 1 & 330.6(--) & 333.24(11) & 8.2(5) & 22.62 & 109.09 & 35(4) & 1.31(20)  \\
FRB 20210617A & FRB 20191106C & 0.66 & 1 & 330.6(--) & 331.91(7) & 9.1(3) & 9.34 & 135.68 & 44(4) & 1.74(22)  \\
FRB 20210822A & FRB 20191106C & 0.66 & 1 & 331.2(--) & 331.09(9) & 8.6(7) & 3.68 & 202.93 & 21.0(2.2) & 1.63(24)  \\
FRB 20211104B & FRB 20191106C & 0.33 & 2 & 331.60(16) & 332.10(8) & 8.1(3) & 6.4 & 317.5 & 23.1(2.4) & 1.95(28)  \\
FRB 20220118B & FRB 20191106C & 0.16 & 2 & 330.56(10) & 331.031(28) & 7.77(13) & 10.84 & 241.25 & 65(7) & 3.4(4)  \\
FRB 20220514C & FRB 20191106C & 0.66 & 2 & 331.38(11) & 331.09(9) & 5.9(6) & 15.37 & 350.34 & 10.1(1.1) & 1.00(15)  \\
FRB 20200118D & FRB 20200118D & 0.66 & 2 & 625.226(21) & 625.253(14) & $<$0.72 & 7.71 & 119.26 & 5.7(6) & 1.51(17)  \\
FRB 20200701A & FRB 20200118D & 0.08 & 4 & 625.290(22) & 625.304(7) & $<$0.32 & 10.62 & 377.32 & 11.5(1.2) & 3.7(4)  \\
FRB 20200120E & FRB 20200120E & 0.08 & 1 & 392.563(4) & 87.853(8) & $<$0.16 & 0.38 & 187.68 & 2.9(3) & 3.6(4)  \\
FRB 20210423G & FRB 20200120E & 0.01 & 1 & 392.83(17) & 87.760 70(29) & $<$0.019 & 0.04 & 242.42 & 1.86(24) & 14.8(1.5)  \\
FRB 20200127B & FRB 20200127B & 0.02 & 2 & 351.32(--) & 351.346 9(7) & $<$0.052 & 0.34 & 400.0 & 11.4(1.2) & 12.5(1.3)  \\
FRB 20200219B & FRB 20200127B & 0.00256 & 2 & 275.61(6) & 351.274 0(4) & $<$0.044 & 0.44 & 400.0 & 17.3(1.8) & 27.7(2.8)  \\
FRB 20230604B & FRB 20200202A & 0.16 & 1 & 734.37(26) & 734.33(4) & $<$1.0 & 2.38 & 117.3 & 7.0(9) & 3.5(4)  \\
FRB 20201014B & FRB 20200202A & 0.16 & 8 & 726.5(--) & 726.433(9) & $<$0.38 & 75.78 & 199.8 & 1.36(14) & 1.2(3)  \\
FRB 20200702C & FRB 20200223B & 0.08 & 4 & 200.355(11) & 201.01(6) & $<$0.99 & 9.24 & 179.86 & 40(4) & 7.2(7)  \\
FRB 20210115C & FRB 20200223B & 0.16 & 4 & 200.521(11) & 200.57(4) & $<$0.15 & 8.12 & 129.81 & 17.8(2.1) & 5.0(7)  \\
FRB 20201022D & FRB 20200619A & 0.66 & 2 & 439.86(--) & 439.773(11) & $<$0.17 & 8.37 & 400.0 & 2.1(5) & 1.5(3)  \\
FRB 20210130E & FRB 20200619A & 0.66 & 1 & 413.64(21) & 440.25(5) & $<$2.1 & 4.91 & 114.57 & -- & --  \\
FRB 20200809E & FRB 20200809E & 0.66 & 1 & 1 702.89(10) & 1 703.04(4) & $<$2.0 & 4.6 & 72.34 & 10.4(1.1) & 2.5(3)  \\
FRB 20201018C & FRB 20200809E & 0.33 & 2 & 1 702.86(3) & 1 703.08(3) & $<$0.43 & 4.86 & 260.41 & 16.8(1.8) & 4.1(5)  \\
FRB 20210208B & FRB 20200809E & 0.66 & 2 & 1 703.358(22) & 1 703.67(17) & $<$0.75 & 9.32 & 169.7 & 7.6(1.0) & 2.0(4)  \\
FRB 20230725D & FRB 20200926A & 0.16 & 3 & 758.63(3) & 758.475(21) & $<$0.74 & 11.85 & 162.27 & 16.0(1.8) & 3.4(4)  \\
FRB 20201125B & FRB 20200929C & 0.08 & 1 & 413.67(7) & 413.740(9) & $<$0.88 & 2.08 & 229.13 & 11.9(1.3) & 5.7(6)  \\
FRB 20201203C & FRB 20200929C & 0.66 & 1 & 413.54(4) & 413.57(3) & $<$1.2 & 2.78 & 113.0 & 4.7(6) & 2.0(4)  \\
FRB 20210313B & FRB 20200929C & 0.16 & 1 & 413.67(11) & 413.599(8) & $<$1.8 & 4.27 & 118.48 & 17.0(1.9) & 4.9(6)  \\
FRB 20210314A & FRB 20200929C & 0.66 & 2 & 413.54(17) & 413.600(25) & $<$1.2 & 7.76 & 169.7 & 11.2(1.6) & 3.8(7)  \\
FRB 20210326B & FRB 20200929C & 0.08 & 1 & 413.75(12) & 413.749(4) & $<$1.6 & 3.83 & 172.04 & 17.7(2.0) & 4.9(6)  \\
FRB 20210930A & FRB 20200929C & 0.66 & 2 & 413.43(19) & 413.775(7) & $<$2.0 & 12.67 & 160.7 & 15.5(1.8) & 2.5(4)  \\
FRB 20220209A & FRB 20200929C & 0.16 & 2 & 413.20(7) & 413.354(11) & $<$1.1 & 7.04 & 148.19 & 14.6(1.8) & 4.4(7)  \\
FRB 20201219A & FRB 20201114A & 0.66 & 2 & 321.31(13) & 321.47(10) & $<$2.2 & 10.57 & 144.28 & 8.0(9) & 1.55(22)  \\
FRB 20201225D & FRB 20201130A & 0.66 & 1 & 287.6(5) & 288.32(7) & $<$4.8 & 11.21 & 274.88 & 18.2(2.0) & 2.6(4)  \\
FRB 20210114B & FRB 20201130A & 0.16 & 1 & 288.27(25) & 288.381(21) & $<$1.7 & 4.03 & 337.44 & 16.8(1.9) & 4.5(6)  \\
FRB 20210117E & FRB 20201130A & 0.66 & 2 & 287.83(10) & 287.96(7) & $<$2.8 & 15.32 & 135.68 & 14.3(1.6) & 2.4(4)  \\
FRB 20210118B & FRB 20201130A & 0.16 & 1 & 288.0(5) & 288.39(3) & $<$2.7 & 6.39 & 310.07 & 20.1(2.3) & 4.1(6)  \\
FRB 20210327F & FRB 20201130A & 0.33 & 1 & 287.68(11) & 287.66(4) & $<$4.1 & 9.61 & 257.28 & 29(3) & 4.2(5)  \\
FRB 20210302E & FRB 20201221B & 0.66 & 3 & 430.43(3) & 510.59(6) & $<$1.6 & 29.32 & 160.7 & -- & --  \\
FRB 20210303F & FRB 20201221B & 0.66 & 2 & 509.72(15) & 510.35(8) & $<$4.6 & 27.27 & 89.54 & 16.5(1.7) & 1.19(16)  \\
\enddata
\tablenotetext{a}{Time resolution at which the burst is analyzed and parameters are reported.}
\tablenotetext{b}{Number of visible sub-bursts at the given time resolution for the burst.}
\tablenotetext{c}{Structure maximizing DM.}
\tablenotetext{d}{DM determined using the \texttt{fitburst} routine.}
\tablenotetext{e}{Scattering time as determined using \texttt{fitburst}. If no visible scattering by-eye, an upper limit on the scattering timescale is set to be equal to the reported \texttt{fitburst} width of the narrowest sub-component. Scattering timescales are referenced to 600 MHz.}
\tablenotetext{f}{Total duration of the burst. }
\tablenotetext{g}{Spectral fluence over the entire 400 to 800 MHz CHIME/FRB band.}
\end{deluxetable*}

\begin{deluxetable}{lcccc}
\tablecaption{Burst Rates\label{Table:Source rates}}
\tablehead{
\colhead{Source} & \colhead{N$_{\textrm{bursts}}$\tablenotemark{a}} & \colhead{Exposure\tablenotemark{b}} & \colhead{Threshold\tablenotemark{c}} & \colhead{Burst Rate\tablenotemark{d}} \\
\colhead{} & \colhead{} &\colhead{(hr)} &\colhead{(Jy ms)} & \colhead{(hr$^{-1}$)} 
}
\startdata
FRB 20201130A & 5 & 43.9 & 16.1 & 0.65$^{+0.87}_{-0.44}$ \\
FRB 20191106C & 0 & 2.0 & 30.6 & 0.0$^{+28.54}$ \\
FRB 20200619A & 4 & 209.0 & 2.3 & 0.0060$^{+0.0094}_{-0.0044}$ \\
FRB 20190804E & 4 & 176.6 & 4.4 & 0.019$^{+0.030}_{-0.014}$ \\
FRB 20190915D & 4 & 178.5 & 10.7 & 0.07$^{+0.11}_{-0.051}$ \\
FRB 20200929C\tablenotemark{e} & -- & -- & -- & -- \\ 
FRB 20201221B & 1 & 1.5 & 16.1 & 4$^{+17}_{-3.7}$ \\
FRB 20200809E & 3 & 914.0 & 10.0 & 0.009$^{+0.018}_{-0.0074}$ \\
FRB 20190609C & 2 & 133.7 & 4.4 & 0.012$^{+0.033}_{-0.011}$ \\
FRB 20200223B & 4 & 1.6 & 22.3 & 24$^{+37}_{-17}$ \\
FRB 20200202A & 2 & 110.7 & 2.1 & 0.005$^{+0.013}_{-0.0043}$ \\
FRB 20190430C & 3 & 57.7 & 6.4 & 0.07$^{+0.14}_{-0.059}$ \\
FRB 20200127B & 0 & 252.5 & 6.7 & 0.0$^{+0.02}$ \\
FRB 20201114A & 0 & 237.9 & 13.4 & 0.0$^{+0.07}$ \\
FRB 20190110C & 0 & 88.6 & 1.7 & 0.0$^{+0.009}$ \\
FRB 20200118D & 1 & 139.1 & 3.8 & 0.005$^{+0.022}_{-0.0047}$ \\
FRB 20191013D & 1 & 126.4 & 12.4 & 0.03$^{+0.14}_{-0.030}$ \\
FRB 20200926A & 5 & 205.4 & 8.9 & 0.058$^{+0.077}_{-0.039}$ \\
FRB 20180910A\tablenotemark{f} & 0 & 7400.7 & 10.7 & 0.0$^{+0.002}$ \\
FRB 20180814A & 8 & 426.6 & 5.1 & 0.019$^{+0.019}_{-0.011}$ \\
FRB 20181030A & 2 & 396.0 & 6.7 & 0.008$^{+0.020}_{-0.0068}$ \\
FRB 20181128A & 9 & 184.2 & 6.7 & 0.075$^{+0.068}_{-0.041}$ \\
FRB 20181119A & 13 & 266.8 & 1.6 & 0.0091$^{+0.0064}_{-0.0042}$ \\
FRB 20190116A & 2 & 127.7 & 9.4 & 0.04$^{+0.11}_{-0.036}$ \\
FRB 20190216A & 4 & 322.0 & 6.7 & 0.019$^{+0.030}_{-0.014}$ \\
FRB 20190222A & 0 & 307.1 & 0.2 & 0.0$^{+0.0001}$ \\
FRB 20190208A & 15 & 173.8 & 4.5 & 0.073$^{+0.048}_{-0.032}$ \\
FRB 20190604A & 3 & 191.1 & 4.6 & 0.014$^{+0.027}_{-0.011}$ \\
FRB 20190212A & 0 & 60.8 & 3.4 & 0.0$^{+0.03}$ \\
FRB 20180908B & 2 & 411.3 & 5.1 & 0.005$^{+0.013}_{-0.0044}$ \\
FRB 20190117A & 15 & 127.7 & 20.1 & 0.95$^{+0.62}_{-0.42}$ \\
FRB 20190303A & 1 & 78.5 & 4.1 & 0.009$^{+0.043}_{-0.0092}$ \\
FRB 20190417A & 12 & 213.5 & 1.6 & 0.0100$^{+0.0074}_{-0.0048}$ \\
FRB 20190907A & 5 & 127.5 & 7.0 & 0.065$^{+0.087}_{-0.044}$ \\
FRB 20200120E & 7 & 315.6 & 3.8 & 0.015$^{+0.016}_{-0.0090}$ \\
\enddata
\tablenotetext{a}{Number of bursts detected from this source that fall within the FWHM at 600 MHz of the corresponding detection beam. If a source has both a lower and upper transit, only bursts within the upper transit are considered.}
\tablenotetext{b}{Total exposure to the position of this source between 2018 August and 2023 September. For sources observed twice a day, this corresponds to the exposure solely in the upper CHIME transit. }
\tablenotemark{c}{Fluence threshold for the source at the 95\% confidence level (only upper transit). }
\tablenotetext{d}{Burst rate (in upper transit; if applicable) scaled to a fluence threshold of 5 Jy ms. }
\tablenotetext{e}{This source's primary transit does not fall within the FWHM of any of CHIME's synthesized beams. Hence, according to the formalism applied in this article, the exposure, number of bursts, and rate for this source is zero.}
\tablenotetext{f}{The particularly high exposure to this source's position is due to its high ($\sim$89\degree) declination.}
\end{deluxetable}

\bibliography{frbrefs, refs}
\bibliographystyle{aasjournal}

\FloatBarrier

\end{document}